\documentclass{aa}

\usepackage{graphicx}
\usepackage{pdflscape}
\usepackage{txfonts}
\usepackage{natbib}
\usepackage{color}
\usepackage{amssymb}
\usepackage{mathrsfs}
\usepackage{multirow}
\usepackage{appendix}

\begin{document}

  \title{Imaging the disc rim and a moving close-in companion candidate in the pre-transitional disc of V1247~Orionis
  \thanks{Based on observations made with the Keck observatory (NASA programme IDs N104N2 and N121N2),
with ESO telescopes at the Paranal Observatory (ESO programme IDs 60.A-9356(A) and 090.C-0904(A)),
and with the Submillimeter Array (SMA programme ID 2012B-S032).}}

   \author{Matthew Willson \inst{1,2}
          \and  
Stefan Kraus \inst{1}
\and
Jacques Kluska \inst{1,3}
\and
John D.\ Monnier \inst{4}
\and
Michel Cure \inst{5}
\and
Mike Sitko \inst{6,7}
\and
Alicia Aarnio \inst{8}
\and
Michael J.\ Ireland \inst{9}
\and
Aaron Rizzuto \inst{10}
\and
Edward Hone \inst{1}
\and
Alexander Kreplin \inst{1}
\and
Sean Andrews \inst{11}
\and
Nuria Calvet \inst{4}
\and
Catherine Espaillat \inst{12}
\and
Misato Fukagawa \inst{13}
\and
Tim J.\ Harries \inst{1}
\and
Sasha Hinkley \inst{1}
\and
Samer Kanaan \inst{5}
\and
Takayuki Muto \inst{14}
\and
David J.\ Wilner \inst{11}
          }

   \institute{$^{1}$~University of Exeter, Astrophysics Group, School of Physics, Stocker Road, Exeter EX4 4QL, UK\\
   $^{2}$~Department of Physics and Astronomy, Georgia State University, GA 30303, USA\\
    $^{3}$~Instituut voor Sterrenkunde (IvS), KU Leuven, Celestijnenlaan 200D, 3001 Leuven, Belgium\\
    $^{4}$~Department of Astronomy, University of Michigan, 311 West Hall, 1085 South University Ave, Ann Arbor, MI 48109, USA\\
    $^{5}$~Instituto de Física y Astronomía, Universidad de Valparaiso. Av. Gran Bretana 1111, Valparaiso, Chile \\
    $^{6}$~Department of Physics, University of Cincinnati, Cincinnati, OH 45221, USA\\
    $^{7}$~Space Science Institute, 475 Walnut St., Suite 205, Boulder, CO 80301, USA\\
    $^{8}$~University of North Carolina Greensboro, Greensboro, NC, 27402,  USA\\
    $^{9}$~Research School of Astronomy and Astrophysics, Australian National University,
Canberra, ACT 2611, Australia\\
    $^{10}$~Department of Astronomy, The University of Texas at Austin, Austin, TX 78712, USA \\
    $^{11}$~Harvard-Smithsonian Center for Astrophysics, 60 Garden Street, Cambridge, MA 02138, USA \\
    $^{12}$~Department of Astronomy, Boston University, 725 Commonwealth Avenue, Boston, MA 02215, USA\\
    $^{13}$~National Astronomical Observatory of Japan, 2-21-1 Osawa, Mitaka, Tokyo 181-8588, Japan \\
    $^{14}$~Division of Liberal Arts, Kogakuin University, 1-24-2, Nishi-Shinjuku, Shinijuku-ku, Tokyo, 163-8677, Japan \\
             }

   \date{Received ; accepted }


  \abstract
   {
    V1247\,Orionis harbours a pre-transitional disc with a partially cleared gap.
    Earlier interferometric and polarimetric observations revealed
    strong asymmetries both in the gap region and in the outer disc.
    The presence of a companion was inferred to explain these asymmetric structures and the ongoing disc clearing.
   }
   {
   Using an extensive set of multi-wavelength and multi-epoch observations we aimed to identify the origin of the previously detected asymmetries.
   }
   {
    We have observed V1247~Ori at three epochs spanning $\sim678$ days using sparse aperture masking interferometry with Keck/NIRC2 and VLT/NACO.
    In addition, we search for signs of accretion through VLT/SPHERE-ZIMPOL spectral differential imaging in H$\alpha$ and R-band continuum.
    Our SMA sub-millimetre interferometry in 880\,$\mu$m continuum and in the CO(3-2) line allows us to constrain the orientation and direction of rotation of the outer disc.
    }
   {
    We find the L'-band emission to be dominated by static features which trace forward-scattered dust emission from the inner edge of the outer disc located to the north-east.
    In H- and K-band, we see evidence for a companion candidate that moved systematically by 45$^{\circ}$ within the first $\sim$345\,days.
    The separation of the companion candidate is not well constrained, but the observed position angle change is consistent with Keplerian motion of a body located on a 6\,au orbit.
    From the SMA CO moment map, the location of the disc rim, and the detected orbital motion, we deduced the three-dimensional orientation of the disc.
    We see no indication of accretion in H$\alpha$ and set upper limits for an accreting companion.
   }
   {
    The measured contrast of the companion candidate in H and K is consistent with an actively accreting protoplanet.
    Hence, we identify V1247\,Ori as a unique laboratory for studying companion-disc interactions and disc clearing.
   }


    \titlerunning{Imaging the disc rim and companion candidate in V1247~Ori}

    \authorrunning{M.\ Willson~et~al.}

  \maketitle
%

\section{Introduction}
\label{sec:intro}

The origins of the formation of planets are thought to lie in the growth of dust particles through steady aggregation within protoplanetary discs surrounding accreting protostars as they near the main sequence \citep{1996Icar..124...62P}.
Early indicators of this process were found within the spectral energy distributions (SEDs) of discs where dramatic drops in the flux at near-infrared (NIR) and mid-infrared (MIR) wavelengths from the disc were seen compared to the flux from a classical T Tauri-type disc \citep{1989AJ.....97.1451S, 1990AJ.....99.1187S}.
This lack of infrared excess was interpreted to be caused by the processing of the dusty material within the inner and mid regions of the disc either through removal or settling to the mid-plane of the dust particles. For objects where the clearing has resulted in an inner disc cavity, the classification of transitional discs is typically used \citep{1989AJ.....97.1451S}, while gapped discs are typically referred to as pre-transitional discs \citep{2007ApJ...670L.135E}. Suggested mechanisms for the clearing include grain growth \citep{2008A&A...480..859B, 2011A&A...525A..11B}, the powerful stellar radiation field driving the photoevaporation of small dust grains \citep{1994ApJ...428..654H, 1997ApJ...480..344G, 2014prpl.conf..475A}, or the dynamical clearing of the orbital path of a forming giant planet through gravitational interactions. Both classes of transitional disc are believed to contain nascent planetary systems still in the process of formation \citep{2014prpl.conf..497E}.

As planetary embryos grow they are expected to influence and be influenced by their parent disc, migrating inwards until trapped by a pressure gradient or when sufficiently massive to carve out gaps and induce their own pressure gradients in the disc \citep{2007prpl.conf..655P, 2006Icar..181..587C, 2012ARA&A..50..211K}.
Once they have become massive enough to clear a gap along the line of their orbit they are still young and hot but also likely still undergoing hydrodynamic collapse and accreting substantial amounts of material from their surroundings through a hot, extended circumplanetary disc onto the core as they undergo oligarchic growth \citep{1996Icar..124...62P,2012MNRAS.427.2597A}. This makes them significantly brighter in the infrared than at later stages in their lives, potentially allowing them to be accessible to observation with high contrast imaging techniques.
For example, an accretion disc around a Jupiter mass protoplanet with accretion rate $\dot M = 10^{-5}\,M_{J}yr^{-1}$ is estimated to be comparable in absolute magnitude to a late M/early L-type brown dwarf \citep{0004-637X-799-1-16}. Even after this initial burst of accretion and clearing of the gap, substantial amounts of material continue to flow from the outer disc onto the central star \citep[$10^{-9}\,M_{\odot}yr^{-1}$;][]{2007MNRAS.378..369N}, forming a potential source of material for continued accretion onto a companion within the disc.

Achieving simultaneously the required spatial resolution and contrasts necessary to directly image such objects remains a challenge. Typically, inner working angles in the 30-50\,milliarcsecond (mas) range are necessary to observe within ten\,astronomical units (au) for most of the nearby young stars, while contrasts of $\sim$6\,mag are needed when observing in near-infrared wavelengths (H-, K-, L'-band) for detecting a moderately accreting circumplanetary disc. Sparse aperture masking (SAM) interferometry has been successful in this respect, first observing and confirming the accreting protoplanet LkCa\,15\,b \citep{2012ApJ...745....5K, 2015Natur.527..342S} and identifying companion candidates in five additional discs (T\,Cha: \citealt{2011A&A...528L...7H}; HD\,142527: \citealt{2012ApJ...753L..38B}; DM\,Tau, LkH$\alpha$\,330, TW\,Hya: \citealt{2016arXiv160803629W}) and additionally identifying structures consistent with emission from a heated disc wall under intermediate inclinations (FL\,Cha: \citealt{2013ApJ...762L..12C}; T\,Cha: \citealt{2013A&A...552A...4O}; FP\,Tau: \citealt{2016arXiv160803629W}).

However, the detection of companions with SAM interferometry can be complicated by the presence of disc-related asymmetries in the brightness distribution that can be difficult to discern from the phase signals of a close-in companion. Such disc-related asymmetries could be caused by spiral arm structures or disc warps created by the presence of an unseen companion \citep{2010A&A...519A..88A,2009ApJ...704L..15M}.

A useful way to distinguish disc asymmetries from an accreting protoplanet is to search directly for on-going accretion within the disc itself. H$\alpha$ emission is one such accretion tracer. A compact object accreting material at a moderate rate is expected to be considerably brighter in H$\alpha$ than in nearby continuum wavelengths and in comparison to the surrounding disc \citep{2012A&A...548A..56R}. In this way, evidence for HD142527\,b \citep{2014ApJ...781L..30C} and LkCa\,15\,b \citep{2015Natur.527..342S} as continued accretors was found.

V1247\,Orionis is part of the Orion OB1\,b association \citep{1981AJ.....86.1057G} whose distance was estimated to 319$\pm$27 \citep{2016yCat.1337....0G}. \citet{2008A&A...485..931C} estimate the age of the star to be 5-10 Myr. \citet{2010A&A...511L...9C} observed two deep UX\,Ori-like occultation events which were attributed to the obscuring effect of disc material passing in front of the star. They additionally identified in the SED a substantial drop in the mid-infrared excess between 3-15$\mu$m, marking V1247\,Ori as a pre-transitional disc. \citet{2013ApJ...768...80K} determined the spectral type to F0V with HARPS spectroscopy.

The pre-transitional disc of \object{V1247\,Ori} is of particular interest due to previous interferometric observations that revealed complex structures within the disc gap extending from within an au out to 37\,au \citep{2013ApJ...768...80K}.
This study found that the unusually extended mid-infrared emission around V1247\,Ori could be explained with the presence of optically thin carbonaceous dust grains located within the gap region, suggesting that the disc is still undergoing the process of clearing its gap.
Complementary SAM observations indicated small-scale asymmetries across multiple wavebands which could not be fitted with a single point-source companion model.
These observations were interpreted as a potential spiral density wave in the dust distribution within the gap.

\begin{table}
\caption{V1247 Ori Parameters}
\label{tab:targlist}
\centering
\vspace{0.1cm}
\begin{tabular}{c c c }
\hline\hline
 Parameter  & Unit  & Value \\
 \hline
 Association            &                   & Orion OB1\,b \\
 Distance               & [pc]              & 319$\pm$27   \\
 Spectral Type          &                   & F0V   \\
 Visual Extinction      &                   & 0.02   \\
 Effective Temperature  & [K]               & 7250  \\
 Stellar Mass           & [M$_{\odot}$]     & 1.86  \\
 Stellar Radius         & [R$_{\odot}$]     & 2.3   \\
 Accretion Luminosity   & [L$_\odot$]       & 1.3   \\
 \hline
\end{tabular}
\tablefoot{References: Association: \citet{1971A&A....14...66S, 1981AJ.....86.1057G}, Distance: \citet{2016yCat.1337....0G}, Extinction: \citet{2003AJ....126.2971V},  Remaining: \citet{2013ApJ...768...80K} }
\end{table}

Observations undertaken by \citet{2016PASJ...68...53O} utilising Subaru/HiCIAO adaptive optics polarimetric imaging detected a strong spiral arm structure to the south of the object in scattered light. The single-armed nature was interpreted as potentially due to shadowing by the rim of the disc at 46\,au. They cast doubt on this explanation, finding the radial decline in flux is not steep enough for a shadowed region. Other potential explanations were proposed such as the trapping of small dust grains within a gas vortex or a single spiral arm launched by a planetary companion.

Recent 870\,$\mu$m continuum images obtained by ALMA \citet{2017ApJ...848L..11K} has revealed more of this spiral arm. Extensively modelling the combined ALMA 870\,$\mu$m continuum images and Subaru/HiCIAO adaptive optics polarimetric images, they propose the arm to be the result of the presence of planet at roughly 100\,au from the central star. Further they identify a bright ring of material between 0$\farcs$15-0$\farcs$25, containing significant substructure, in particular to the south-east and north-west.

We present new SAM observations of this target in H, K, and L' wavebands at two additional epochs to those presented by \citet{2013ApJ...768...80K} and attempt to understand the origin of the detected asymmetries. Our observations cover 678\,days and make use of the Keck-II/NIRC2 and VLT/NACO instruments \citep{1998SPIE.3354..606L,2000SPIE.4007...72R,2010SPIE.7735E..1OT}. We attempt to directly detect accretion signatures via observations in the H$\alpha$ line with VLT/SPHERE-ZIMPOL \citep{2008SPIE.7014E..18B,2018arXiv180805008S}. To examine the position, orientation and kinematics of the outer disc we present 880$\mu m$ SMA \citep{2004ApJ...616L...1H} data that covered the $^{12}$CO(3-2) line. In Section \ref{sec:obser}, we detail our observations. We layout our methodology for interpreting the sparse aperture masking data in Section \ref{sec:modelling} and the procedure using these data to produce reconstructed images in Section \ref{sec:imgrec}. In Section \ref{sec:results} we display and examine the results of our observations. We discuss our findings in Section \ref{sec:discussion} and finally present our conclusions in Section \ref{sec:conclusions}.

%

\section{Observations}
\label{sec:obser}

\begin{table*}
\centering
\caption{Observation log}
\label{tab:obslog}
\begin{tabular}{c c c c c c}
\hline\hline
 Instrument & Date & Spectral setup & Configuration & Calibrator & DIT (s) $\times$ NDIT\\
            & [dd/mm/yyyy] & 		& 	   &    &  \\
\hline

 Keck/NIRC2 & 10/01/2012 & H  & 9-Hole & \object{HD\,36811}, \object{HD\,37331} & 20$\times$48  \\
 VLT/NACO   & 18/12/2012 & H  & 7-Hole & \object{HD\,37295} & 60$\times$17 \\
 Keck/NIRC2 & 09/01/2012 & K' & 9-Hole & HD\,37331, \object{HD\,38406} & 20$\times$48 \\
 Keck/NIRC2 & 10/01/2012 & K' & 9-Hole & HD\,36811 & 40$\times$12 \\
 VLT/NACO   & 18/12/2012 & Ks & 7-Hole & \object{HD\,290679}, \object{HD\,290735}, \object{HD\,37634} & 84$\times$34 \\
 &&&&\object{HD\,38406}   \\
 Keck/NIRC2 & 20/10/2013 & K' & 9-Hole & HD\,37331, HD\,37634 & 20$\times$24 \\
 Keck/NIRC2 & 10/01/2012 & L' & 9-Hole & HD\,37634, HD\,38406 & 20$\times$42 \\
 VLT/NACO   & 18/12/2012 & L' & 7-Hole & HD\,36696 & 30$\times$32 \\
 Keck/NIRC2 & 16/11/2013 & L' & 9-Hole & \object{HD\,290685}, HD\,37634, HD\,38406 & 15$\times$72 \\
 	\hline
 SPHERE/ZIMPOL	& 04/12/2014	& B\_H$\alpha$/CntHa	& Coron.\ mask	& --- & 40 $\times$ 15\\
 SPHERE/ZIMPOL	& 06/12/2014	& B\_H$\alpha$/CntHa	& Coron.\ mask	& --- & 120 $\times$ 18 \\
 SPHERE/ZIMPOL	& 06/12/2014	& B\_H$\alpha$/CntHa	& No mask	& --- & 40 $\times$ 21 \\
 SPHERE/ZIMPOL	& 09/12/2014	& B\_H$\alpha$/CntHa	& Coron.\ mask	& ---& 40 $\times$ 21\\
 & & & & & 160 $\times$ 7 \\
 SPHERE/ZIMPOL	& 09/12/2014	& B\_H$\alpha$/CntHa	& No mask	& --- & 40 $\times$ 36 \\
 \hline
 SMA    &    27/11/2012	& 880\,$\mu$m	& COM	& 0532+075, 0607-085, Uranus, 3c84 \\
 SMA    &    30/12/2012	& 880\,$\mu$m	& EXT	& 0532+075, 0607-085, Uranus, 3c84 \\
 SMA    &    04/02/2013	& 880\,$\mu$m	& VEX	& 0532+075, 0607-085, Callisto, 3c279 \\
 \hline
 \end{tabular}
\end{table*}

\subsection{VLT/NACO + Keck-II/NIRC2 Sparse Aperture Masking Interferometry}

Our observations were obtained with the NIRC2 instrument at the 10\,m Keck-II telescope (Keck Programme IDs, N104N2 \& N121N2) situated on the summit of Mauna Kea, Hawaii and the NACO (ESO Programme ID, 090.C-0904(A)) instrument on the 8.2\,m UT4 at the Paranal Observatory in Chile \citep{2003SPIE.4839..140R, 2003SPIE.4841..944L}. We utilise SAM interferometry to remove atmospheric phase noise via the construction of closure phases. This allows us to resolve structures at or even below the diffraction limit of the telescope while maintaining adequate contrasts down to 6\,mag within 100\,mas in the near infrared under ideal conditions.

Our data sets were obtained on five nights during three observing campaigns in January 2012 (NIRC2), November 2012 (NACO), and November 2013 (NIRC2), as summarised in Table \ref{tab:obslog}. We utilised the nine-hole and seven-hole masks with NIRC2 and NACO respectively, which provide a good compromise between sensitivity and $uv$-coverage.
Details about the hole geometry can be found in \citet{2010SPIE.7735E..1OT} for the NACO mask and on the NIRC2 website (\url{https://www2.keck.hawaii.edu/inst/nirc2/nonRedundantApMask.pdf}) for the NIRC2 instrument mask (See Figure~\ref{fig:keck9hMask}).
We observed with K'/Ks and L' filters at all three epochs with additional observations being conducted in the H-band during the first two epochs. The K' and Ks filters have very similar transmission profiles, namely $\lambda_{0,\textrm{K'}}=2.12\mu$m, $\Delta\lambda_{\textrm{K'}}=0.35\mu$m and $\lambda_{0,\textrm{Ks}}=2.18\mu$m, $\Delta\lambda_{\textrm{Ks}}=0.35\mu$m. 
Also the L'-band filters employed by NACO ($\lambda_{0,\textrm{L'}}=3.80\mu$m, $\Delta\lambda_{\textrm{L'}}=0.62\mu$m) and NIRC2 ($\lambda_{0,\textrm{L'}}=3.776\mu$m, $\Delta\lambda_{\textrm{L'}}=0.700\mu$m) cover similar wavelength bands and can therefore be combined for our interpretation. Similarly,
the H-band filters of NACO ($\lambda_{0}=1.66\mu$m, $\Delta\lambda=0.33\mu$m) and NIRC2 ($\lambda_{0}=1.633\mu$m, $\Delta\lambda=0.296\mu$m) approximately cover the same spectral window.

The best seeing was obtained on the night of 18/12/2012 with mean seeing around 0$\farcs$5. Mean seeing conditions of 0$\farcs$7 were obtained on 09/01/2012 and 16/11/2013 while conditions on 10/01/2012 and 20/10/2013 were inferior with mean conditions of 1$\farcs$0 and $0\farcs9$. The Keck-II/NIRC2 K' data set acquired on the night of 10/01/2012 is not included within our analysis due to the poor seeing conditions on that night and the low number of recorded frames. When combined with the K' data set obtained from the night before, detection limits were not improved.

In order to correct for atmospheric effects, we monitor the instrument transfer function by bracketing our science observations with observations on unresolved calibrator stars. We observed multiple calibrator stars, which allows us to still calibrate our data even in the case when a calibrator star is found to be resolved and needed to be rejected (i.e.\ due to a previously unknown binary companion).

To extract the visibility amplitudes and phases from the NIRC2 and NACO data, we use the data reduction pipeline previously described in \citet{2008ApJ...678L..59I}, \citet{2013ApJ...768...80K} and \citet{2016arXiv160803629W}. SAM interferometry uses relatively long integration times compared to long baseline interferometry, which results in a poor calibration of the transfer function for the visibility amplitudes. To correct for this effect, we renormalise the visibilities by fixing the shortest baselines to a squared visibility of 1.

\subsection{VLT/SPHERE spectral differential imaging}
\label{subsec:sphere}

We attempt to detect on-going accretion onto a companion through high resolution imaging in H$\alpha$ with ZIMPOL \citep{2010SPIE.7735E..4BR} from the VLT/SPHERE instrument mounted on the 8\,m UT4 at the Paranal Observatory, Chile. The observations were obtained as part of SPHERE science verification (ESO programme ID 60.A-9356(A), P.I.\ S.\ Kraus) and spanned three nights (2014-12-04, 2014-12-06, 2014-12-09) with the majority of data taken on 2014-12-09.
In order to detect accretion signatures we observed with ZIMPOL's spectral differential imaging (SDI) mode simultaneously in H$\alpha$ and in the continuum around the line using the \texttt{B\_Ha} ($\lambda_0=$ 656.6nm, $\Delta\lambda=$ 5.5nm) and \texttt{CntHa} ($\lambda_0=$ 644.9nm, $\Delta\lambda=$ 4.1nm) filters.
We observe V1247\,Ori in pupil-stabilised mode both with (8$^\circ$ of field rotation) and without a coronagraphic mask (18$^\circ$ of field rotation). Coronagraphic images (with the \texttt{V\_CLC\_M\_WF} Lyot coronagraph; diameter 155\,mas) were taken on all three nights and AO-assisted images without mask were taken on the second and third night (2014-12-06 and 2014-12-09).
The dicroic beam-splitter was used.

We achieved the best seeing conditions on 2014-12-09 with an average seeing $\lesssim 1\farcs0$. Average seeing on the remaining two nights were between $1\farcs0$ and $2\farcs0$, resulting in inferior quality data sets. The poor AO correction on these nights produced noisy data sets which did not improve the detection limits when stacked with the data set acquired on the 'good' night.  We thus present here the results on the best seeing night (i.e.\ 2014-12-09).

Our SPHERE images were reduced using the Esorex SPHERE pipeline v0.22 \citep{2010SPIE.7740E..22M}. After cleaning (bias, flat, bad pixels) and derotating each frame, we analysed the data using customised scripts separating the images with and without the coronagraph.
The H$\alpha$ and continuum frames were stacked after de-rotating for sky rotation(see Figure\,\ref{fig:ZIMPOLimages}, top-left).
The H$\alpha$ images were re-scaled so that wavelength dependent location of their speckle halos match the ones in the continuum images.
Then, the H$\alpha$ images were normalised to the continuum images by finding the optimal flux factor that minimises the differences between the two filters in a circular area with a size of two PSF.
The continuum images were then subtracted from the H$\alpha$ ones.
To normalise the coronagraphic images we used an area with a radius of 250\,mas around the coronagraphic mask.
In the resulting image without coronagraphic mask we found a negative point source corresponding to a positive emission in the CntHa filter (Figure\,\ref{fig:ZIMPOLimages},bottom-left).
The contrast of this point source is about 10$^3$ and it is located at $\sim$110\,mas from the central star (see Figure \ref{fig:ZIMPOLimages}). No such feature is present in the images with coronagraphic mask (see Figure\,\ref{fig:ZIMPOLimages}, right). This feature is likely to be a ghost and was already seen in other ZIMPOL observations (H.\ M.\ Schmid, private communication).

\subsection{SMA Sub-millimetre Interferometry}
\label{sec:SMA}

V1247\,Ori was observed using the Submillimeter Array (SMA) interferometer at Mauna Kea, Hawaii in the compact (2012-11-28), extended (2012-12-31), and very extended (2013-02-05) configurations, with baseline lengths from 8--509\,m (programme 2012B-S032; PI S.~Kraus).  The correlator was configured to process two 2\,GHz-wide IF (intermediate frequency) bands centred at $\pm$5 and 7\,GHz from the LO (local oscillator) frequency of 340.8\,GHz (880\,$\mu$m).  The CO $J$=3$-$2 transition at 345.796\,GHz was centred in the upper sideband of the lower IF band, with a channel spacing of 0.35\,km s$^{-1}$.  Observations of V1247\,Ori were interleaved with regular visits to \object{J0532+075} and \object{J0607-085} for use in gain calibration.  Additional observations of the bright quasars 3C 279 and \object{3C\,84}, as well as \object{Uranus}, \object{Callisto}, and \object{Titan} were made (depending on availability and the array configuration) for bandpass and flux calibration (see Table~\ref{tab:obslog}).

The raw visibility data were calibrated using standard procedures in the {\tt MIR} software package (\url{https://www.cfa.harvard.edu/~cqi/mircook.html}).  The calibrated visibilities were then Fourier inverted assuming natural weighting, deconvolved with the {\tt clean} algorithm \citep{1974A&AS...15..417H}, and restored with the synthesised beam ($0\farcs7\times0\farcs6$ at PA = 45\degr).  The resulting continuum map has an root-mean squared (RMS) noise level of 0.85\,mJy beam$^{-1}$; the CO channel maps have an RMS of 55\,mJy beam$^{-1}$ in each channel.

%

\section{Image Reconstruction}
\label{sec:imgrec}

To retrieve the inner brightness distribution of V1247\,Ori from our SAM interferometry in a model-independent fashion, we employed image reconstruction techniques developed for long-baseline infrared interferometry. The image reconstruction from a set of discrete aperture masking measurements can be considered an inverse problem, where the aperture masking measurements represent points in the Fourier plane. Given that there are more pixels to be reconstructed than available measurements, there exists no unique solution to the problem and so we must use a Bayesian approach to find the best result. To this end we minimise a global cost function ($\mathscr{F}$) defined as:
\begin{eqnarray}
\label{eqn:cost}
\mathscr{F} = \mathscr{F}_\mathrm{data} + \mu \mathscr{F}_\mathrm{rgl},
\end{eqnarray}
where $\mathscr{F}_\mathrm{data}$ is the likelihood term (here the $\chi^2$), $\mathscr{F}_\mathrm{rgl}$ is the regularisation term and $\mu$ the regularisation weight \citep[for more information, see][]{Thiebaut2008,Renard2011}. The likelihood term ensures that the resulting image reproduces the data, while the regularisation term interpolates between the gaps in the Fourier plane in a way that is determined by the user. In addition, the regularisation term aids the algorithm with the convergence to the most likely a-posteriori estimate of the image.

The MiRA algorithm \citep{Thiebaut2008} was chosen to perform our image reconstructions. This algorithm utilises a downhill gradient method to minimise the cost function $\mathscr{F}$. In order to image the environment around V1247\,Ori we modelled the regions within 1\,au as a point source and reconstruct only an image of the environment, similar to the approach described in \citet{Kluska2014}.

We define the images to have 128$\times$128 pixels each and choose the pixel sizes as 5, 7 and 11\,mas for H, K and L'-bands respectively. We selected the quadratic smoothness regularisation \citep{Renard2011} and utilised the L-curve method \citep[see][for more details]{Renard2011,Kluska2014} to weight the regularisation term. We then averaged the regularisation weight of all L-curves and found a value of $\mu=10^9$.

To define the fraction of the stellar flux in the parametric model, we made a grid of reconstructions with different flux ratios for the star. As we are minimising the global cost function $\mathscr{F}$, we should have chosen the image with the minimum $\mathscr{F}$ value. Because of the regularisation effects, these images still have flux at the star position which is not physical. Therefore we decided to keep the flux ratio for which the image has the smallest likelihood term ($\mathscr{F}_\mathrm{data}$). These images do not differ significantly from the images with smaller $\mathscr{F}$ except in correcting this effect (see Fig.\,\ref{fig:img_f}).

%

\section{Sparse Aperture Masking Model Fitting}
\label{sec:modelling}

\subsection{Companion Model}
\label{subsec:modellingComp}

To search for a companion in our data, we fit a binary model comprised of two point sources \citep{2012ApJ...745....5K}, where we treat the separation ($\rho$), positional angle (PA), and contrast ($f$) as free parameters. Instead of fitting to a combination of closure and squared visibilities, we decided to fit to the closure phase data alone. This is motivated by the much smaller uncertainties in the measured closure phases, which makes them better-suited when searching for close-in companions. The best-fit parameters for the H- and K'-band data sets are listed in Table~\ref{tab:binresultsHK} and the L-band are shown in Table~\ref{tab:binresultsL}.

The binary model is only adequate if the brightness distribution resembles two point sources. More complex systems, for instance with multiple companions or extended disc emission will deviate the closure phase signal away from the simple binary case and result in general in a poorer binary fit. Constructing significance maps allows us to determine when the simple binary scenario is formally no longer an effective description of the observed brightness distribution.

In order to construct the significance maps, we fit the following equation across a grid of positions, 400$\times$400\,mas in RA and Dec with a resolution of 1\,mas and the primary star located in the centre of frame:

\begin{equation}
    \centering
    \label{eq:BinaryComplexVisibilities}
    V(u, v) = - \frac{1 + f \exp(- 2 \pi i (u \alpha + v \beta))}{1 + f},
\end{equation}

where $f$ corresponds to the flux ratio of the model companion over the primary star, $u$ and $v$ are the Fourier plane coordinates (x/$\lambda $ and y/$\lambda$ respectively) and $\alpha$ and $\beta$ are the angular coordinates of the companion within the model \citep{2007NewAR..51..576B}.  We find the best fit contrast at each position and calculate a significance for that fit, enforcing positive fluxes with a contrast ratio of less than 1.

\begin{table*}
\centering
\caption{Sparse aperture masking binary fit results - H+K band}
\label{tab:binresultsHK}

\begin{tabular}{c c c c c c | c c c c c}
\hline\hline
 Filter & Date & Separation & PA & Contrast & Significance & \multicolumn{4}{c}{Detection limits (99\%)} \\
 & [dd/mm/yy] & [mas] & [$^{\circ}$] & [mag] & [$\sigma$] & 20-40 & 40-80 & 80-160 & 160-240  \\
\hline
 H  & 10/01/2012 &  59$\pm$4  &  93$\pm$4 & 6.0$\pm$0.3 & 3.20 & 5.60 &  5.64 &  5.51 &  5.47 \\
 K' & 09/01/2012 &  44$\pm$4  & 308$\pm$3 & 5.1$\pm$0.2 & 7.01 & 4.58 &  5.44 &  5.30 &  5.14 \\
 \hline
 H  & 18/12/2012 &  40$\pm$5  & 353$\pm$3 & 4.5$\pm$0.2 & 6.91 & 4.20 &  4.96 &  4.73 &  4.55 \\
 Ks & 18/12/2012 &  81$\pm$4  &  23$\pm$3 & 5.3$\pm$0.2 & 5.06 & 2.75 &  4.60 &  4.53 &  2.69 \\
 \hline
 K' & 20/10/2013 &  75$\pm$5  &  38$\pm$3 &  6.1$\pm$0.3  & 3.83 & 5.05 &  5.89 &  5.78 &  5.69 \\
 \hline
\end{tabular}
\tablefoot{Detection limits are calculated for a range of annuli centred on the central star, namely for radii 20-40\,mas, 40-80\,mas, 80-160\,mas, and 160-240\,mas. Limits are quoted in units of $\Delta$mag between the central star and a hypothetical companion.}
\end{table*}

\begin{table*}
\centering
\caption{Sparse Aperture Masking Binary Fit Results - L band}
\label{tab:binresultsL}

\begin{tabular}{c c c c c c | c c c c c}
\hline\hline
 Filter & Date & Separation & PA & Contrast & Significance & \multicolumn{4}{c}{Detection Limits (99\%)} \\
 & [dd/mm/yy] & [mas] & [$^{\circ}$] & [mag] & [$\sigma$] & 20-40 & 40-80 & 80-160 & 160-240  \\
\hline
 L' & 10/01/2012 & 114$\pm$9  & 329$\pm$4 &  6.0$\pm$0.3  & 3.30 & 3.73 &  5.37 &  5.74 &  5.61 \\
 L' & 18/12/2012 & 117$\pm$4  &  22$\pm$2 & 5.34$\pm$0.15 & 8.28 & 3.96 &  5.44 &  6.11 &  6.14 \\
 L' & 16/11/2013 & 118$\pm$7  &  26$\pm$3 &  5.8$\pm$0.3  & 4.16 & 3.76 &  5.39 &  5.79 &  5.71 \\
 \hline
\end{tabular}
\tablefoot{Detection limits are calculated for a range of annuli centred on the central star, namely for radii 20-40\,mas, 40-80\,mas, 80-160\,mas, and 160-240\,mas. Limits are quoted in units of $\Delta$mag between the central star and a hypothetical companion.}
\end{table*}

In some cases we find the best-fit position to lie within the diffraction limit of the telescope. Here we enter a regime where the phase signals of a companion become more difficult to constrain due to degeneracies that can occur between different separation-contrast parameter combinations, where a lower contrast point-source can produce an equally good fit as lower separation/higher contrast point-source (the profile of degeneracy can be seen in Figure~\ref{fig:PotCompanions}, third column). Hence we define this as the degenerate region. The ability to constrain the separation/contrast parameter space, and hence the inner working angle, depends on the SNR of the closure phase data recorded in a specific observation. We quantified and described this effect within \citet{2016arXiv160803629W} and provided a method for deriving the profile of the degeneracy. Fortunately, the degeneracy affects only the best-fit separation value, but not the position angle when determining the position of a potential companion.

\subsection{Companion + Disc Model}
\label{sec:compDisk}

Previous SAM observations of V1247\,Ori showed significant asymmetries in the K- and L'-band that were inconsistent with a simple companion model \citep{2013ApJ...768...80K}. Long-baseline observations from the same study indicated the presence of the inner edge of the outer disc to be located on similar spatial scales as the detected asymmetry. To explore the possibility that the inconsistency could be due to the presence of a disc wall we implemented a companion + disc model.

We model the disc wall geometrically as forward scattering from the near side of a disc wall as described in \citet{2016arXiv160803629W} with the addition of a companion. We take the best-fit position found from the simple binary model fit as the initial parameters for the companion and the disc properties found by \citet{2013ApJ...768...80K} for the initial disc parameters and then adjust the parameters by hand to achieve a best fit.

%

\section{Results}
\label{sec:results}

\subsection{Sparse Aperture Masking}
\label{subsec:SAM}

\subsubsection{Disc Rim Detection in L' Band}
\label{subsubsec:LBandDisc}

\begin{figure*}
 \begin{center}
\scriptsize
$\begin{array}{ @{\hspace{-1.0mm}} c @{\hspace{-6.0mm}} c @{\hspace{-6.0mm}} c }
   \includegraphics[height=4cm, angle=0]{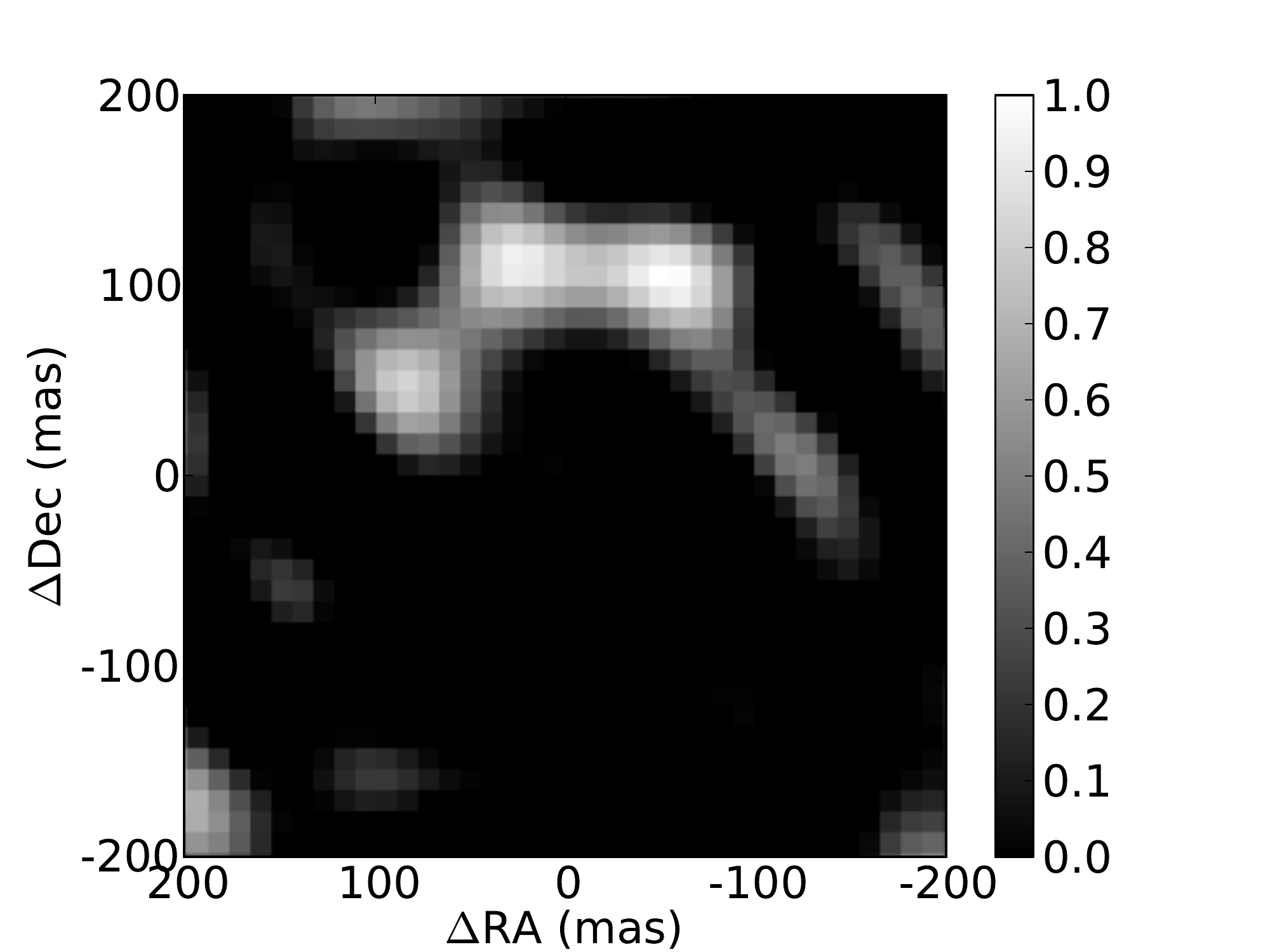} & \includegraphics[height=4cm, angle=0]{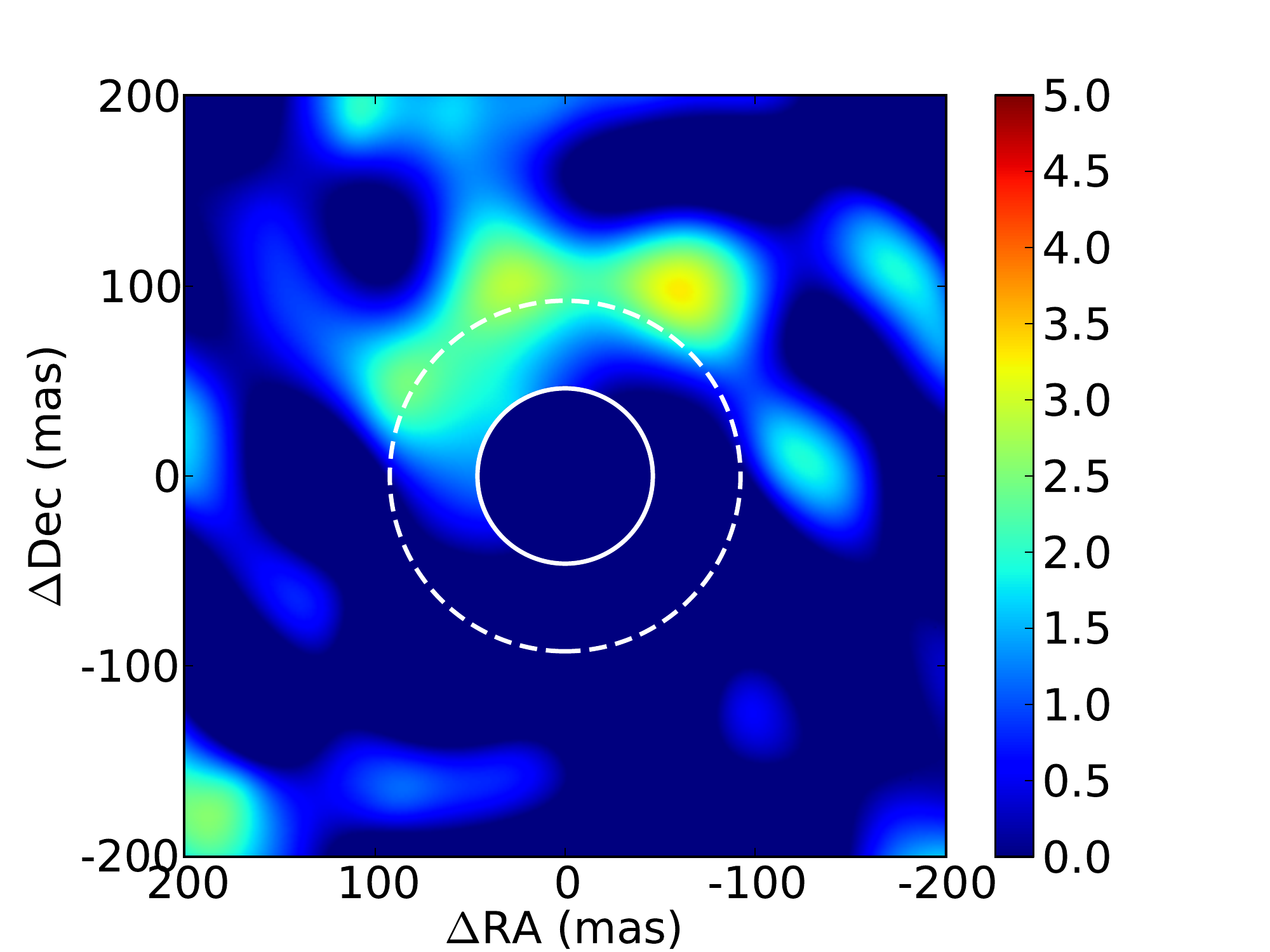} &   \includegraphics[height=4cm, angle=0]{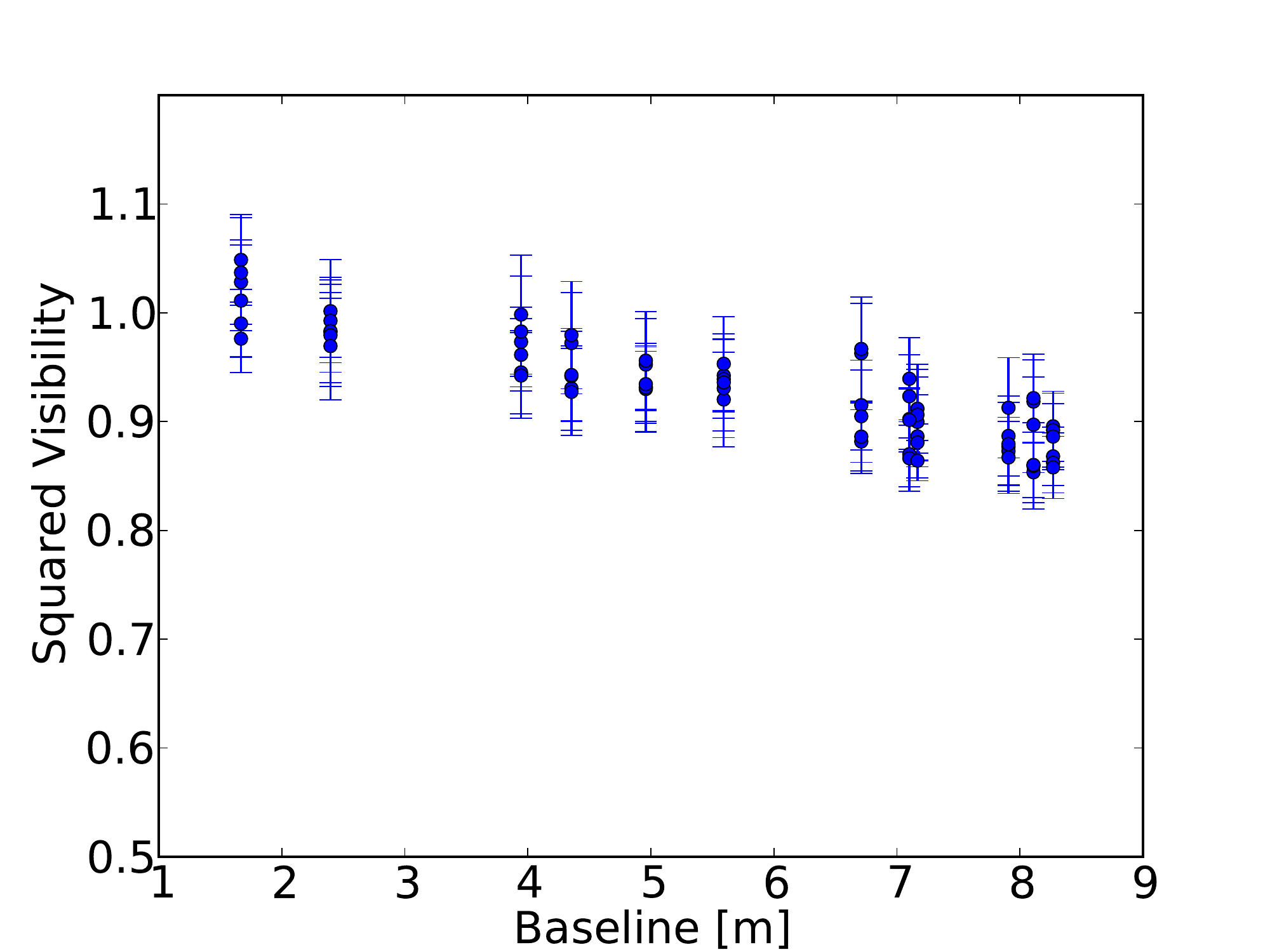}
   \\
   \includegraphics[height=4cm, angle=0]{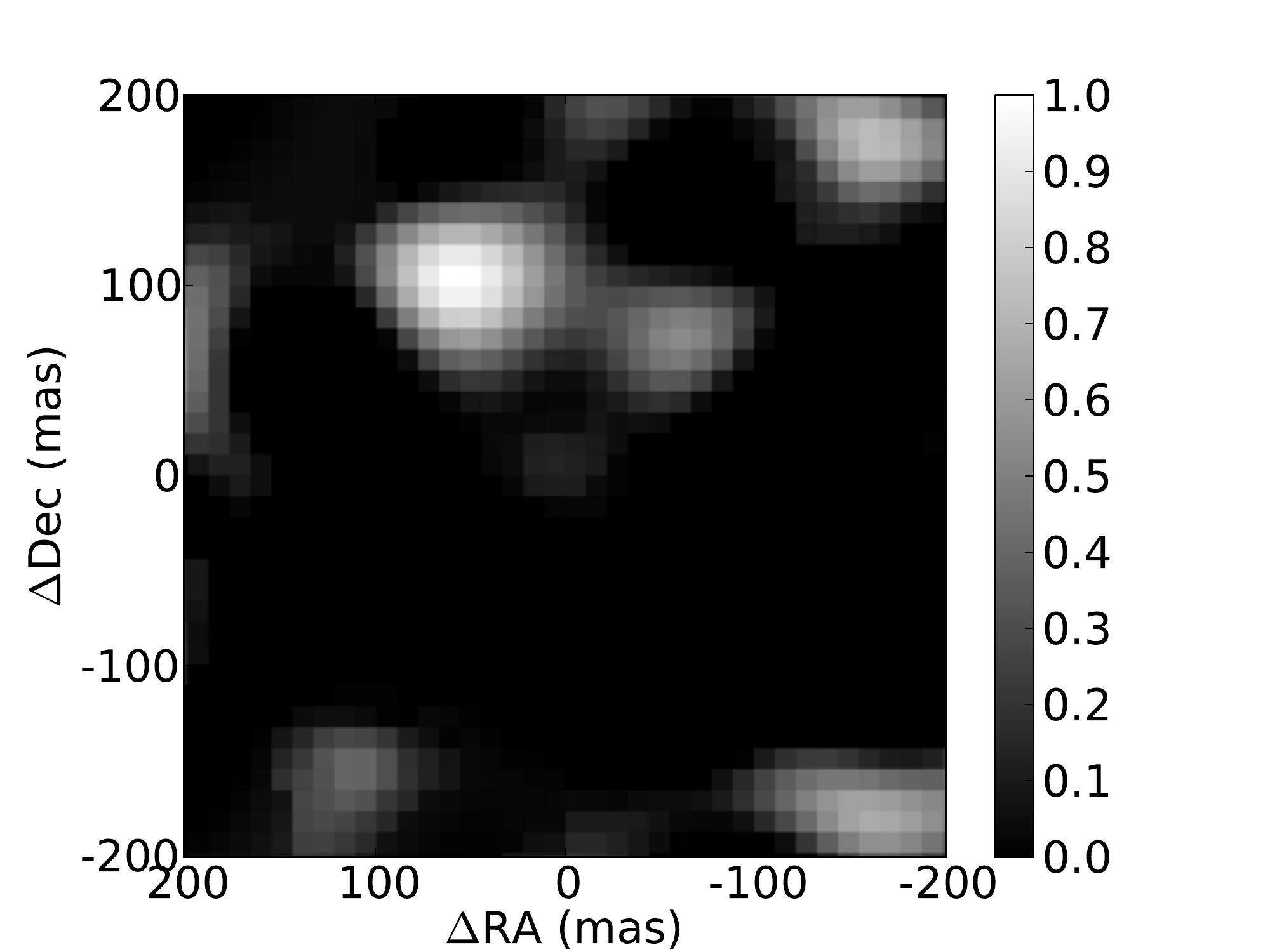} & \includegraphics[height=4cm, angle=0]{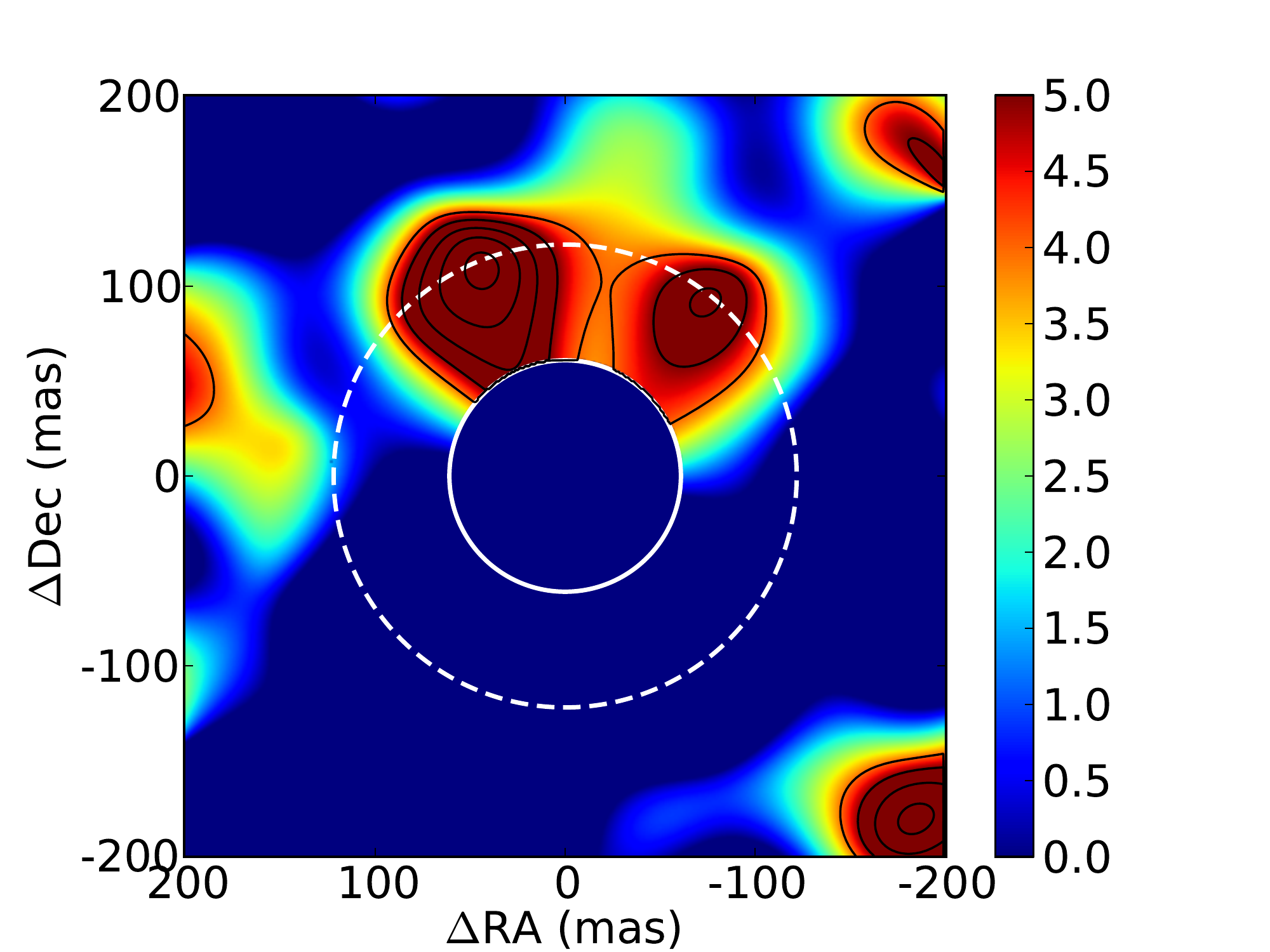} &   \includegraphics[height=4cm, angle=0]{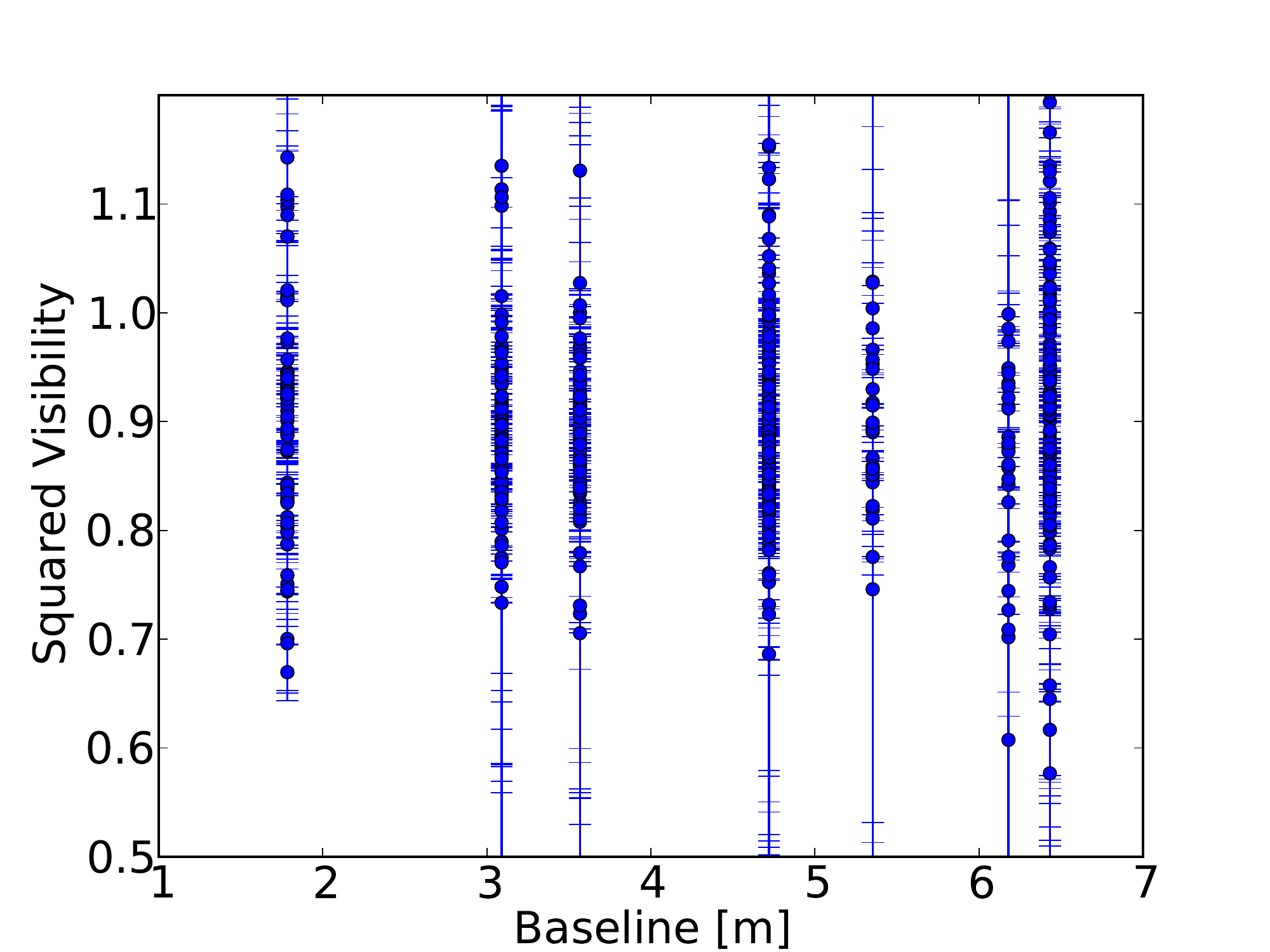}
   \\
   \includegraphics[height=4cm, angle=0]{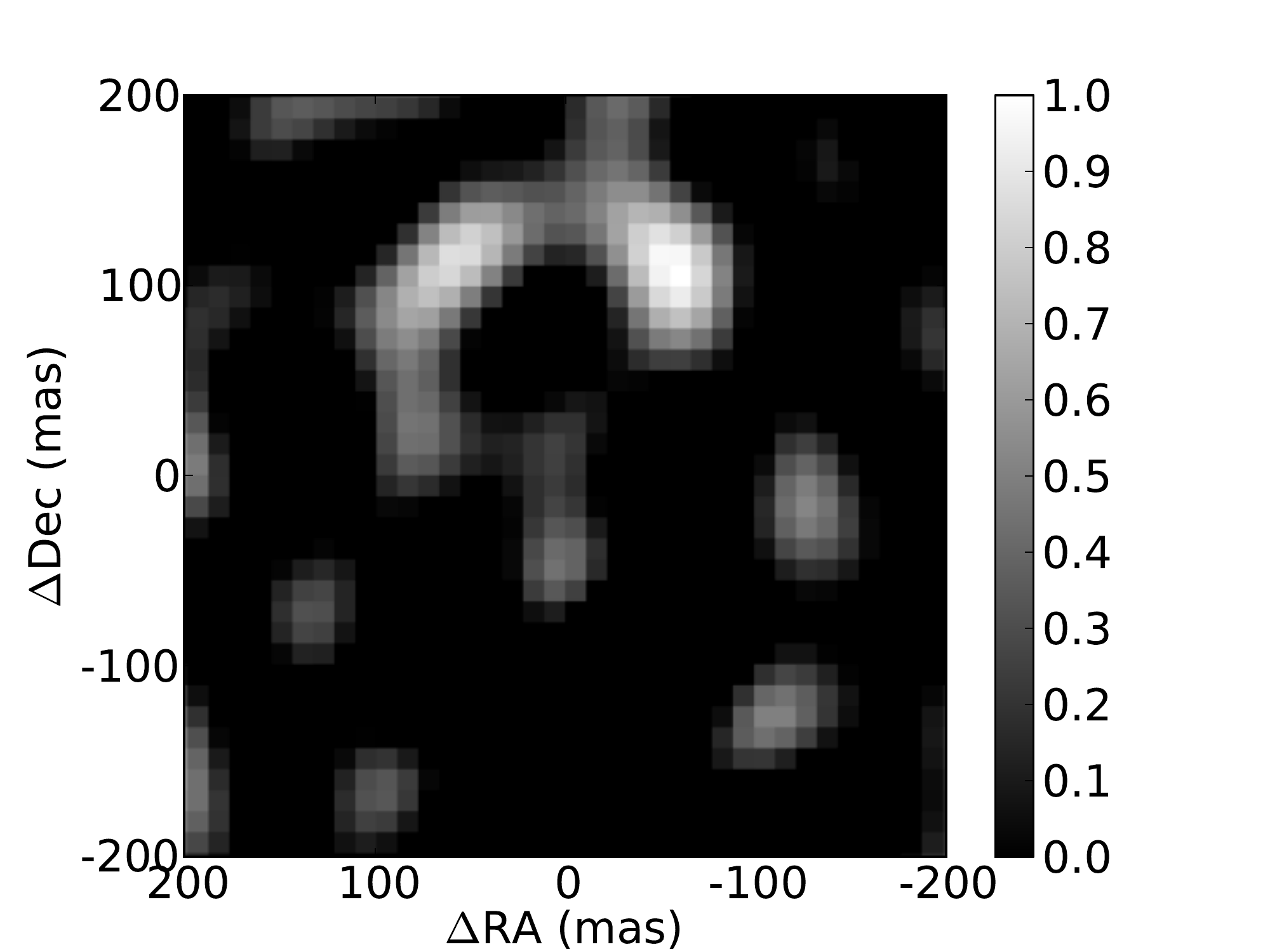} & \includegraphics[height=4cm, angle=0]{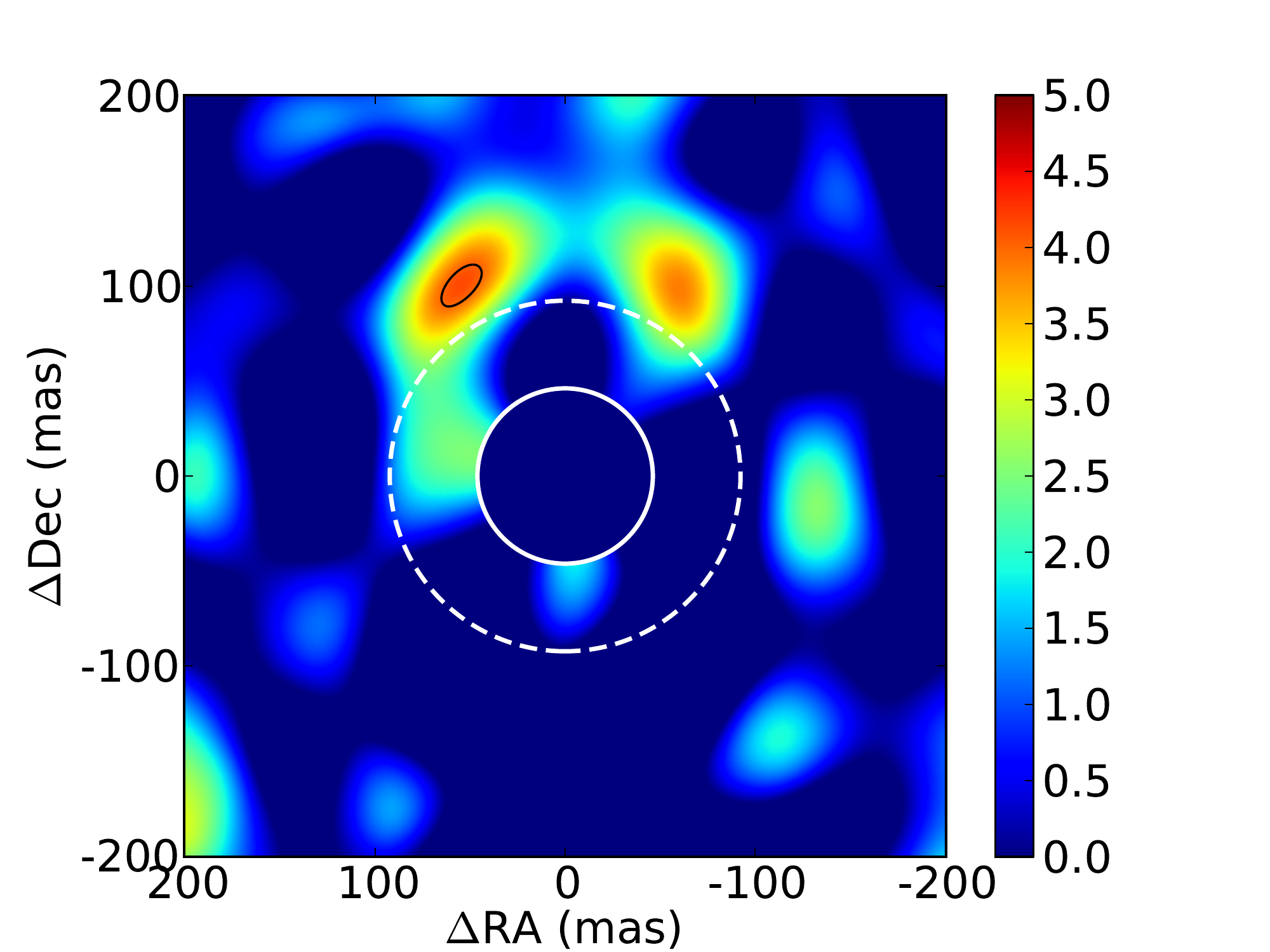} & \includegraphics[height=4cm, angle=0]{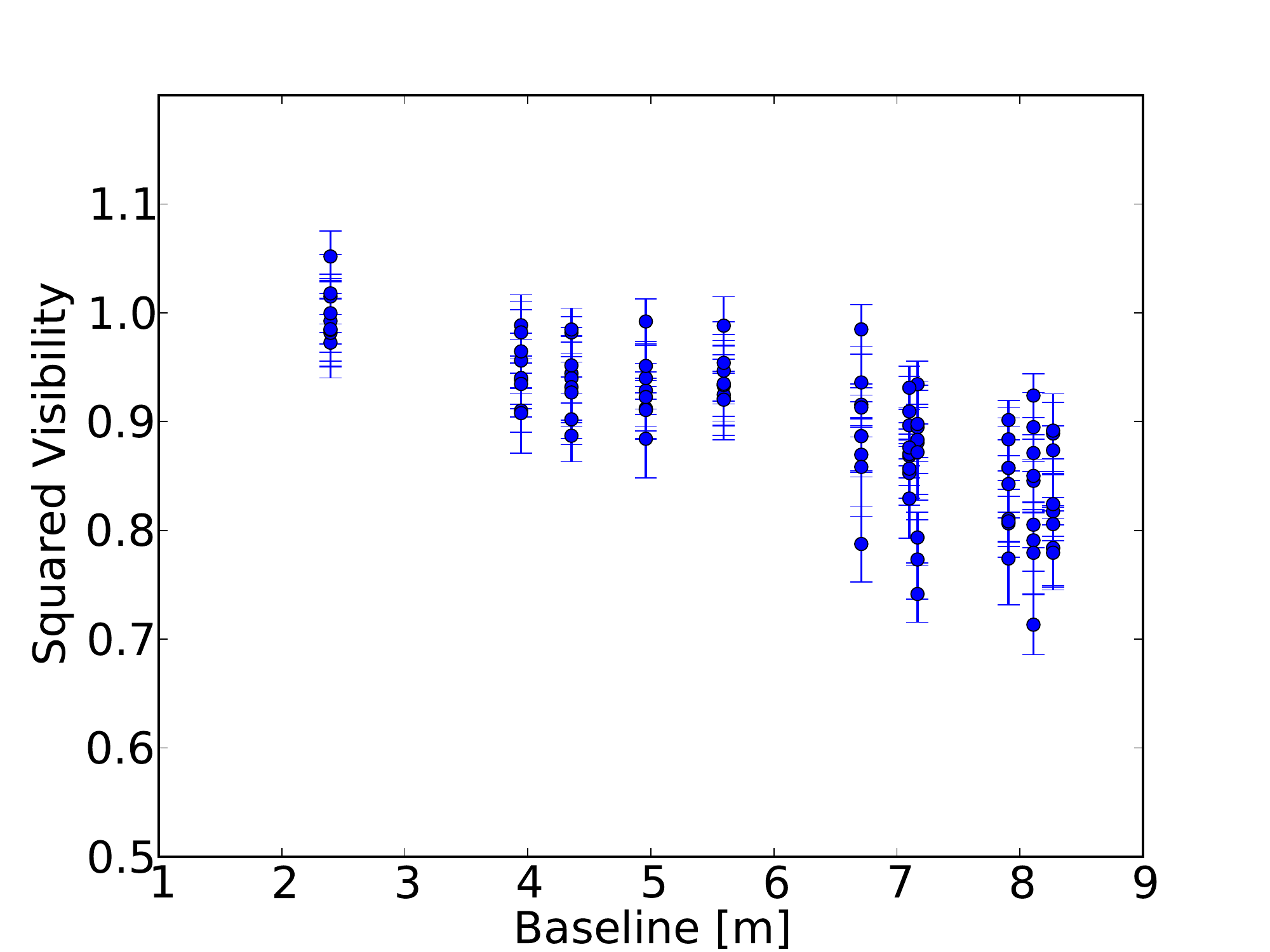}
   \\
   \\
        \end{array}$
\end{center}
\caption{Figure showing L'-band observations and the combined observations. \textbf{Left:}  Reconstructed Images \textbf{Middle:} Computed significance map \textbf{Right:} Squared visibilities; \textbf{First row:} 2012-01-10; \textbf{Second row:} 2012-12-18; \textbf{Third row:} 2013-11-16. We interpret the structures seen in all three epochs as caused by an illuminated disc rim. The colour bars in the reconstructed images indicate the flux contained in each pixel as a fraction of the peak flux in the image after subtracting the central source. The colour bars in the significance maps represent significance of the binary fit to the data at each pixel location ($\sigma$).}
\label{fig:PotDiskFeatures}
\end{figure*}

In the L'-band, we measure a similar brightness distribution at all three epochs and with a similar significance level. Based on the pair-like morphology, the separation, and static nature, we interpret these L'-band structures as a disc rim located to the north of the star (see Figure \ref{fig:PotDiskFeatures}).

\begin{figure*}
 \begin{center}
\scriptsize
$\begin{array}{ @{\hspace{-0.0mm}} c @{\hspace{-4.8mm}} c @{\hspace{-4.8mm}} c }
    \includegraphics[height=4.5cm, angle=0,trim={0.5cm 0.1cm 0.65cm 1.25cm},clip]{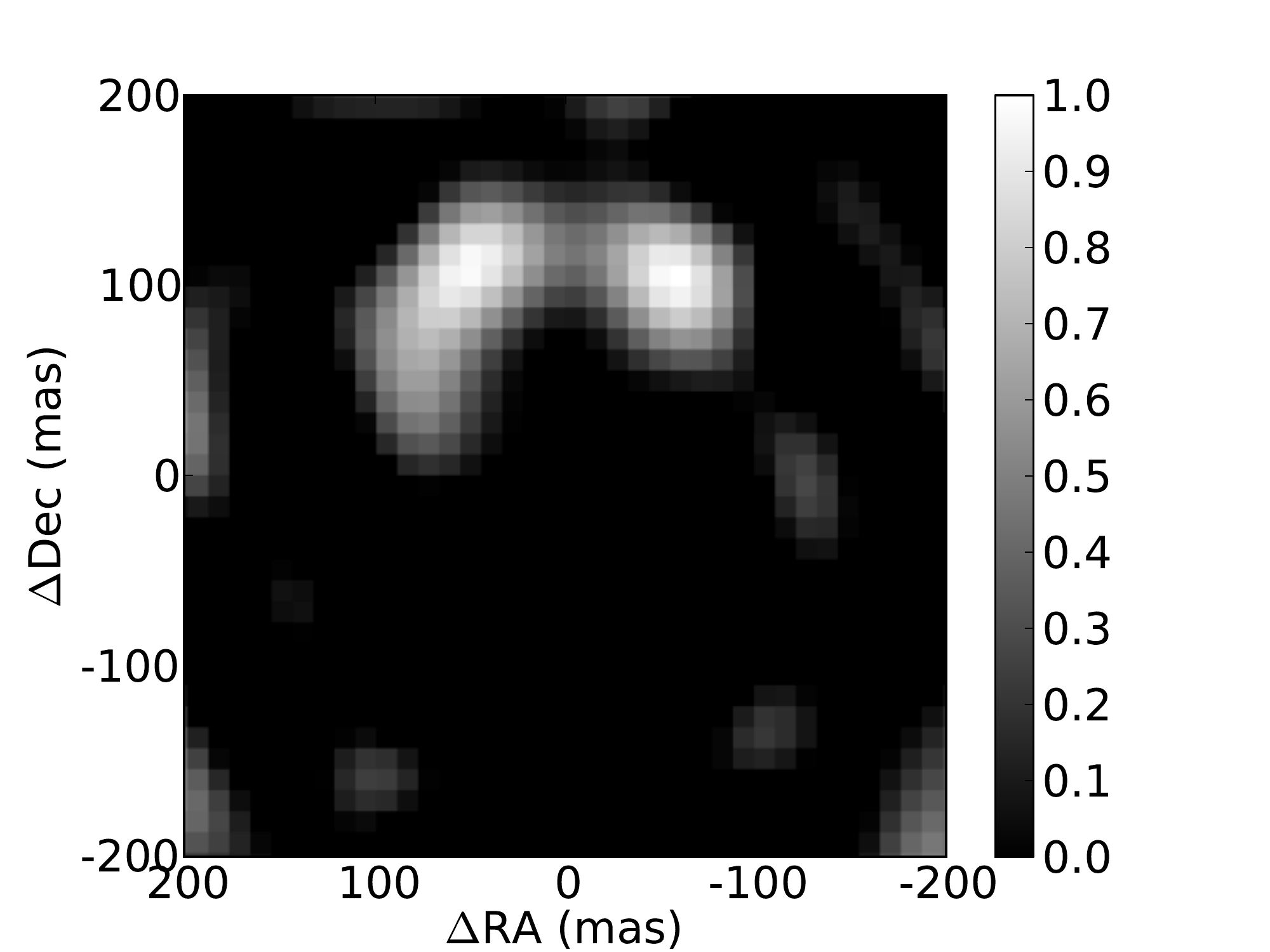} &
    \includegraphics[height=4.5cm, angle=0,trim={0.5cm 0.1cm 0.65cm 1.25cm},clip]{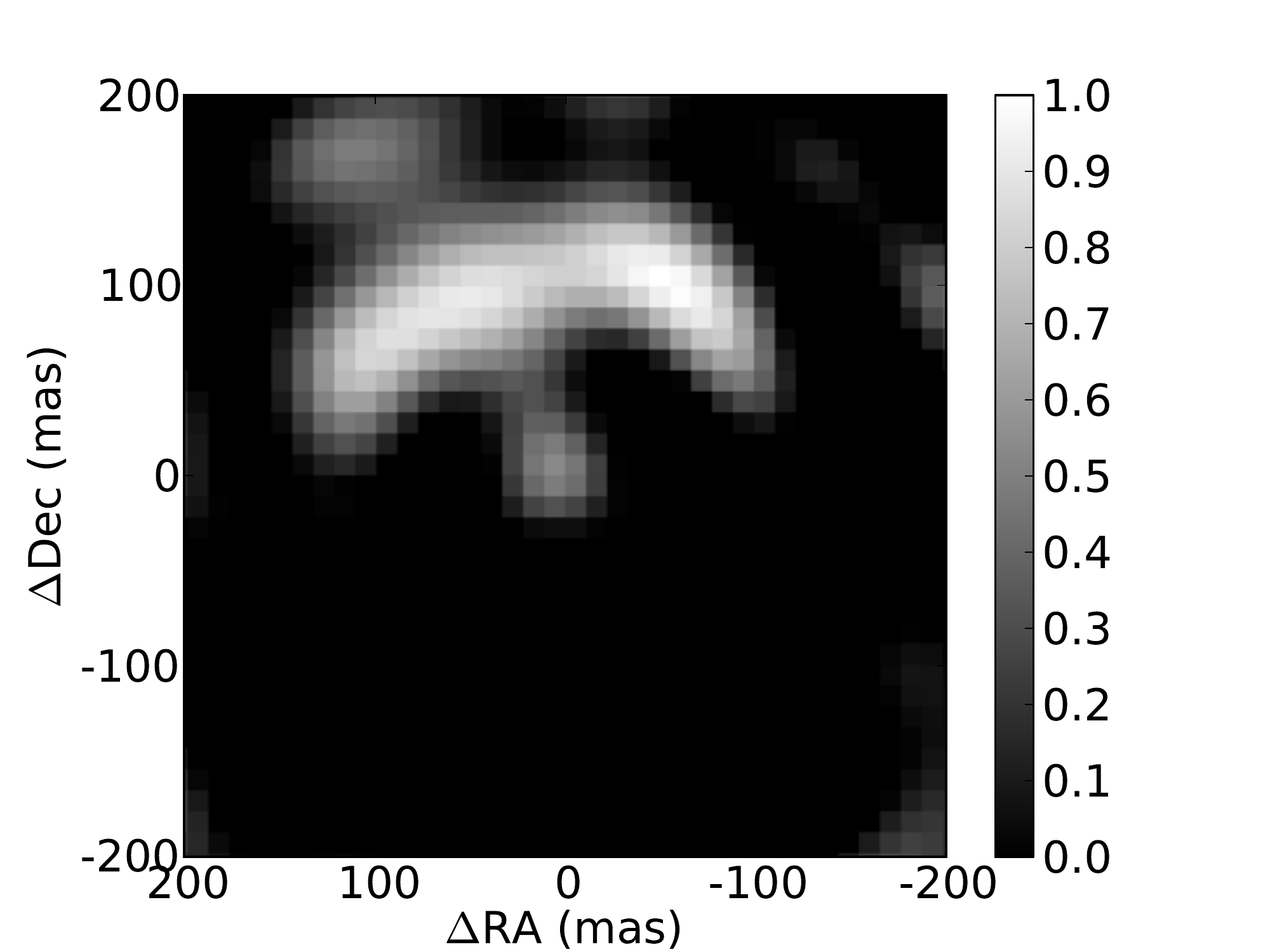} &
    \includegraphics[height=4.5cm, angle=0,trim={0.15cm 0.15cm 0.0cm 0.05cm},clip]{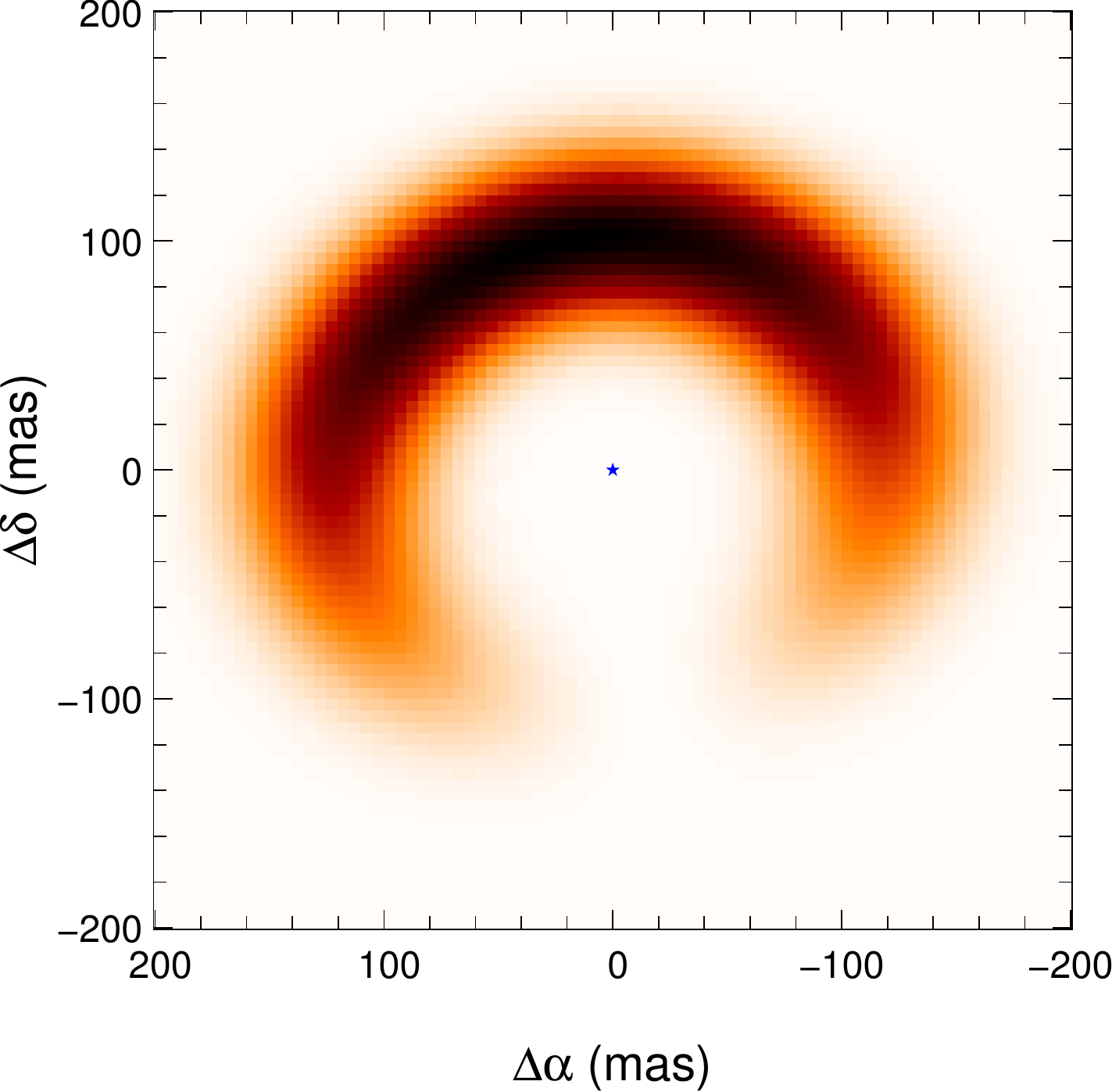}
   \\
        \end{array}$
\end{center}
\caption{Combined L'-band SAM observations. From left to right are the reconstructed image from the combined data set of all three L-band data sets, the reconstructed image from the best fit geometric model using the uv-coverage from the combined data set, and finally the geometric model used. The improved uv-coverage creates an image with improved fidelity. The disc rim is clearly resolved with the artefacts substantially reduced. We find the total unresolved flux fraction to be 0.95 located centrally which is removed from the reconstructed images to allow better inspection of the resolved structures.
The colour scale in the images refers to the flux as a fraction of the peak flux after subtracting the central source.}
\label{fig:DiskFeaturesModel}
\end{figure*}

For a disc rim, any dynamical changes should occur on timescales much longer than the orbital timescale, which is $\sim$25 years at the separation of $\sim$100\,mas, where we resolve the L'-band emission.
This is consistent with our observation that the size and orientation of the resolved L'-band structures is constant between the three epochs. Therefore, we also combined the data sets from all epochs to improve the $uv$-coverage for our image reconstruction. The resulting image is shown in Figure \ref{fig:DiskFeaturesModel} (left).
We resolve the disc as an arc with two regions of higher flux (to the north-east at a position angle of $\sim$30$^\circ$ and to the north-west at a position angle $\sim$315$^\circ$).
We model the disc rim as a skewed ring with a radial Gaussian profile and derive an inclination of $i=30\pm10^{\circ}$, a disc position angle of $95\pm10^{\circ}$, and a disc semi-major axis of 110$\pm$10\,mas, which are consistent with the inner edge of the outer disc deduced by \citet{2013ApJ...768...80K} from N-band interferometry data.
We define skewness as the ratio between the amount of flux attributed to opposite sides of the ring along the minor axis. We set this value to 0.8 and the disc width (50\,mas) to represent radiative transfer effects such as thermal emission or scattered light, but these parameters were found to have only a marginal affect on the model closure phases.
The resulting best-fit model is shown in Figure~\ref{fig:DiskFeaturesModel} (right).

We determine the disc flux contribution based on the amount of flux outside the central star in the best reconstructed images and find that it contributes 5\% of the total L'-band flux. This value is determined through fitting to a grid of flux contributions and computing a $\chi^2$ for each. We then select the best value.

\subsubsection{Companion Scenario}
\label{subsubsec:CompOnly}

Within the shorter-wavelength (H and K-band) data sets we see phase signals that are more consistent with point-source emission than the L'-band data. In this section we investigate the possibility of these signals being consistent with an orbiting companion, assuming that there are no contributions due to disc emission, but introduce and examine more complex scenarios later in Sections~\ref{subsubsec:RimOnly} and \ref{subsubsec:CompDisc}. In these sections we attempt to explain the observed asymmetries with a disc rim model, as in Section~\ref{subsubsec:LBandDisc}, or through a companion\,+\,disc rim model where the disc rim position and orientation matches that of the disc rim seen in the L'-band observations.

\begin{figure*}
 \begin{center}
$\begin{array}{ @{\hspace{-8.0mm}} c @{\hspace{-5.0mm}} c @{\hspace{-5.5mm}} c @{\hspace{-5mm}} c }
    \includegraphics[height=4.0cm, angle=0,trim={0.5cm 0.1cm 0.65cm 1.25cm},clip]{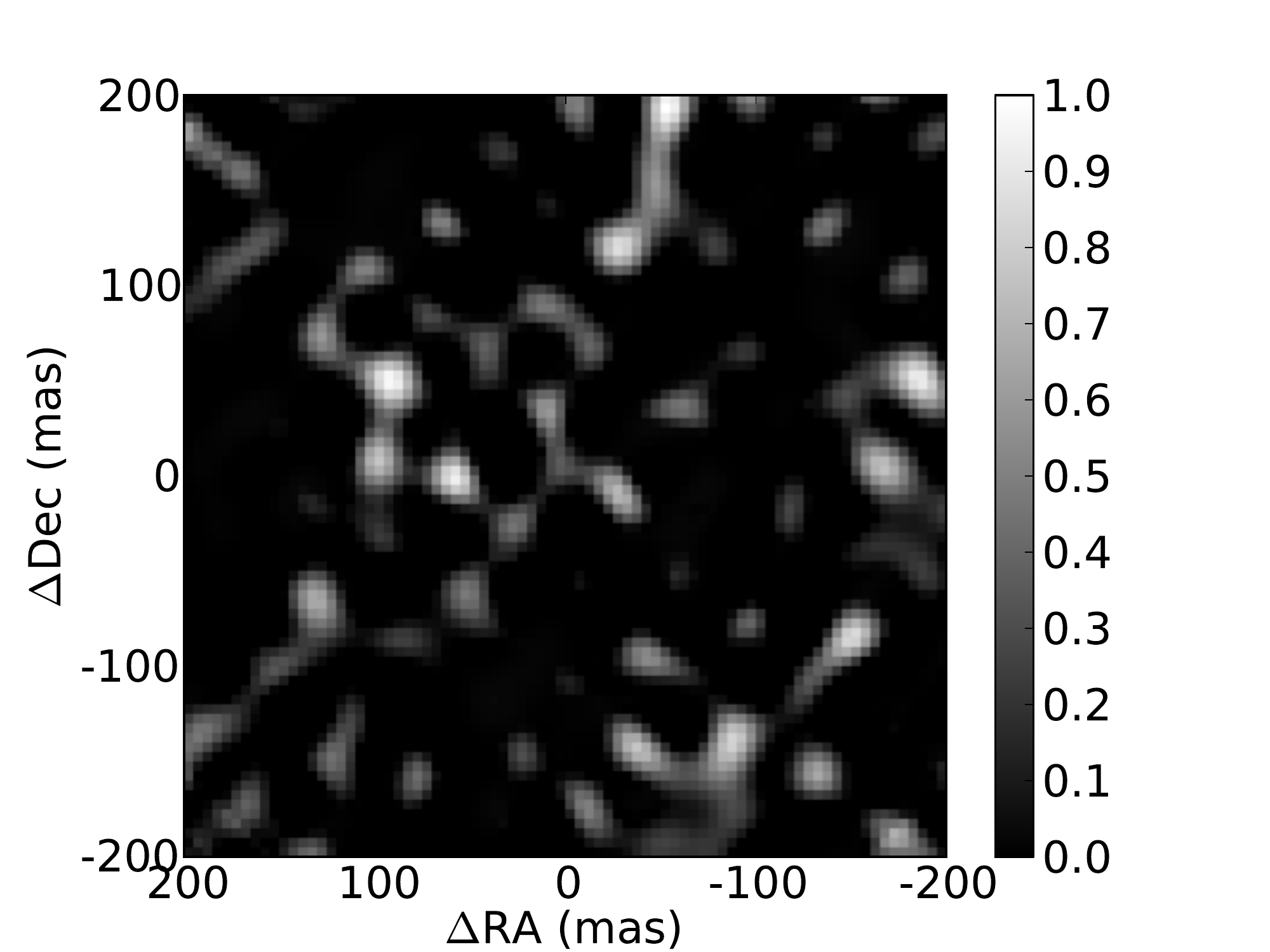} & \includegraphics[height=4.0cm, angle=0,trim={0.5cm 0.1cm 0.65cm 1.25cm},clip]{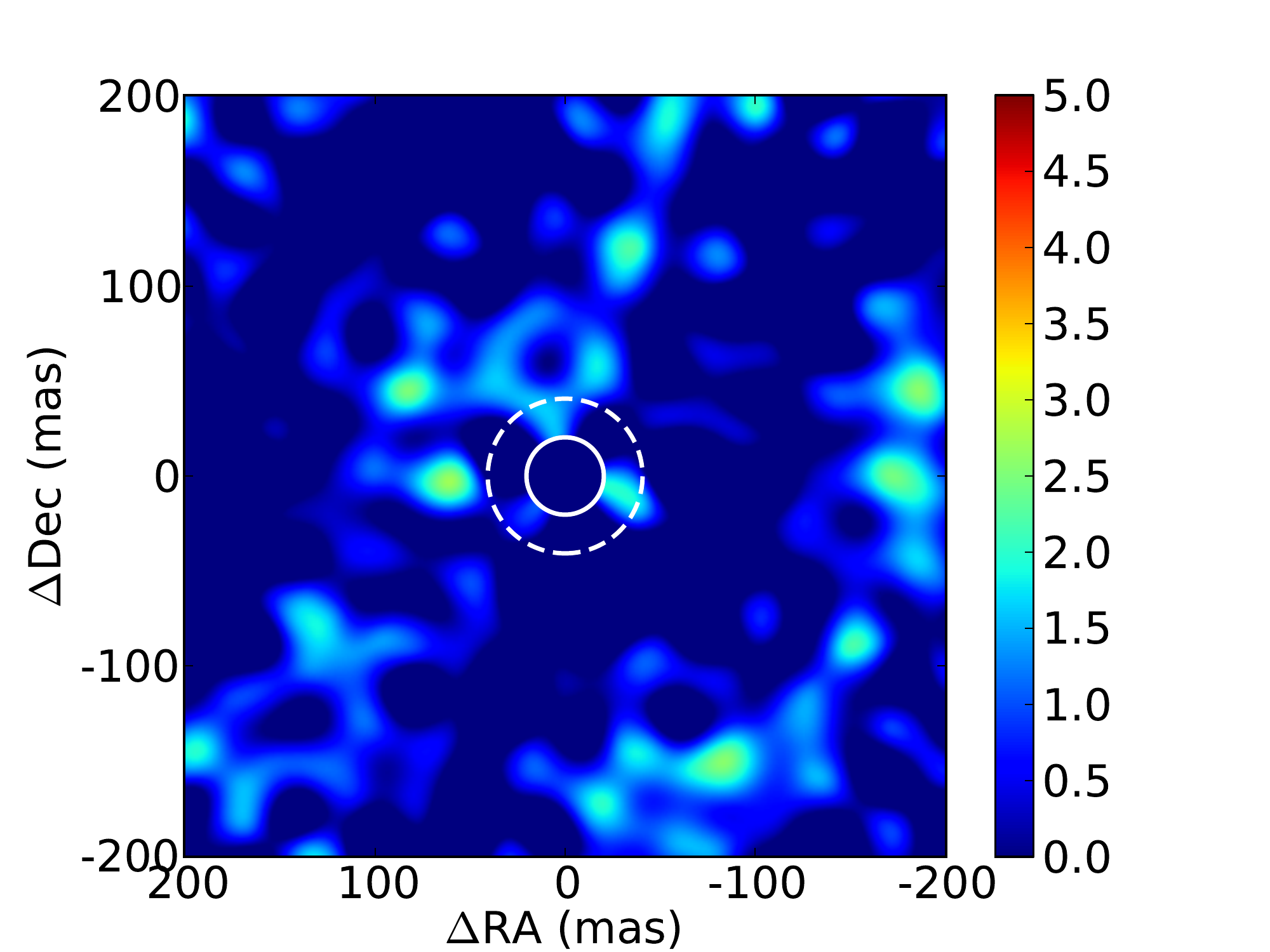} & \includegraphics[height=4.0cm, angle=0,trim={0.9cm 0.1cm 0.65cm 1.25cm},clip]{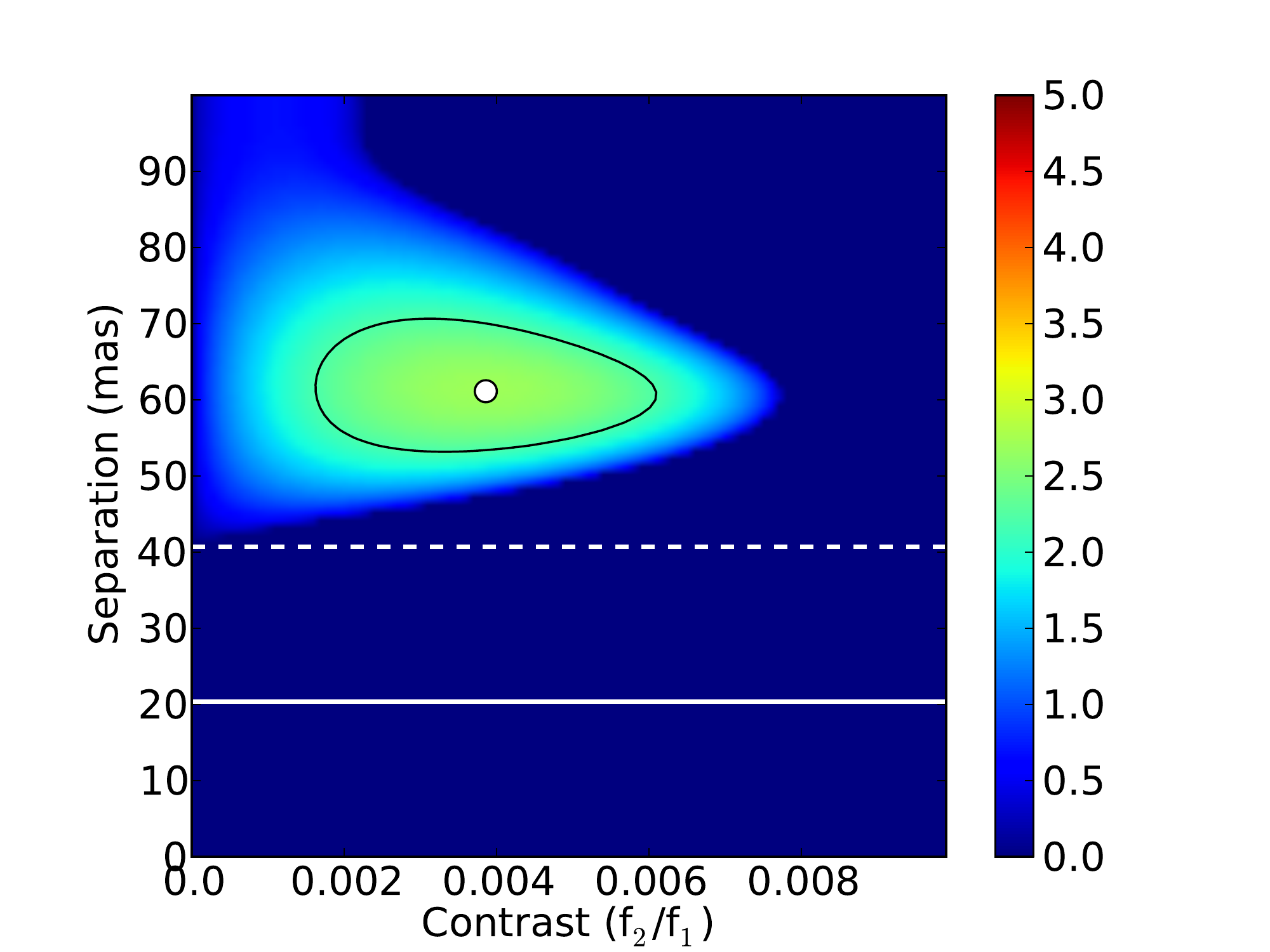} & \includegraphics[height=4.0cm, angle=0,trim={0.5cm 0.1cm 0.65cm 1.25cm},clip]{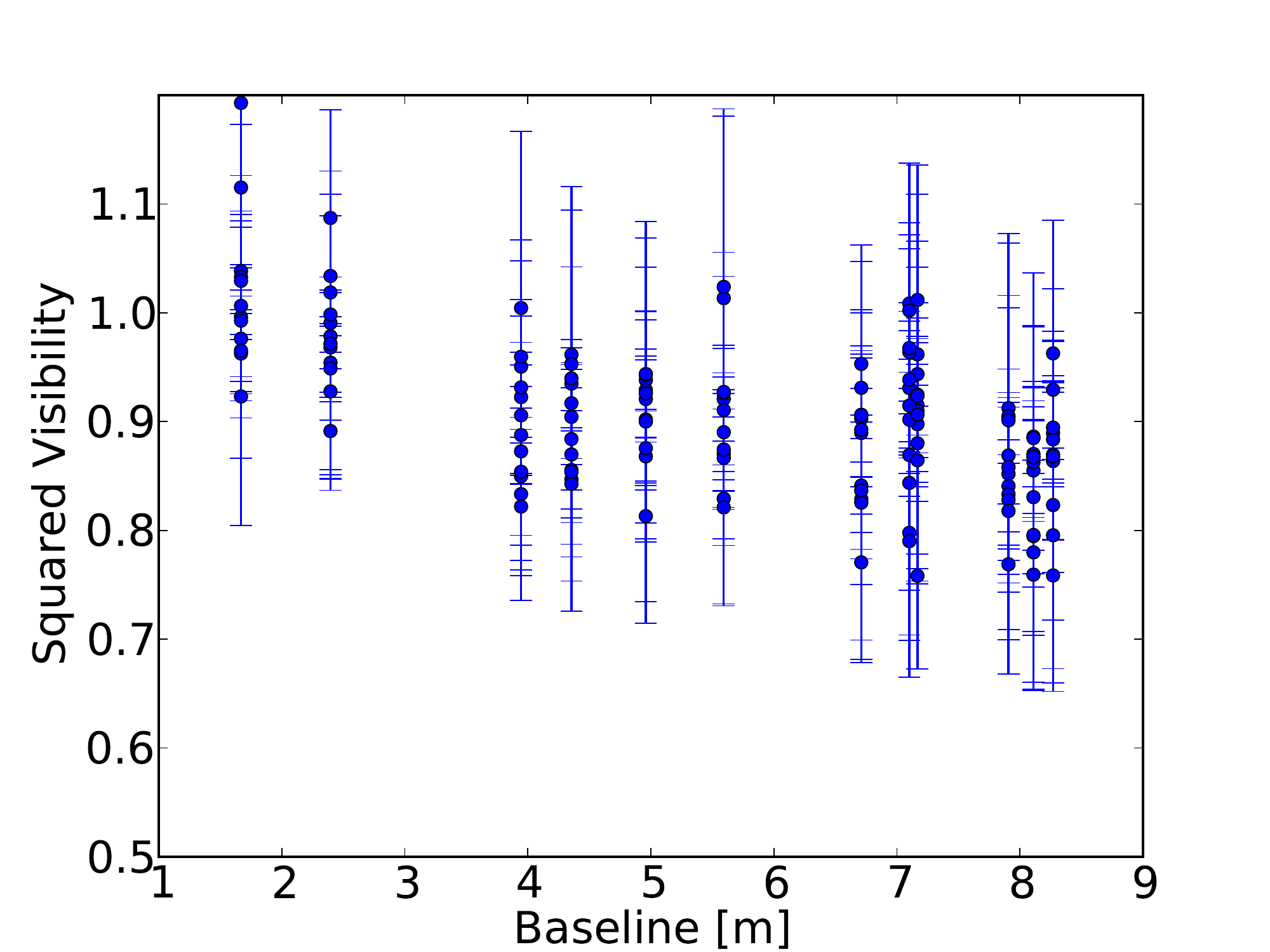} \\
   \includegraphics[height=4.0cm, angle=0,trim={0.5cm 0.1cm 0.65cm 1.25cm},clip]{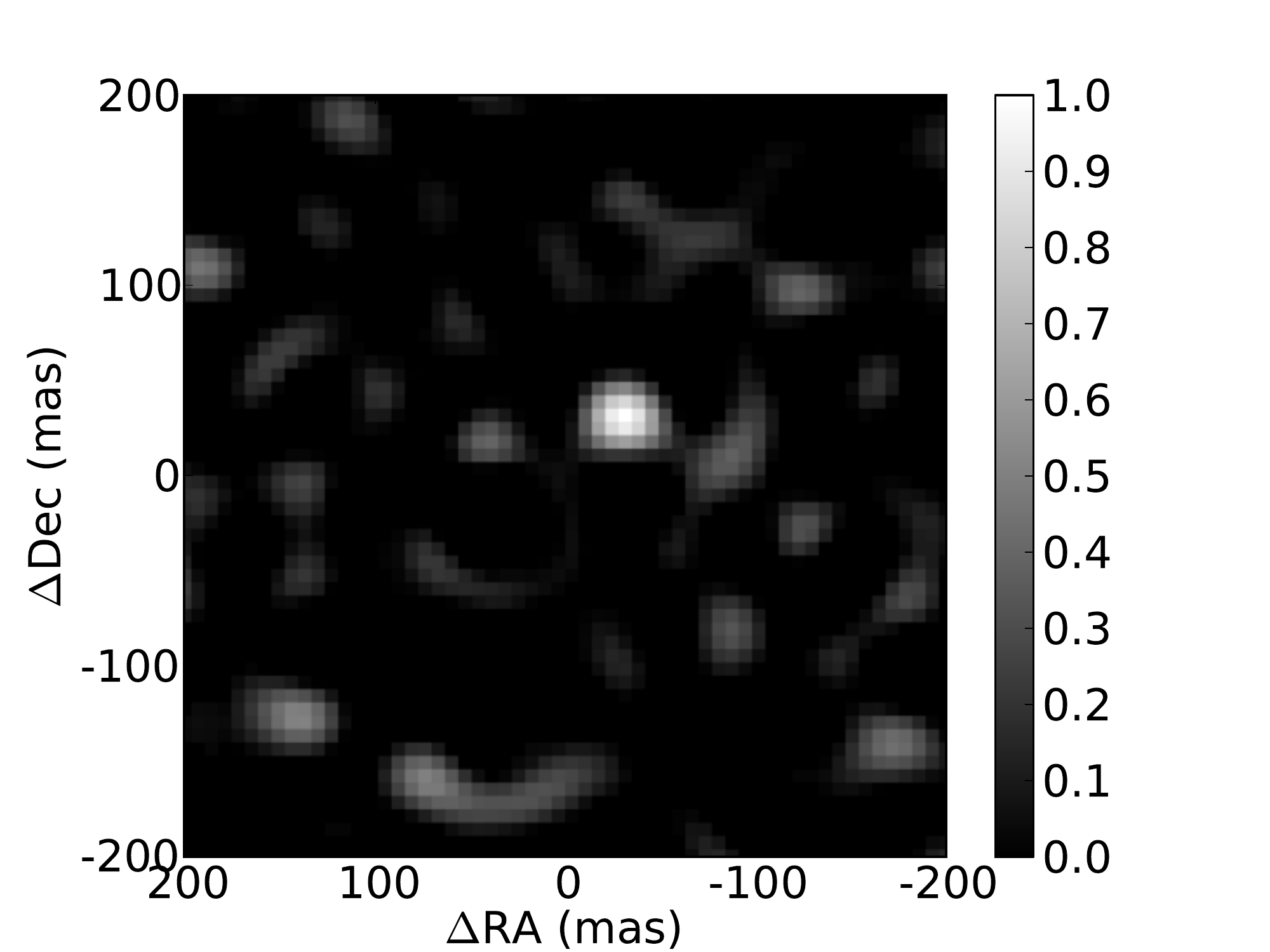} & \includegraphics[height=4.0cm, angle=0,trim={0.5cm 0.1cm 0.65cm 1.25cm},clip]{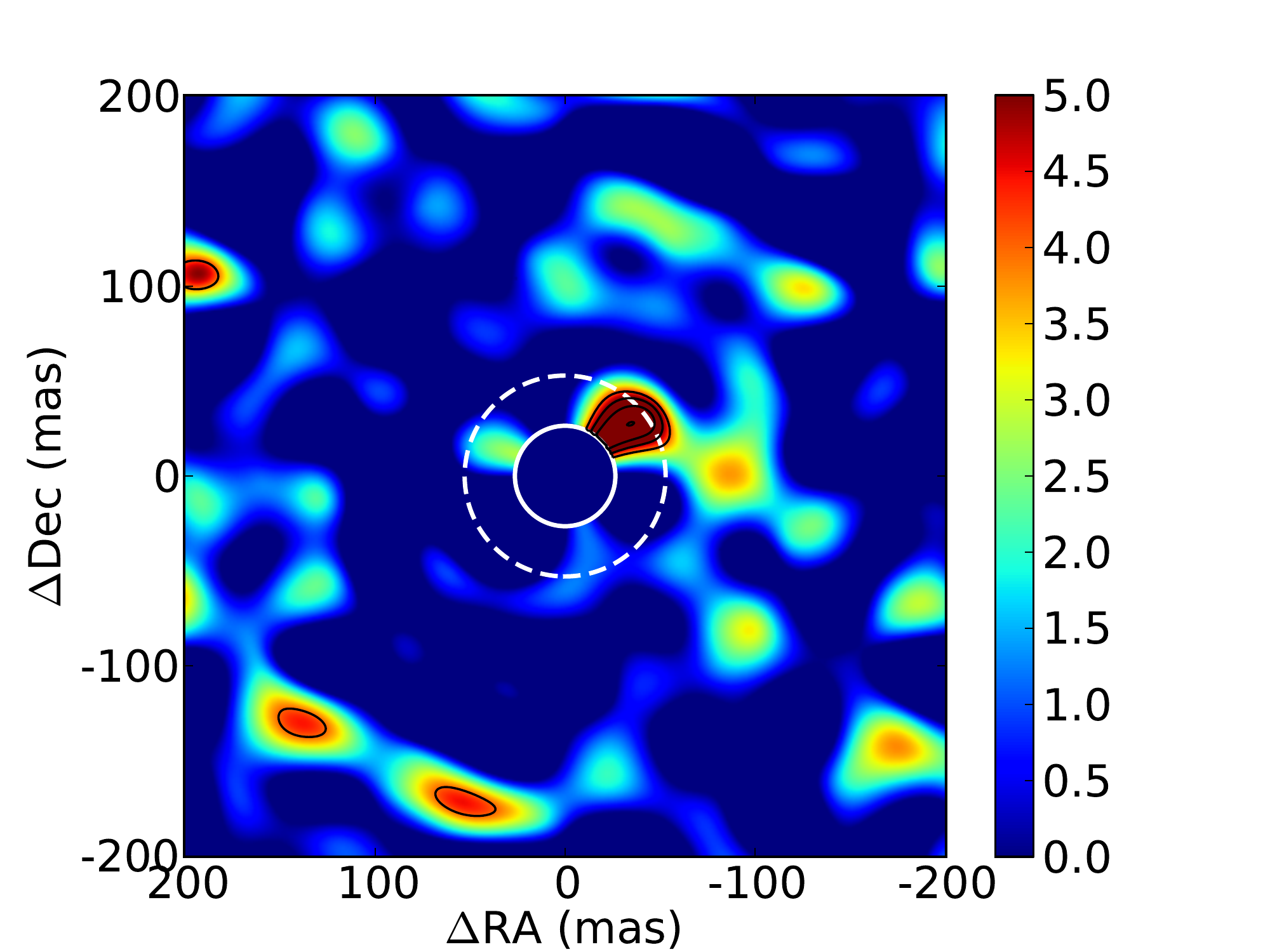} & \includegraphics[height=4.0cm, angle=0,trim={0.9cm 0.1cm 0.65cm 1.25cm},clip]{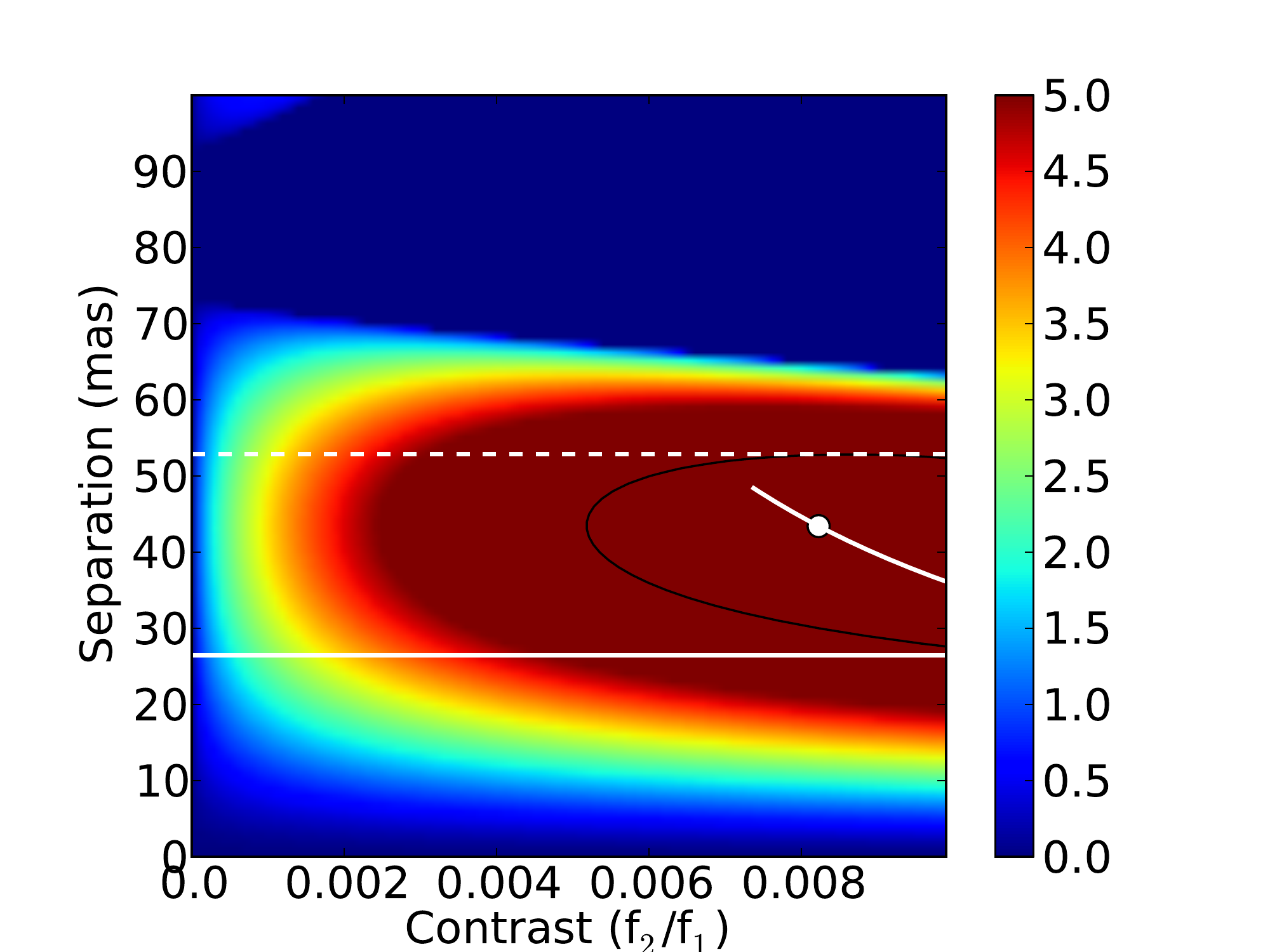} & \includegraphics[height=4.0cm, angle=0,trim={0.5cm 0.1cm 0.65cm 1.25cm},clip]{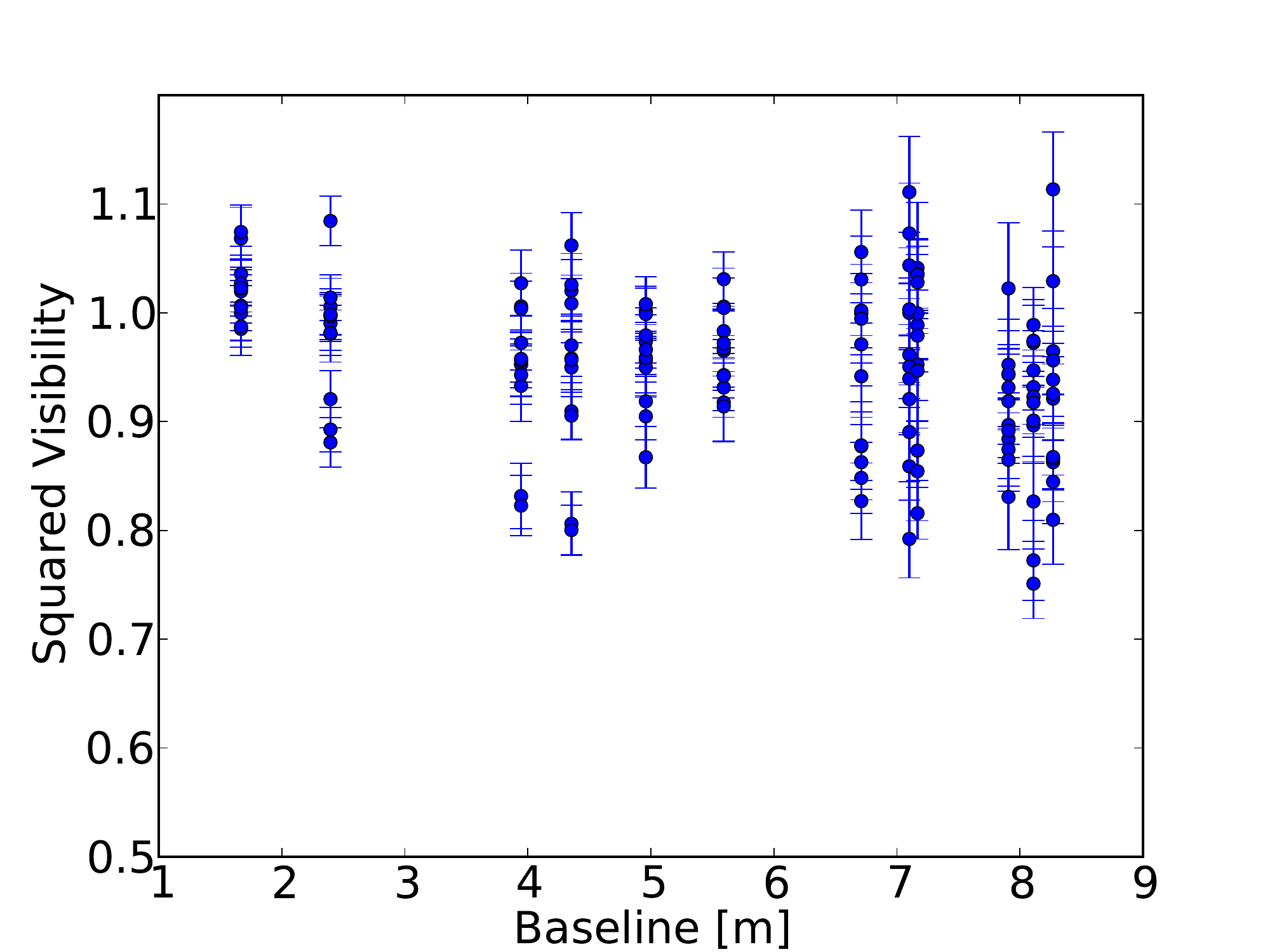} \\
    \includegraphics[height=4.0cm, angle=0,trim={0.5cm 0.1cm 0.65cm 1.25cm},clip]{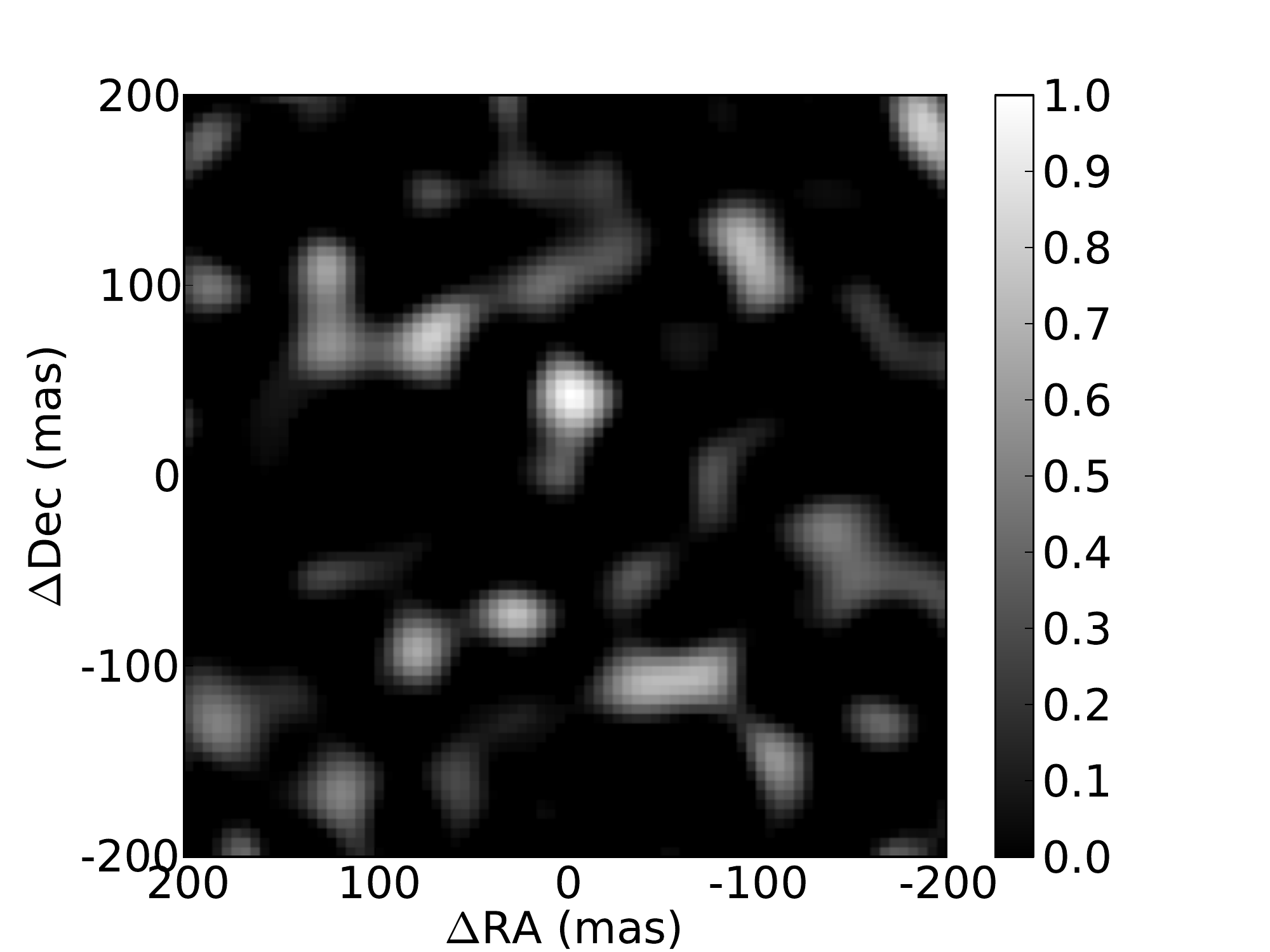} & \includegraphics[height=4.0cm, angle=0,trim={0.5cm 0.1cm 0.65cm 1.25cm},clip]{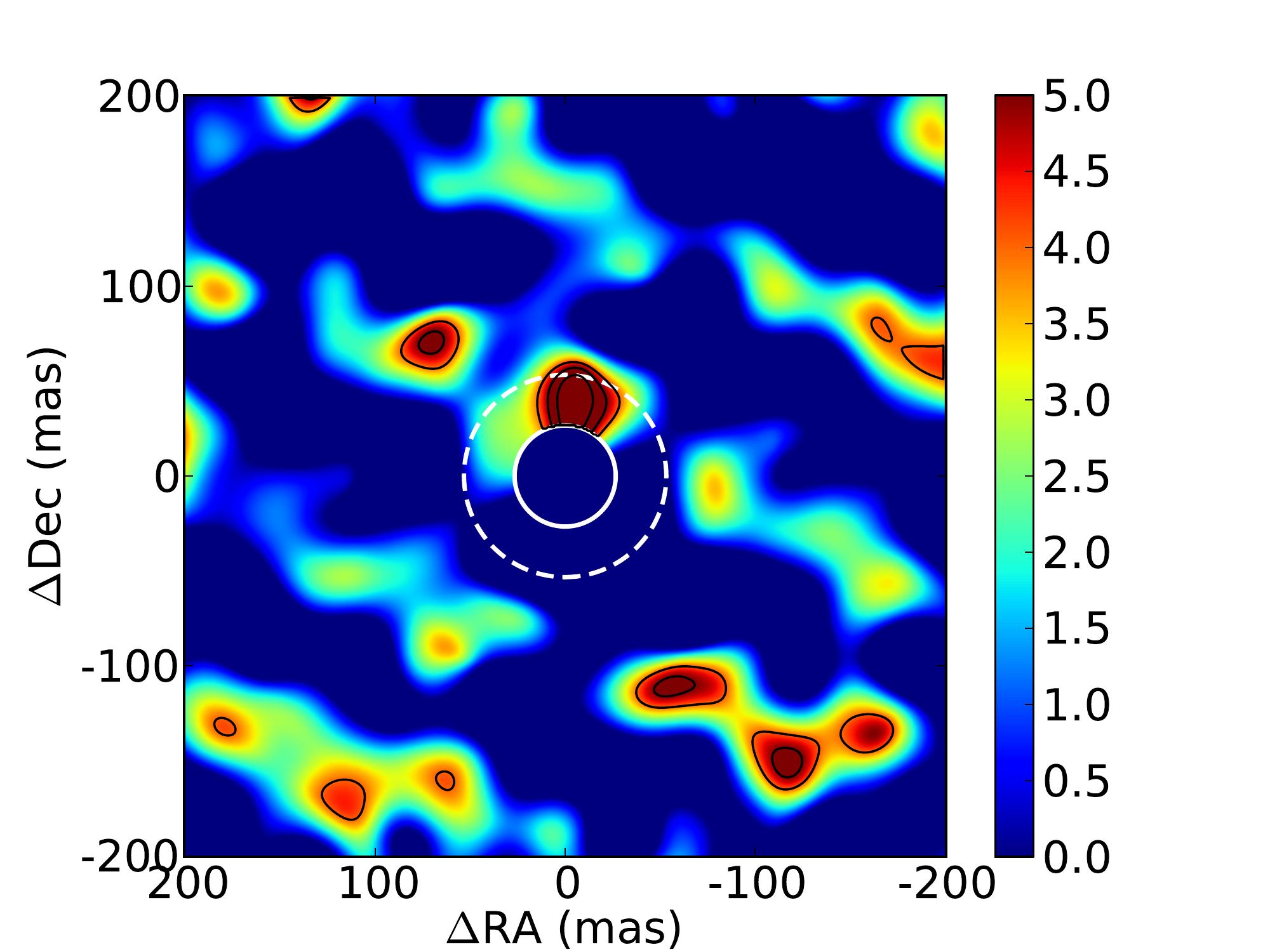} & \includegraphics[height=4.0cm, angle=0,trim={0.9cm 0.1cm 0.65cm 1.25cm},clip]{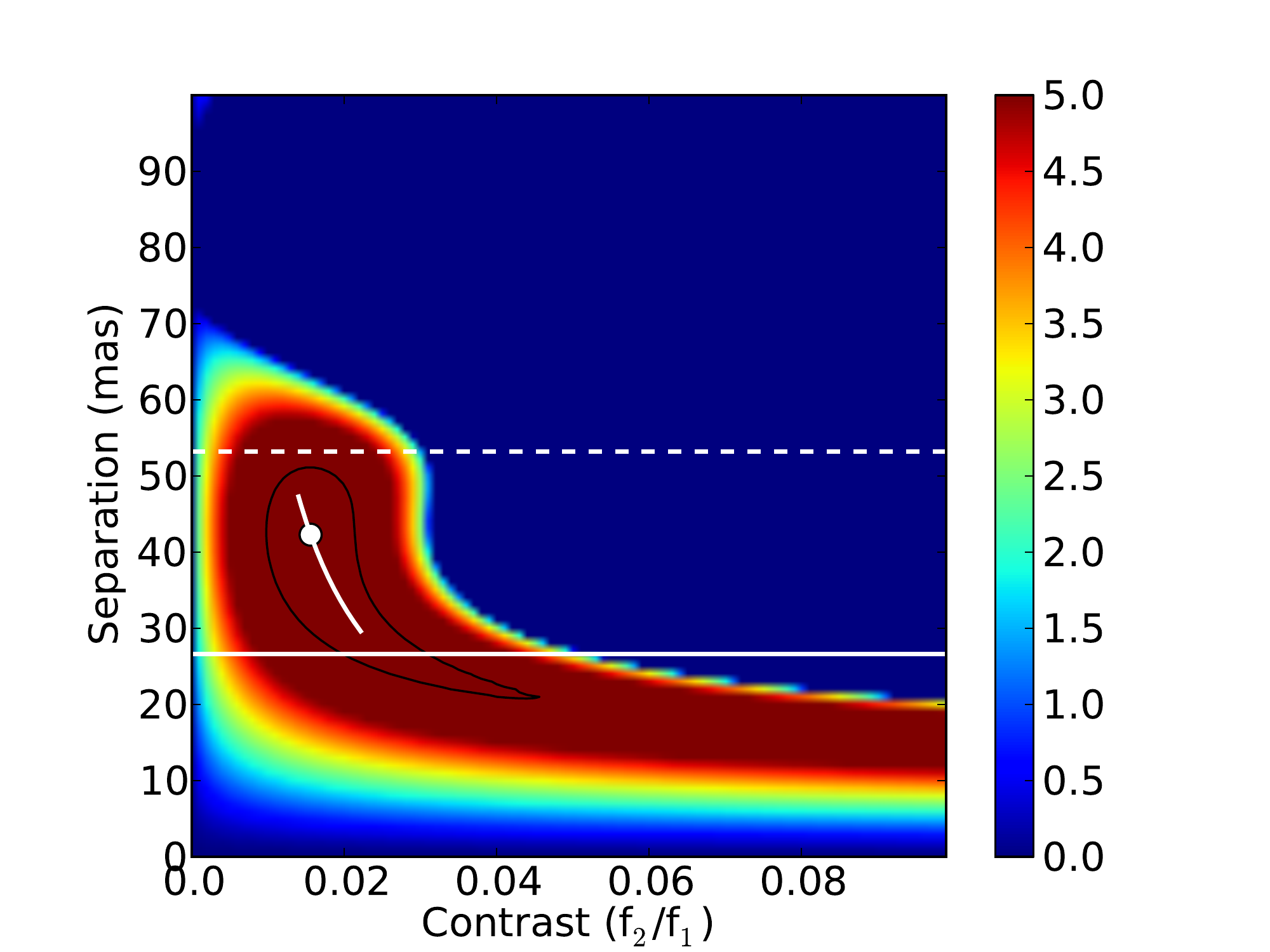} & \includegraphics[height=4.0cm, angle=0,trim={0.5cm 0.1cm 0.65cm 1.25cm},clip]{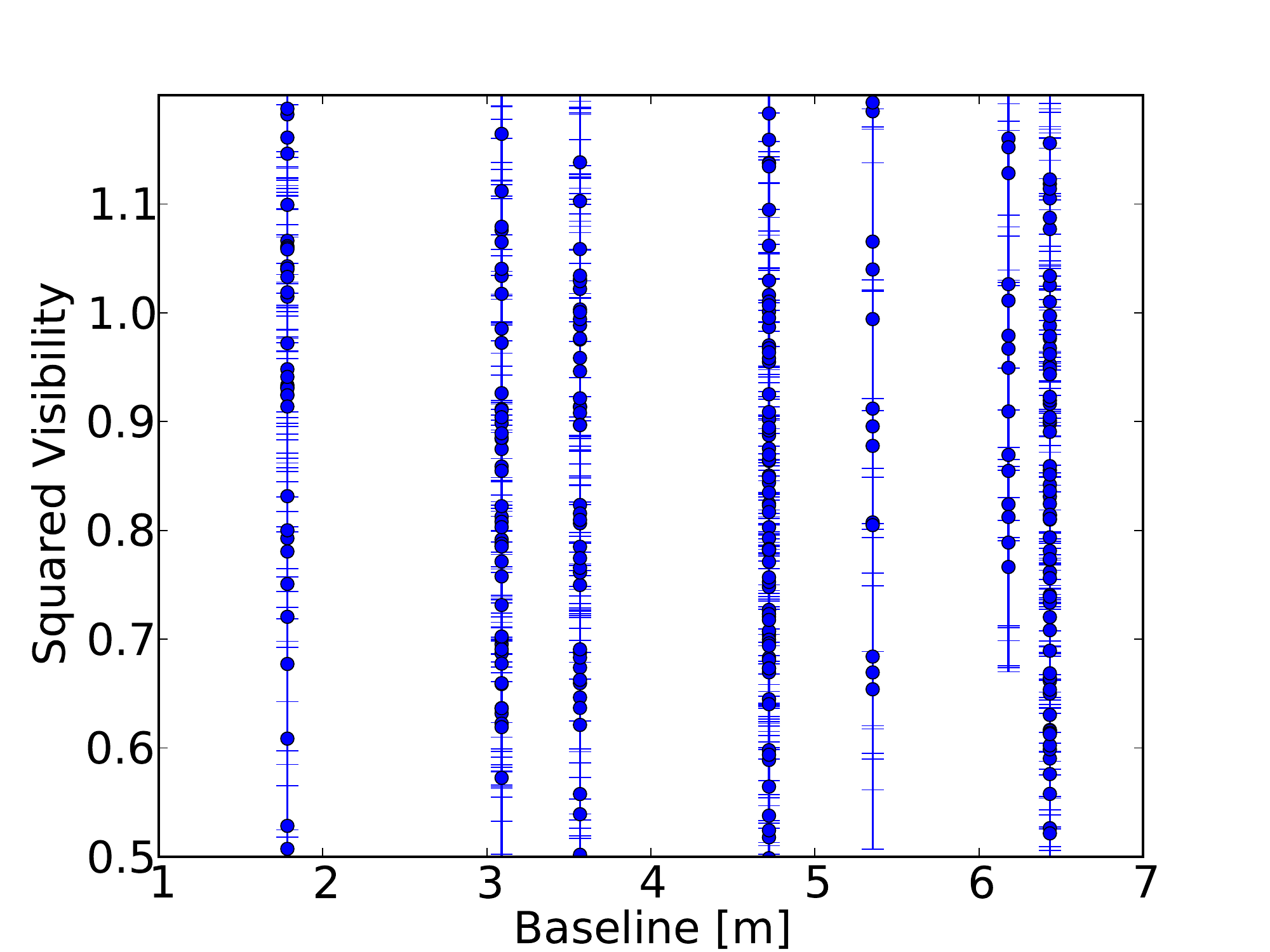} \\
   \includegraphics[height=4.0cm, angle=0,trim={0.5cm 0.1cm 0.65cm 1.25cm},clip]{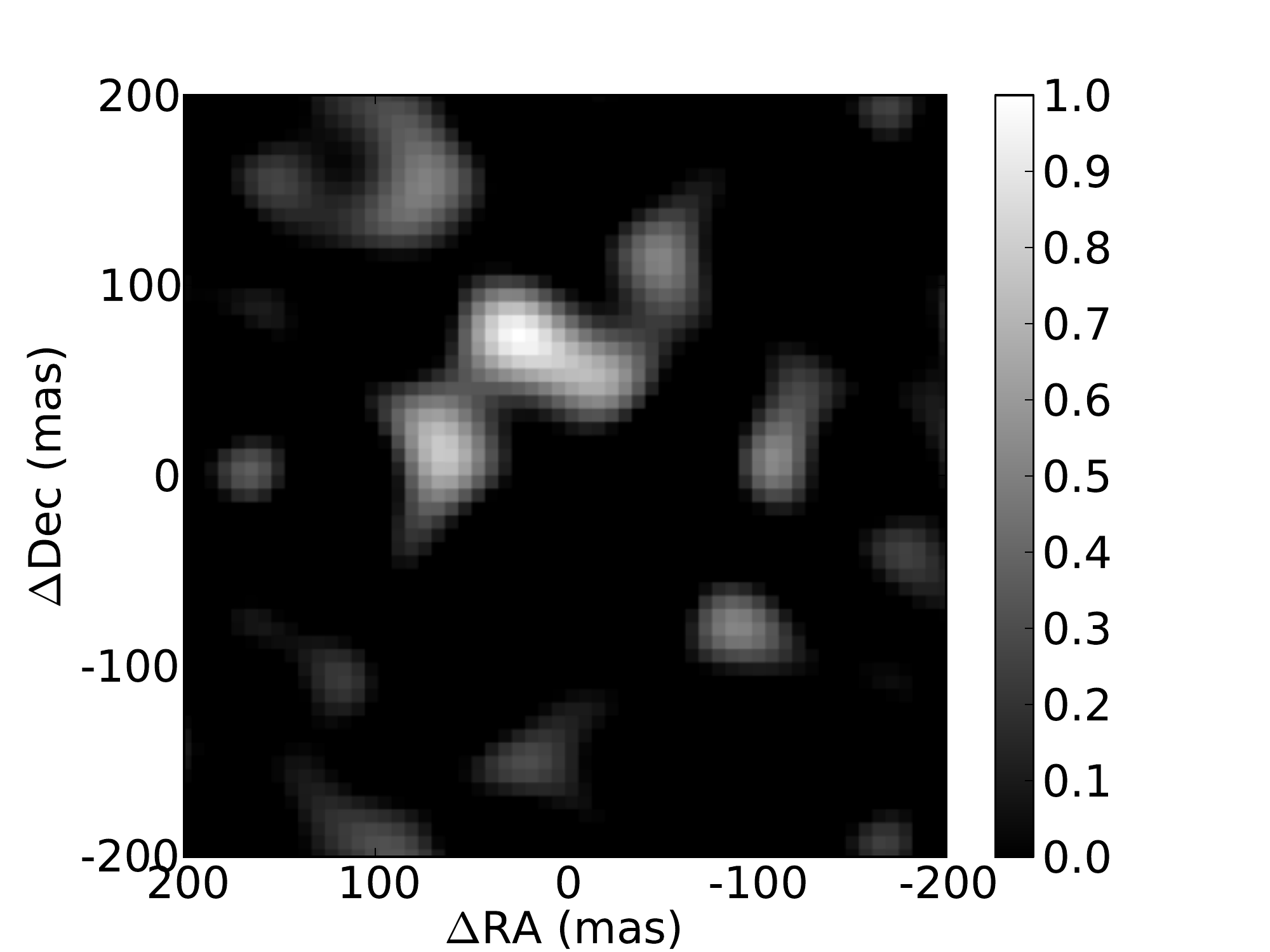} & \includegraphics[height=4.0cm, angle=0,trim={0.5cm 0.1cm 0.65cm 1.25cm},clip]{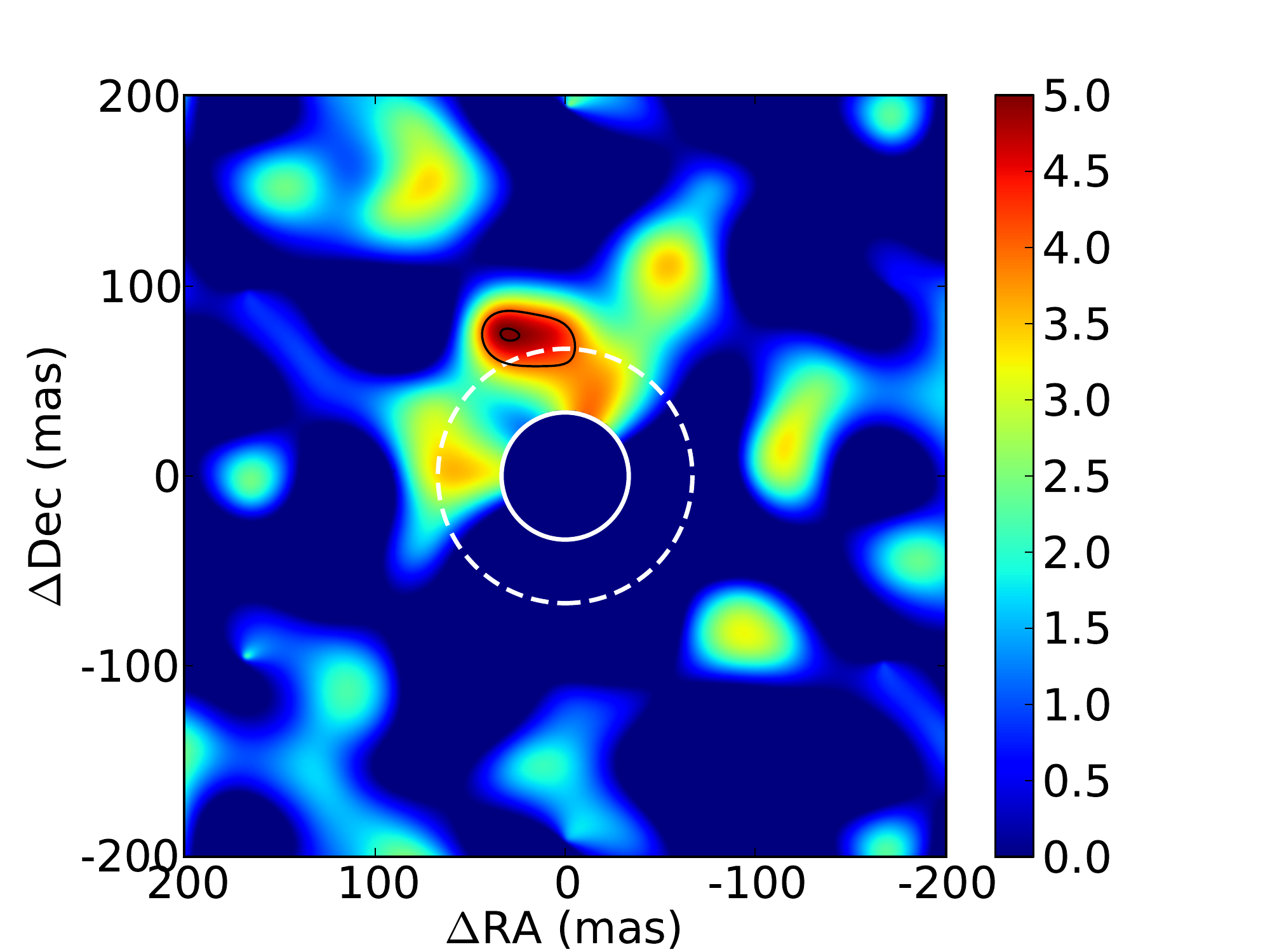} & \includegraphics[height=4.0cm, angle=0,trim={0.9cm 0.1cm 0.65cm 1.25cm},clip]{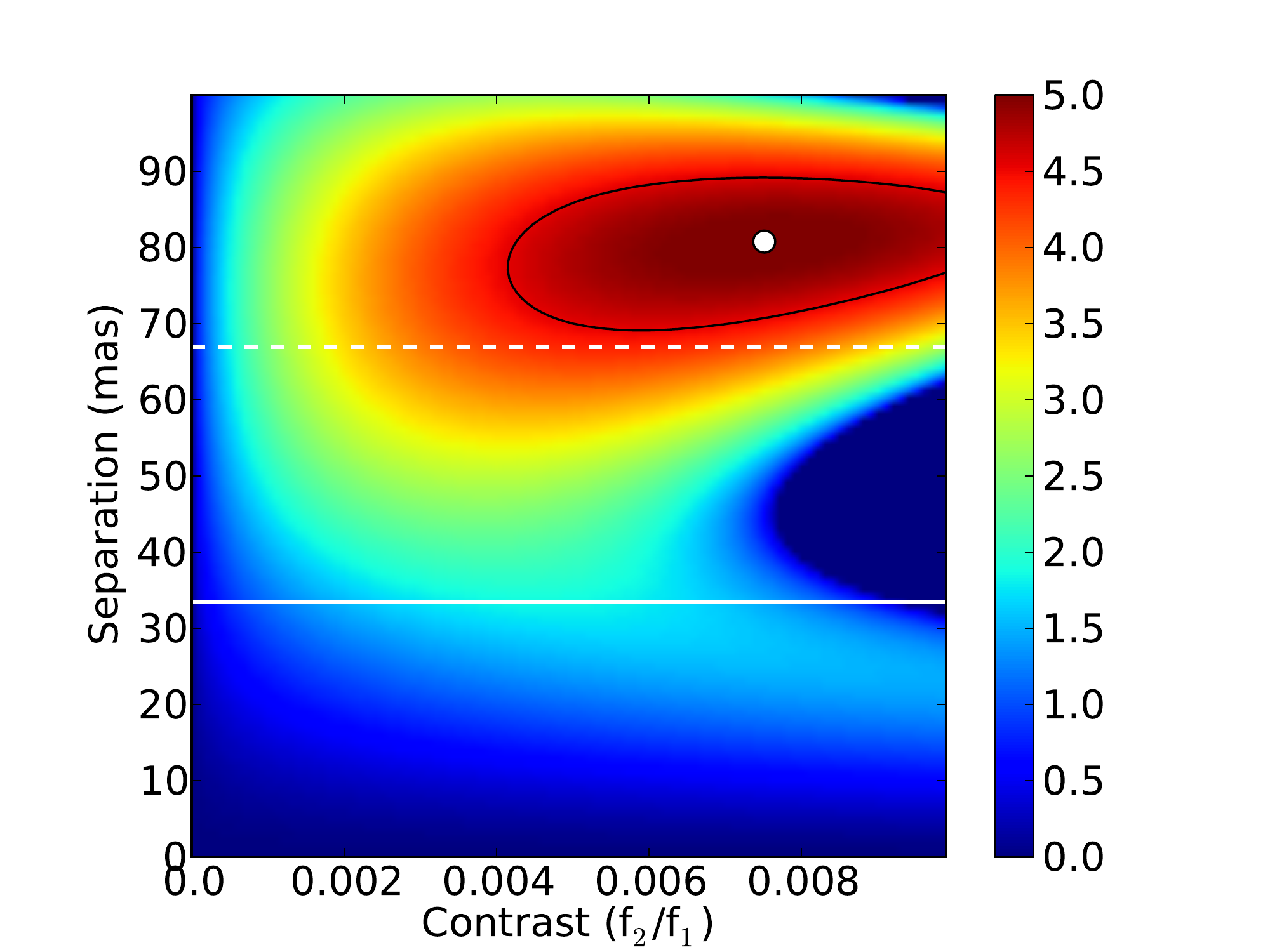} & \includegraphics[height=4.0cm, angle=0,trim={0.5cm 0.1cm 0.65cm 1.25cm},clip]{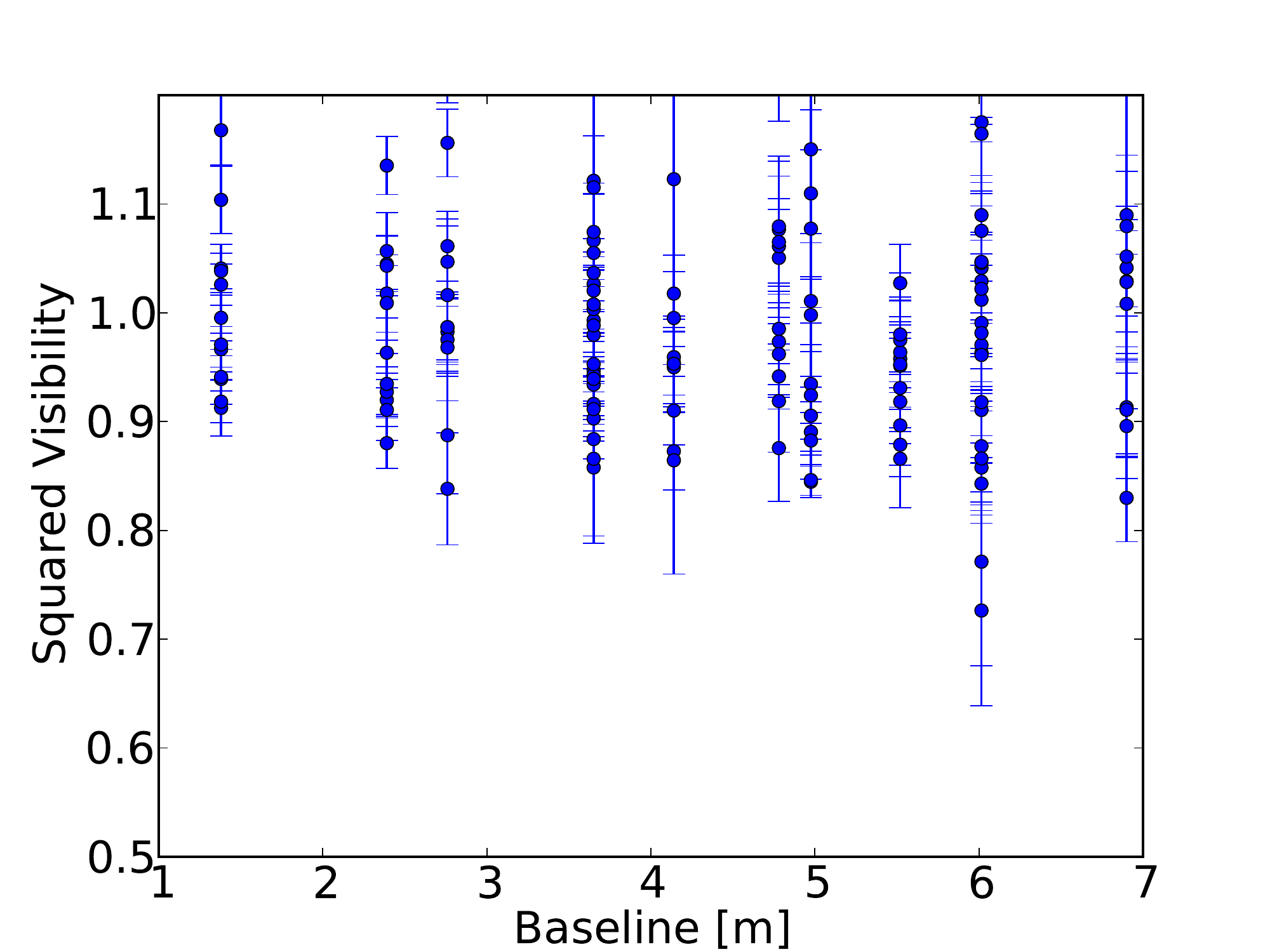} \\
   \includegraphics[height=4.0cm, angle=0,trim={0.5cm 0.1cm 0.65cm 1.25cm},clip]{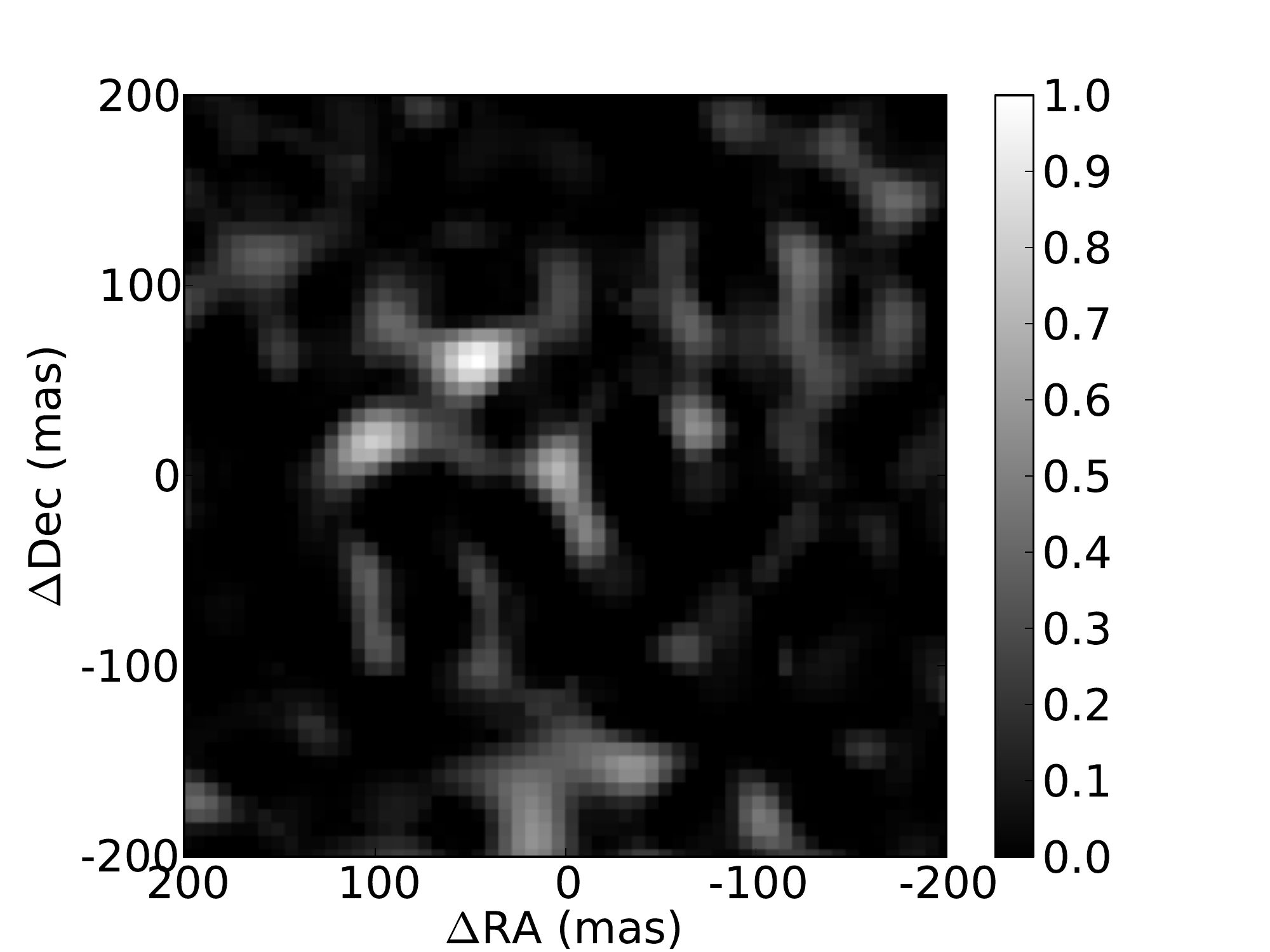} & \includegraphics[height=4.0cm, angle=0,trim={0.5cm 0.1cm 0.65cm 1.25cm},clip]{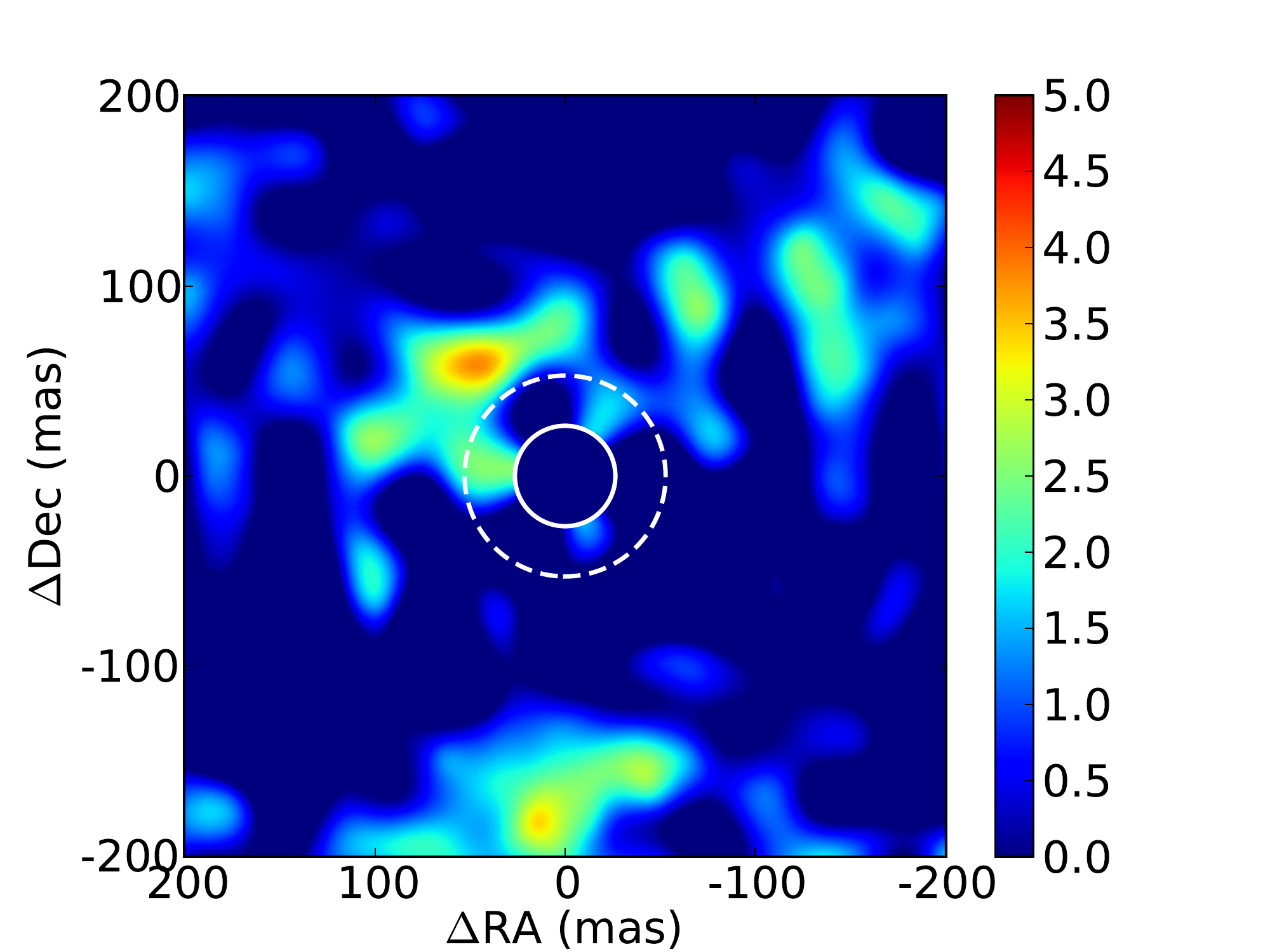} & \includegraphics[height=4.0cm, angle=0,trim={0.9cm 0.1cm 0.65cm 1.25cm},clip]{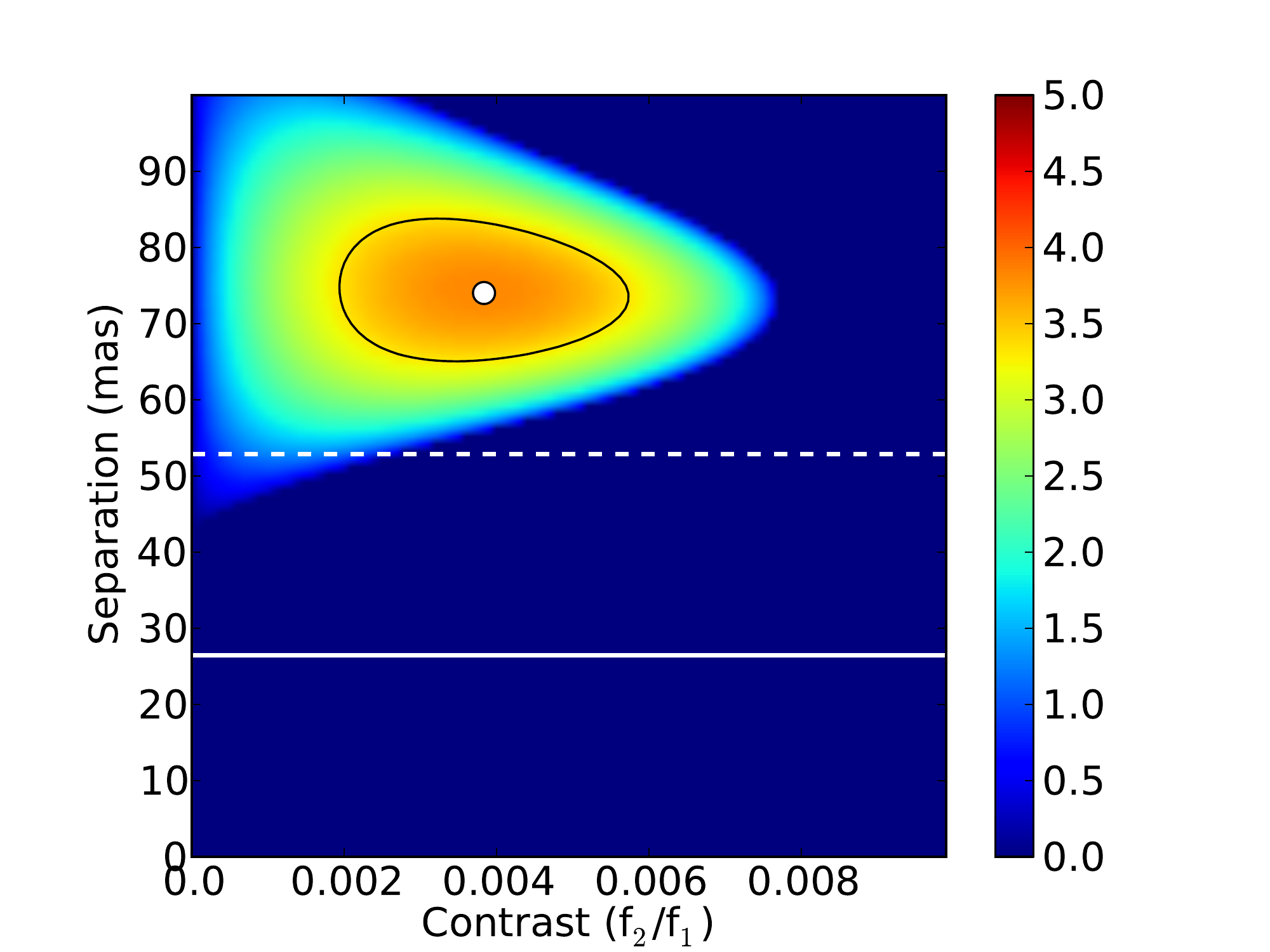} & \includegraphics[height=4.0cm, angle=0,trim={0.5cm 0.1cm 0.65cm 1.25cm},clip]{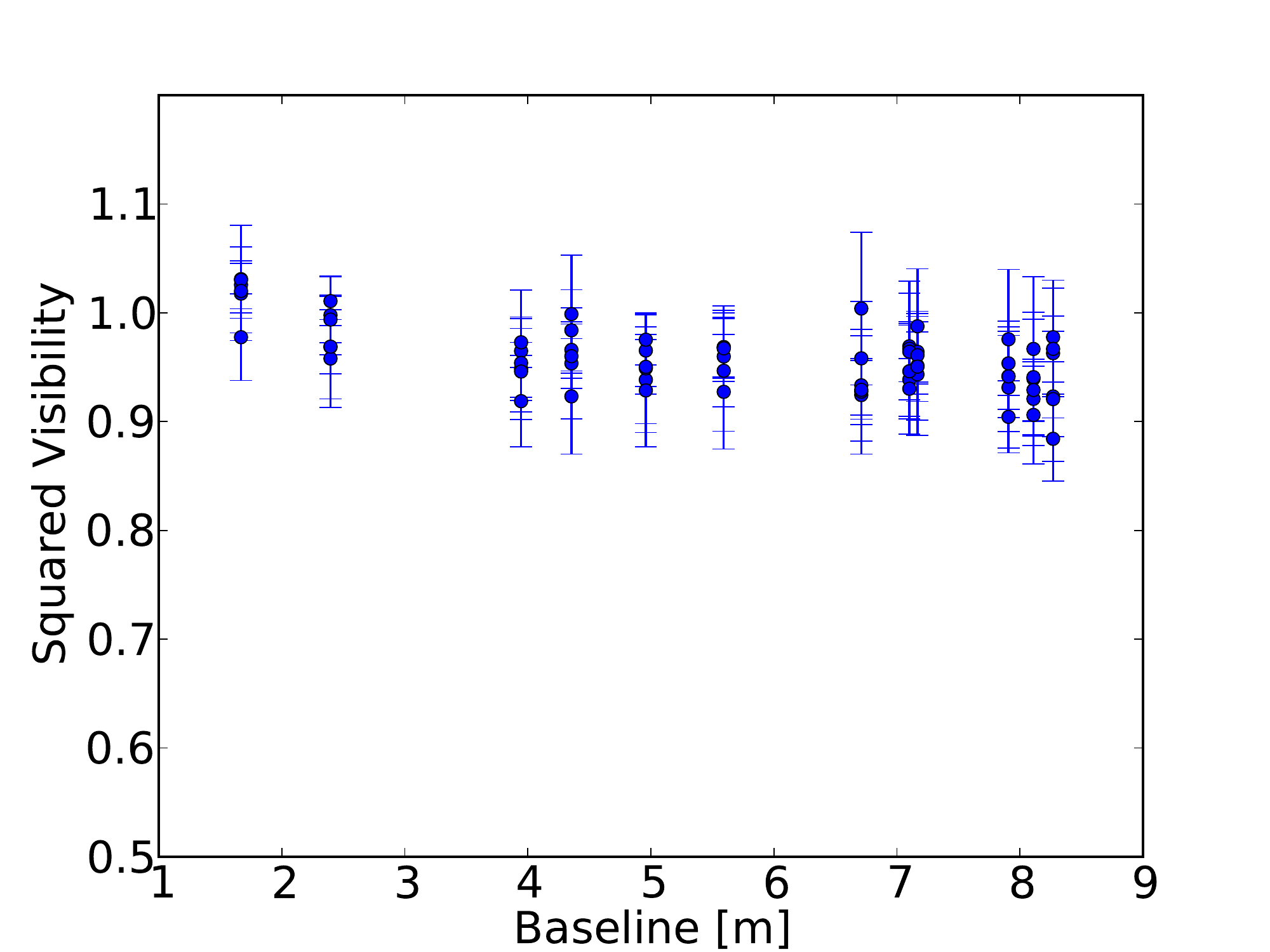} \\
 \end{array}$
\end{center}
\caption{Figure showing H- and K- band observations where we see significant structure. \textbf{Left:}  Reconstructed Images \textbf{Middle left:} Computed significance maps. \textbf{Middle right:} Degeneracy plots. \textbf{Right:} Squared visibilities. \textbf{First row:} 2012-01-10, H-band, \textbf{Second row:} 2012-01-09, K'-band, \textbf{Third row:} 2012-12-18, H-band,\textbf{Fourth row:}, 2012-12-18, Ks-band, \textbf{Fifth row:}, 2013-10-20, K'-band. The 2012-01-09 K-band data (second row) shows a companion candidate north-west of the star, which appears to have moved to north by 2012-12-18 (H-band data set, third row). This motion appears to be too rapid for Keplerian motion of a companion at the estimated ($\sim 45$\,mas) separation. In Section~\ref{subsubsec:CompDisc} we interpret this is due to an separation over-estimation caused by the presence of scattering from the outer disc rim. The dashed and solid white lines within the degeneracy plots represent separations corresponding to $\lambda/B$ and $\lambda/2B$, where $B$ is the longest baseline of the array. The black contours represent the region of $\Delta\sigma$=\,1 around the best-fit value.
Colour bars in the reconstructed images represent flux as a fraction of the peak flux within the frame after subtracting the central source while those in the significance maps represent significance ($\sigma$) of the binary fit to the data at each pixel location.}
\label{fig:PotCompanions}
\end{figure*}

We observe phase signals that are consistent with a point-source at one K-band epoch (2012-01-09) and at a second H-band epoch (2012-12-18; see Figure \ref{fig:PotCompanions}). The measured phase at the remaining two K-band epochs are not as well reproduced by a simple binary fit and appear to be more complex (see Figure \ref{fig:PotCompanions}).
In the 2012-12-18 K-band observations we see two close-in structures. The inner structure ($\sim$40\,mas) is displaced from the point source-like signal in the H-band observations taken on the same night. A position angle difference was found to be 11$\pm$6$^\circ$ as measured between the peak of both structures. Determining radial distance is more difficult as both structures lie within the region in which the separation and contrast are not as well constrained resulting in large uncertainties.
The second structure is located outside the degenerate region (indicated with a dashed circle in Figure~\ref{fig:DiscCompComparisons}) and does not appear in the H-band observations.
In the 2013-10-20 K-band data we see again a single, weak, point source-like structure outside the degenerate region.

Based purely on the quality of the fit to the interferometric data, the strong detections ($>5\sigma$) in the H- and K-band are consistent with a companion at $>45$\,mas separation from the star.
If we de-redden the measured contrasts using the method outlined by \citet{1989ApJ...345..245C}, we compute absolute magnitudes of 5.1 in H-band and 4.3 in K-band.
According to the circumplanetary disc models by \citet{0004-637X-799-1-16}, the SEDs of an optically thick accretion disc is determined by the parameters $M \dot M$ (the product of the planet mass and disc accretion rate) and the inner disc radius.
The measured magnitudes and H-K colour would imply a $M \dot M$ value of 10$^{-3}$ M$^2_J$yr$^{-1}$ when assuming a circumplanetary disc with an inner radius of 2R$_\textrm{Jup}$ viewed under a similar inclination as the surrounding circumstellar disc ($\sim$30$^\circ$). The inner radius is chosen to be the same as that assumed in previous observations of companion candidates to facilitate comparison \citep{2015Natur.527..342S,2016arXiv160803629W}. The exact properties of the circumplanetary disc are unknown however and hence such estimates should be treated with caution.

However, we find that the apparent motion between the two epochs is too large to be consistent with a physical Keplerian orbit solution. Under the assumption of a co-planar, circular orbit with the outer disc, a distance of $\sim$17\,au between the central star and the companion candiate is derived. The change in position angle between the epochs however implies a distance $\sim$6\,au which cannot be reconciled with the fitted separation of the companion candidate. Also, this scenario cannot explain why the companion candidate was not detected in H-band during the first epoch and in the K-band during the third epoch. Hence we find the probable cause of the strong asymmetries not to be a single companion at $\sim 40$-50\,mas separation. We explore within the next two subsections alternative scenarios, namely a stationary disc rim at 40\,mas or a bright point source on a close orbit ($\sim$20\,mas) influenced by the presence of the disc rim observed in L'-band.

\subsubsection{Disc Rim Scenario}
\label{subsubsec:RimOnly}

Inspired by the detection of a disc rim in L'-band (Section~\ref{subsubsec:LBandDisc}), we test whether the asymmetries detected in H+K band (and their apparent temporal changes) might also be explained with emission from a static disc structure, without invoking the presence of a companion.
Previous work \citep{2013ApJ...762L..12C,2016arXiv160803629W} has shown that the forward-scattered light of an inclined disc rim can cause asymmetries in the brightness distribution that can mimic companion-like phase signals in aperture masking observations.
The presence of such a disc rim with a similar position angle and inclination as the outer disc but with a semi-major axis close to 40\,mas, combined with the differing $uv$ coverages between the epochs could produce a point source-like asymmetry whose position appears to move.
\citet{2013ApJ...762L..12C} found this to be the case when investigating the asymmetry detected in the disc around FL\,Cha, where they too observed apparently non-Keplerian motion.

To test this scenario we constructed a model similar in geometry and orientation to the one described in Section~\ref{subsubsec:LBandDisc}, but leaving the radius of the ring as a free parameter (i.e. an inner ring, co-planar to the ring seen in the L-band data). The ring radius which was found to produce the lowest $\chi^2$ value to the data was selected (110\,mas). We then generated a synthetic data set from these model images for our particular $uv$-coverage. The reconstructed images did not resemble those of a point source nor the structures seen in our real data. In addition, our binary fits to the model data sets found no significant change in the position angle of the the best fit positions between the two epochs and uv-coverages.

Therefore we conclude that static disc models, such as employed in this specific simulation, do not provide a satisfactory fit to our multi-epoch H+K band data. Any static disc asymmetry should also have resulted in a strong closure phase signal in the third epoch (2013-10-20), where we observe no significant signal neither in H or K.

\subsubsection{Companion + Disc Scenario}
\label{subsubsec:CompDisc}

The binary fits presented in Section~\ref{subsubsec:CompOnly} indicate companion candidate separations of $\sim 45$\,mas, that is near the $\lambda/B$ resolution limit. In this regime, the determined parameters can suffer from a degeneracy between separation and contrast ratio (see Section~\ref{subsec:modellingComp}). Also, the parameters could potentially be affected by asymmetric emission from the disc wall that we detected in the L'-band data sets, which could lead to an over-estimation of the fitted companion candidate's separation. Therefore, we explore whether the companion+disc model introduced in Section~\ref{sec:compDisk} could explain the temporal evolution that we see in our aperture masking data between the different epochs.

\begin{figure}
 \begin{center}
\scriptsize
    \includegraphics[width=8cm, angle=0]{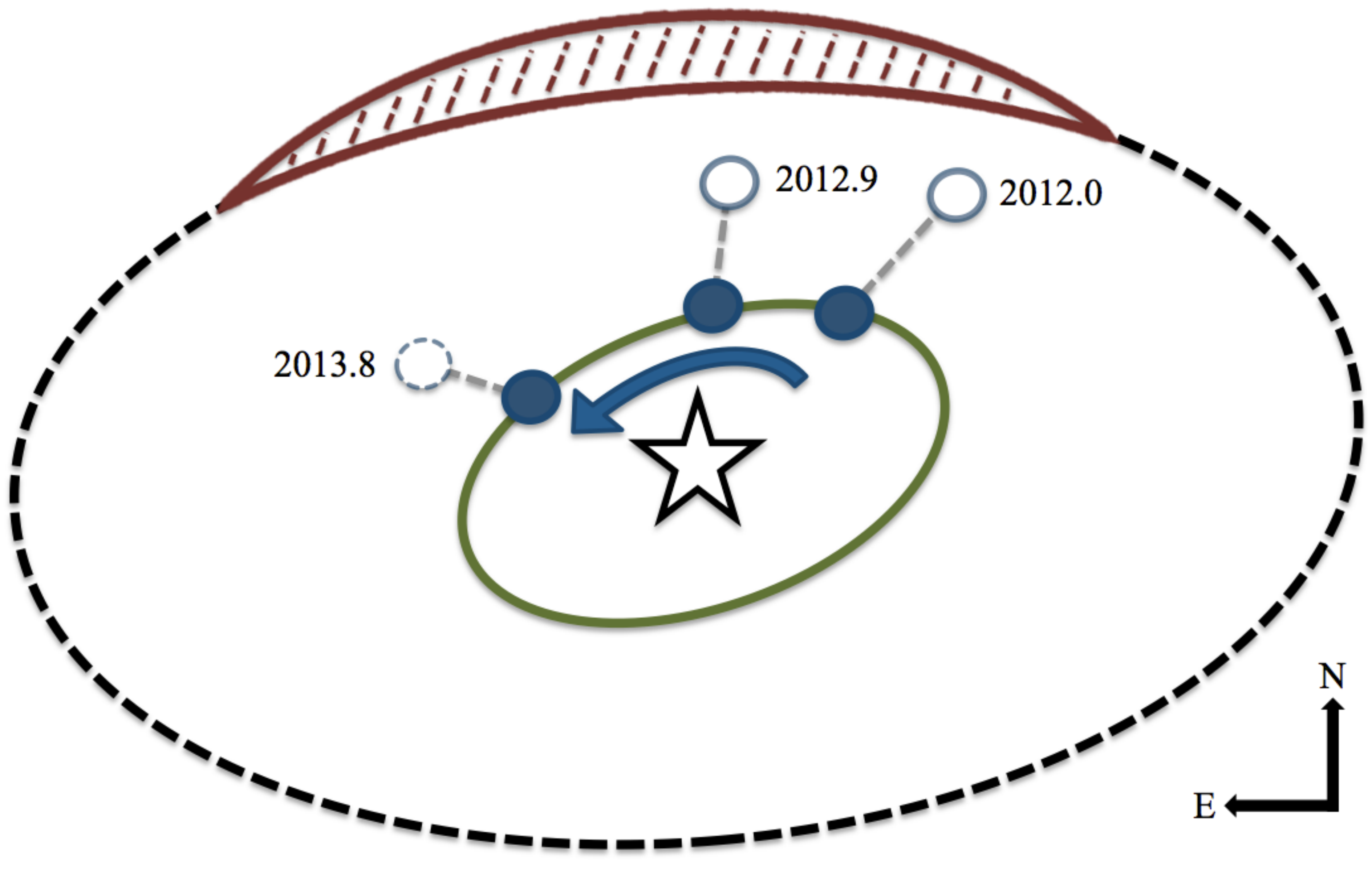} \\
\end{center}
\caption{Simple sketch of the scenario we propose to explain the nonphysically fast change in position angle that is retrieved with simple companion-only fits. Our simulations show that for a close-in companion (solid circles), the best-fit angular separation derived with SAM interferometry (hollow circle) can be over-estimated, if the companion is roughly aligned with asymmetric disc emission (red dashed region).
The strength of this over-estimation is related to several parameters, in particular the disc radius and the difference between the companion candidate position angle and disc position angle.
An unfavourable alignment between the companion and the peak of the disc rim emission or a too-small separation result in no overestimation effect in the derived separation and the companion candidate remains undetected. This seems to be the case at the 2013.8 epoch, where the absence of extended disc emission along the same position angle causes no overestimation. Here, the orbit of the companion candidate and the arc of the disc is misaligned to better represent the position angles found within the previous disc observations of the outer disc ($\sim120^\circ$, green ring) and the position angle of the disc rim from the L'-band observations presented here ($\sim95^\circ$, black ring).}
\label{fig:DiscCompSket}
\end{figure}

A sketch of the proposed scenario is shown in Figure~\ref{fig:DiscCompSket}.
The separation of the companion candidate, as seen in the reconstructed images and derived with interferometry fitting routines, is over-estimated in those epochs, where the position angle of the companion is roughly aligned with the asymmetry introduced by the disc rim.
This is the case for the first two epochs allowing the companion candidate to be observed as a strong asymmetry in the fits and reconstructed images. In the third, the orbital motion of the companion candidate alters the alignment between the companion candidate and peak disc rim emission such that the position angles between the two are no longer such to produce an over-estimation. Here, no over-estimation in the fit to the separation occurs and hence the companion candidate remains undetected, providing an explanation for its absence in the third epoch.

In order to test this scenario and to understand how the presence of a disc rim affects the retrieved parameters in a binary fit, we construct a suite of models that contain a disc wall similar to the one found in the L'-band observations (semi-major axis\,=\,150\,mas, semi-major axis PA\,=\,95$^\circ$, disc contrast ratio = 0.04) and with a bright point source similar in contrast ratio and position angle to the companion candidate found within the 2012-01-09 K-band observations (see Table~\ref{tab:binresultsHK}). However, we place the point source at a separation that would be consistent with a Keplerian orbit ($\rho = 20\,\textrm{mas}$) between the first and second epoch in our new data set. We keep the point source position angle the same as the fitted position angle, 308$^\circ$. We use the $uv$-coverage and noise from the 2012-01-09 K-band observations to generate simulated data sets. We then vary within our models a single parameter of either the disc or point source, namely either the disc major axis, major-axis position angle, disc inclination, disc width, disc flux ratio, disc skewness, or point-source flux ratio.

\begin{figure*}
 \begin{center}
\scriptsize
$\begin{array}{ @{\hspace{-1.0mm}} c @{\hspace{-5.0mm}} c @{\hspace{-5.0mm}} c}
    \includegraphics[height=7cm, angle=0]{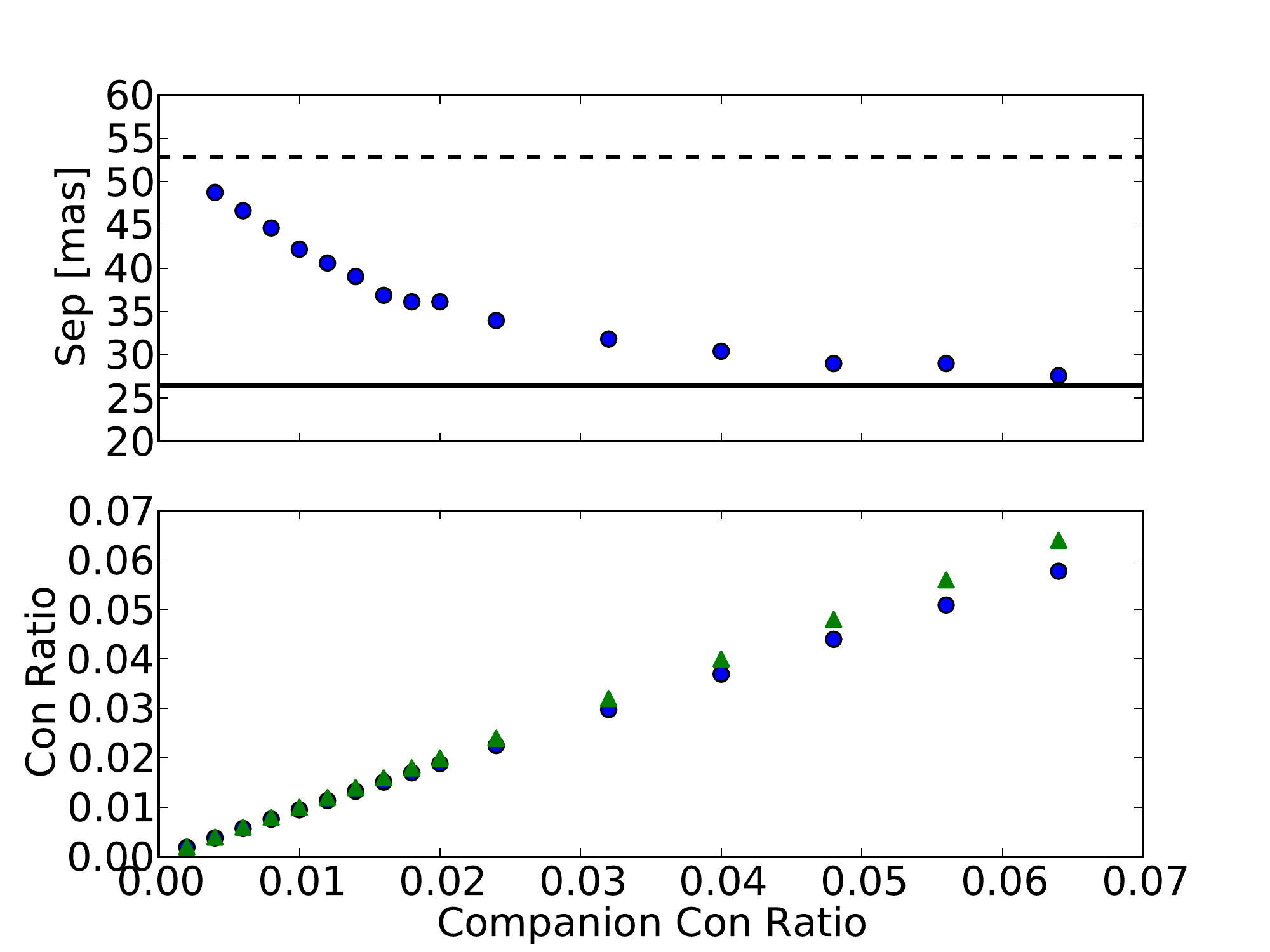} & \includegraphics[height=7cm, angle=0]{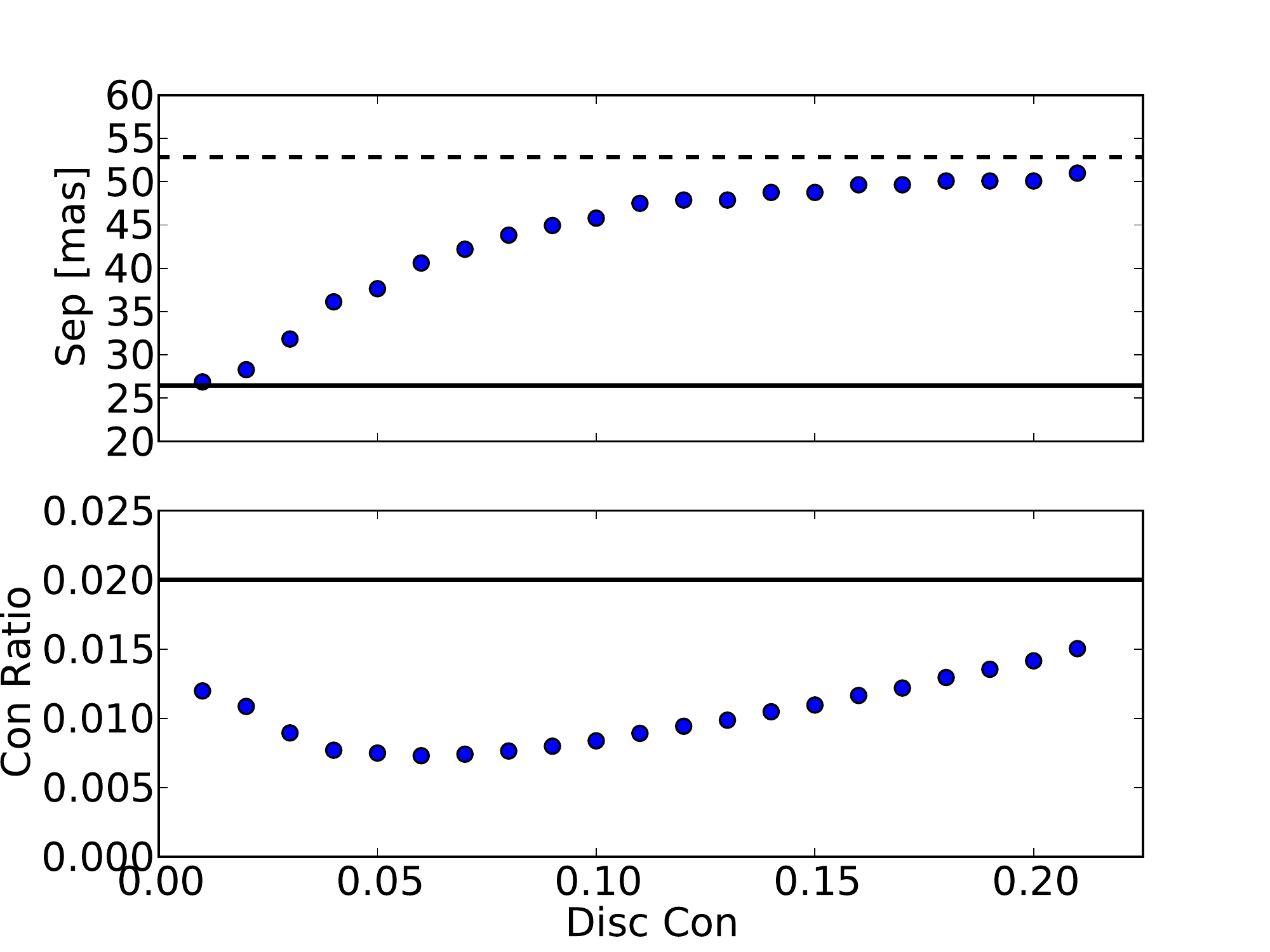} \\ \includegraphics[height=7cm, angle=0]{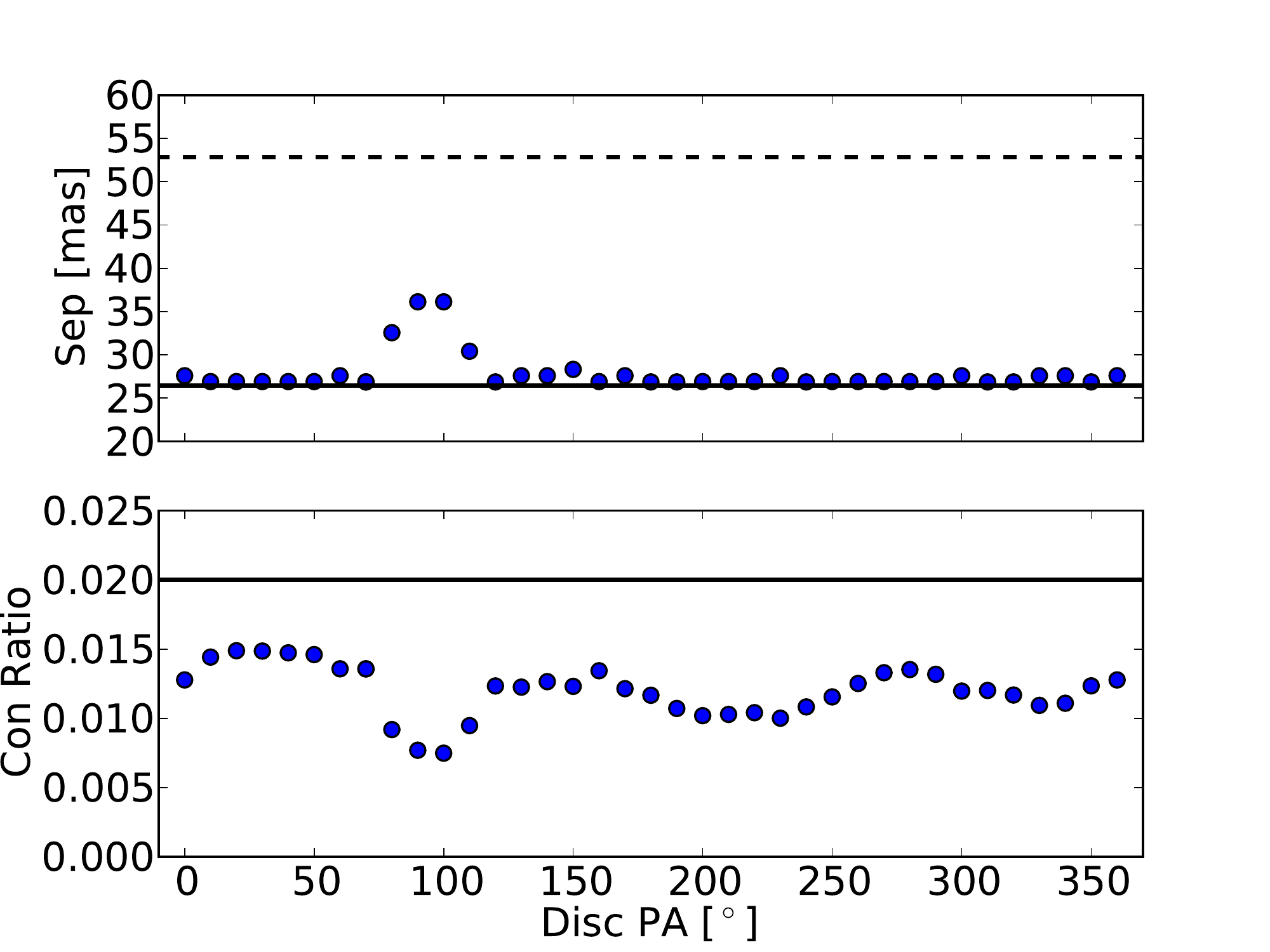} &
    \includegraphics[height=7cm, angle=0]{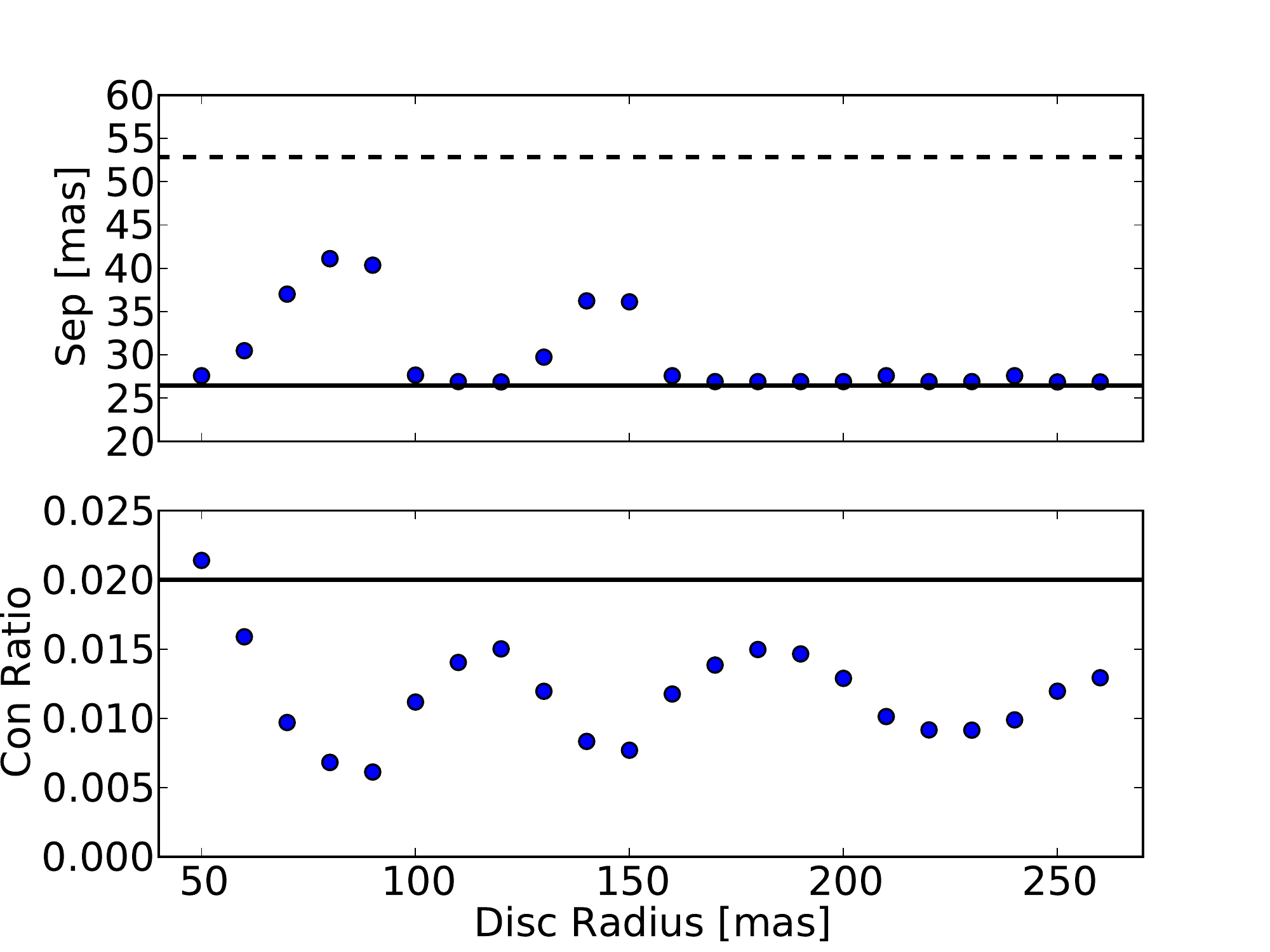} \\
 \end{array}$
\end{center}
\caption{Simulated data sets to investigate the effect on the position and contrast ratio of a point source at 20\,mas. Each contains an upper and lower panel displaying how the fitted separation (upper) and contrast ratio (lower) depends upon different disc parameters. \textbf{Top left:} Varying point source contrast ratio (model - blue circles, best fit - green triangles). \textbf{Top right:} Varying disc contrast, $f_{\mathrm{disc}}/f_{\mathrm{total}}$. \textbf{Bottom left:} Varying disc major-axis position angle. \textbf{Bottom right:} Disc semi-major axis. In the upper panels, the dashed and solid black lines represent $\lambda/B$ and $\lambda/2B$ resolution limits of the uv-plane respectively. $\lambda/2B$ ($\sim$26\,mas) represents the minimum separation for which we search for a companion. Within the lower panels, the solid black line represents the true contrast ratio of the point source in the absence of a disc. We never find a point source at this contrast ratio as a result of the minimum separation limits we place on the fitting routine. At smaller separations the algorithm struggles to find a minimum in the $\chi^2$. We find that the separation is over-estimated in cases where the position angle of the point source and of the disc semi-major axis are approximately orthogonal and where the disc radius is such to produce an enhancement to the closure phase signal at short baselines.}
\label{fig:DiscCompSim}
\end{figure*}

\begin{figure*}
 \begin{center}
\scriptsize
$\begin{array}{ @{\hspace{-1.0mm}} c @{\hspace{-0.5mm}} c @{\hspace{-4.5mm}} |  c @{\hspace{-4.5mm}} | c }
   \includegraphics[height=4cm, angle=0,trim={0.15cm 0.15cm 0.0cm 0.05cm},clip]{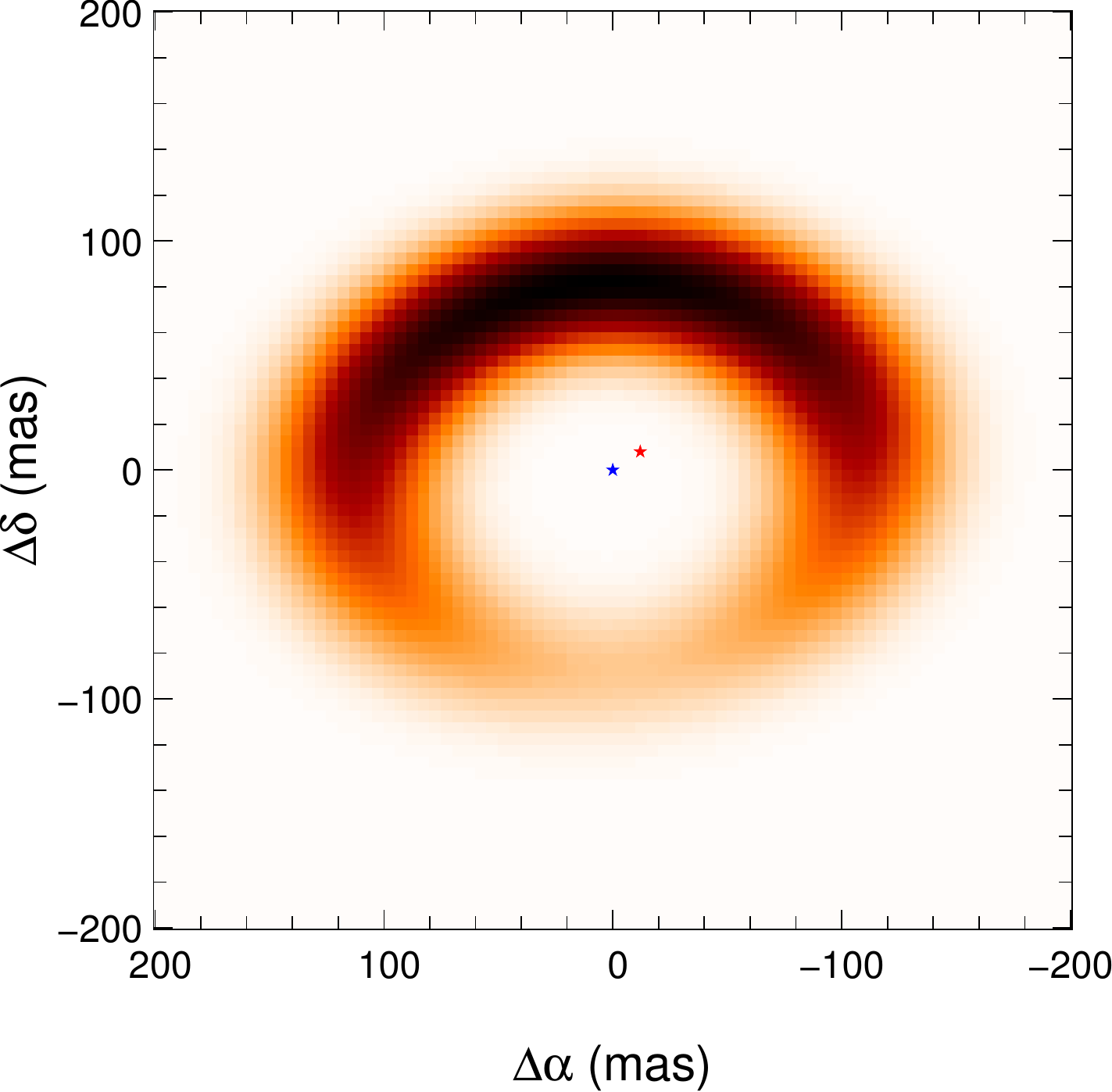} &
   \includegraphics[height=4cm, angle=0,trim={0.5cm 0.1cm 0.65cm 1.25cm},clip]{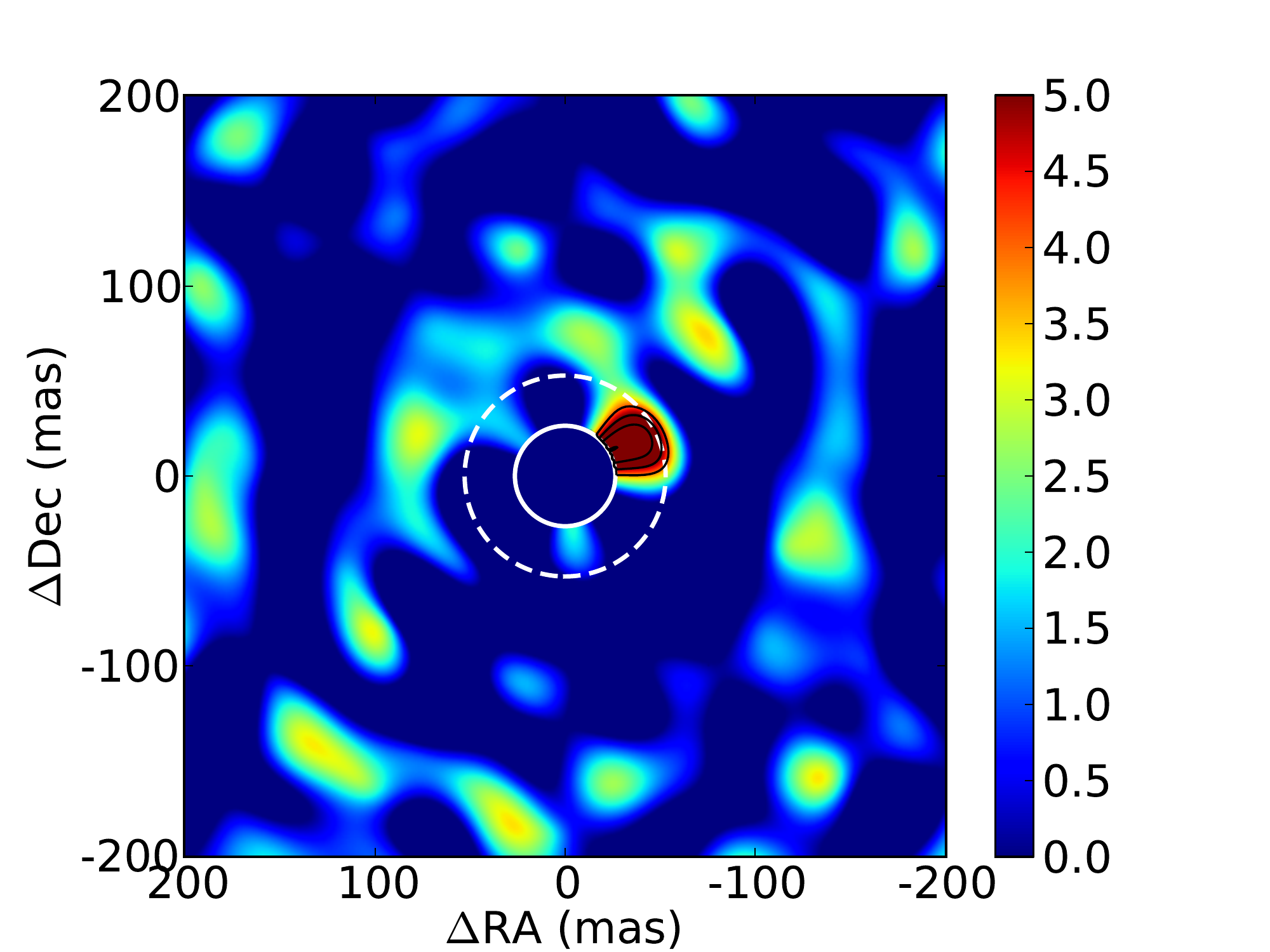} &
   \includegraphics[height=4cm, angle=0,trim={0.5cm 0.1cm 0.65cm 1.25cm},clip]{V1247Ori_KBand_120108_sigMap_phys.pdf} & \includegraphics[height=4cm, angle=0,trim={0.5cm 0.1cm 0.65cm 1.25cm},clip]{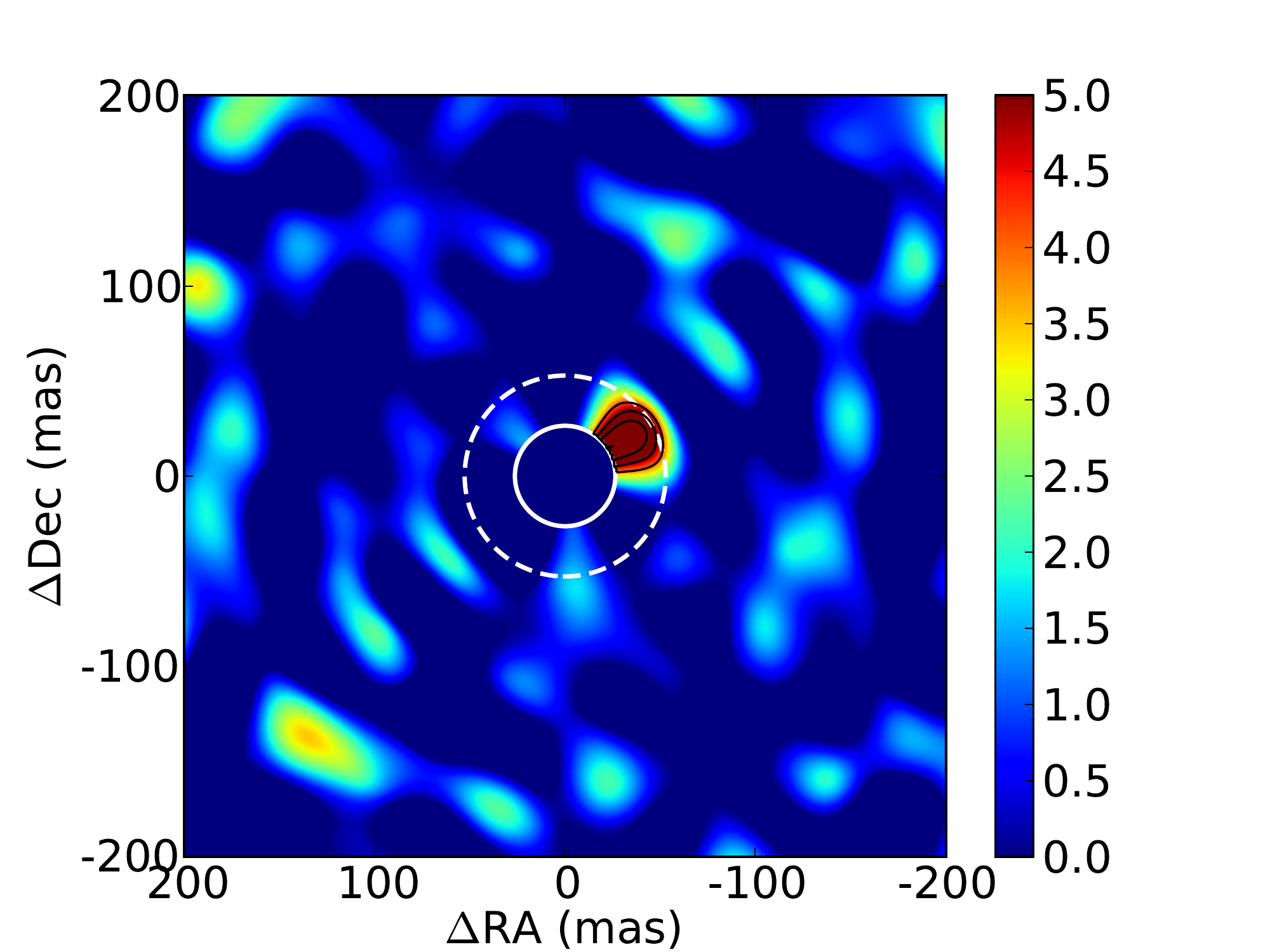}\\
   \includegraphics[height=4cm, angle=0,trim={0.15cm 0.15cm 0.0cm 0.05cm},clip]{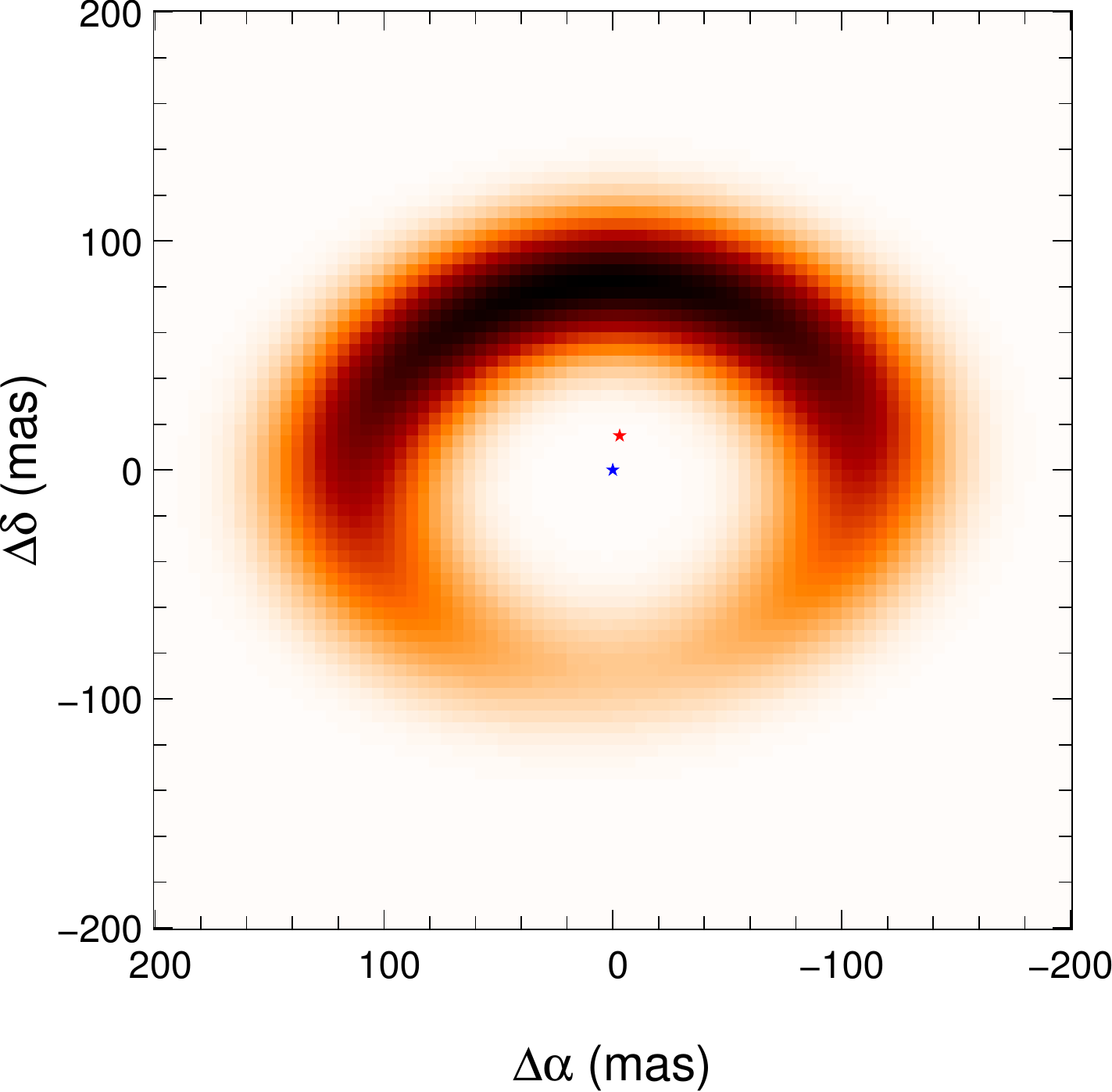} &
   \includegraphics[height=4cm, angle=0,trim={0.5cm 0.1cm 0.65cm 1.25cm},clip]{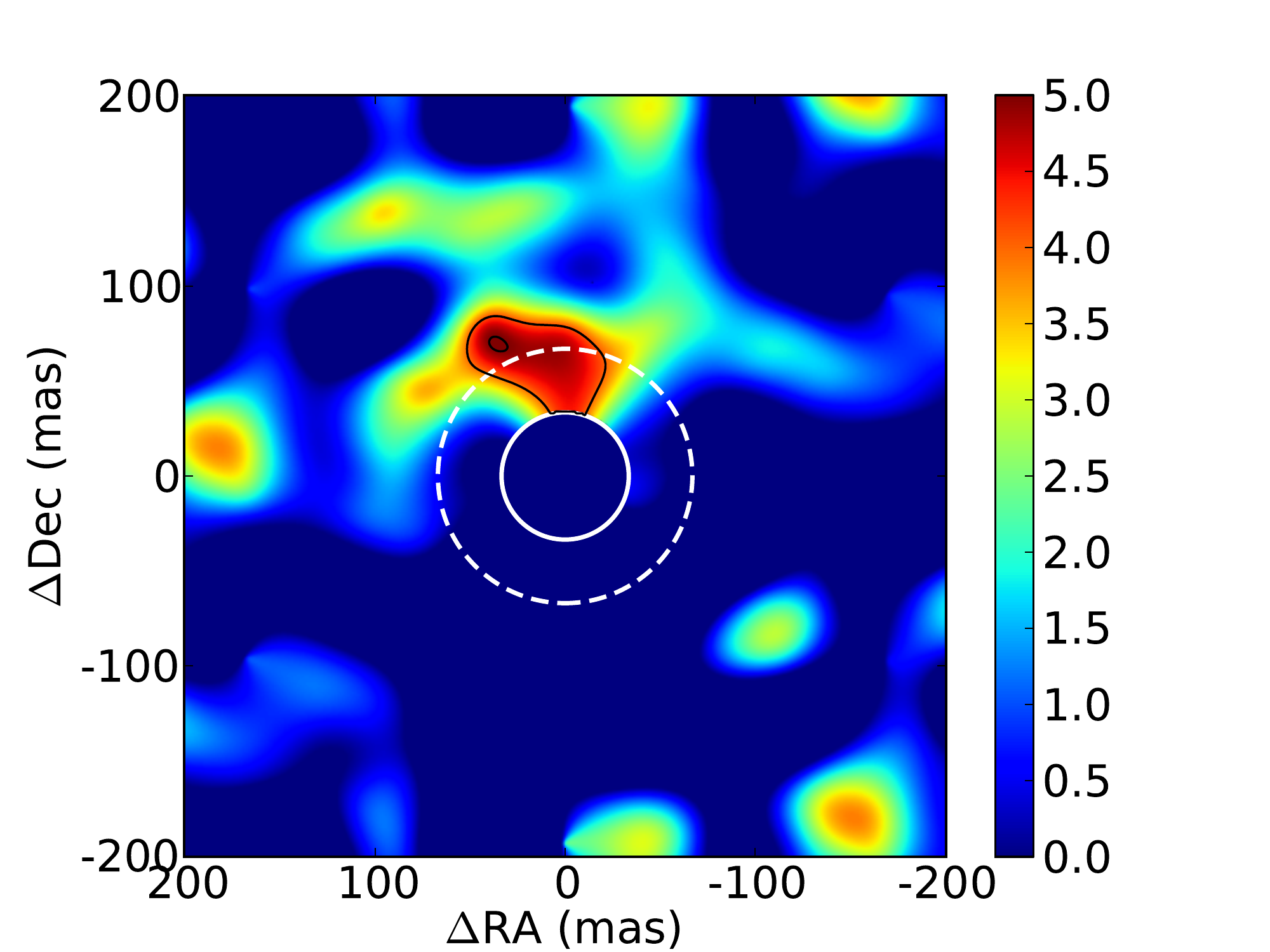} &
   \includegraphics[height=4cm, angle=0,trim={0.5cm 0.1cm 0.65cm 1.25cm},clip]{V1247Ori_KBand_121218_sigMap_phys.pdf} & \includegraphics[height=4cm, angle=0,trim={0.5cm 0.1cm 0.65cm 1.25cm},clip]{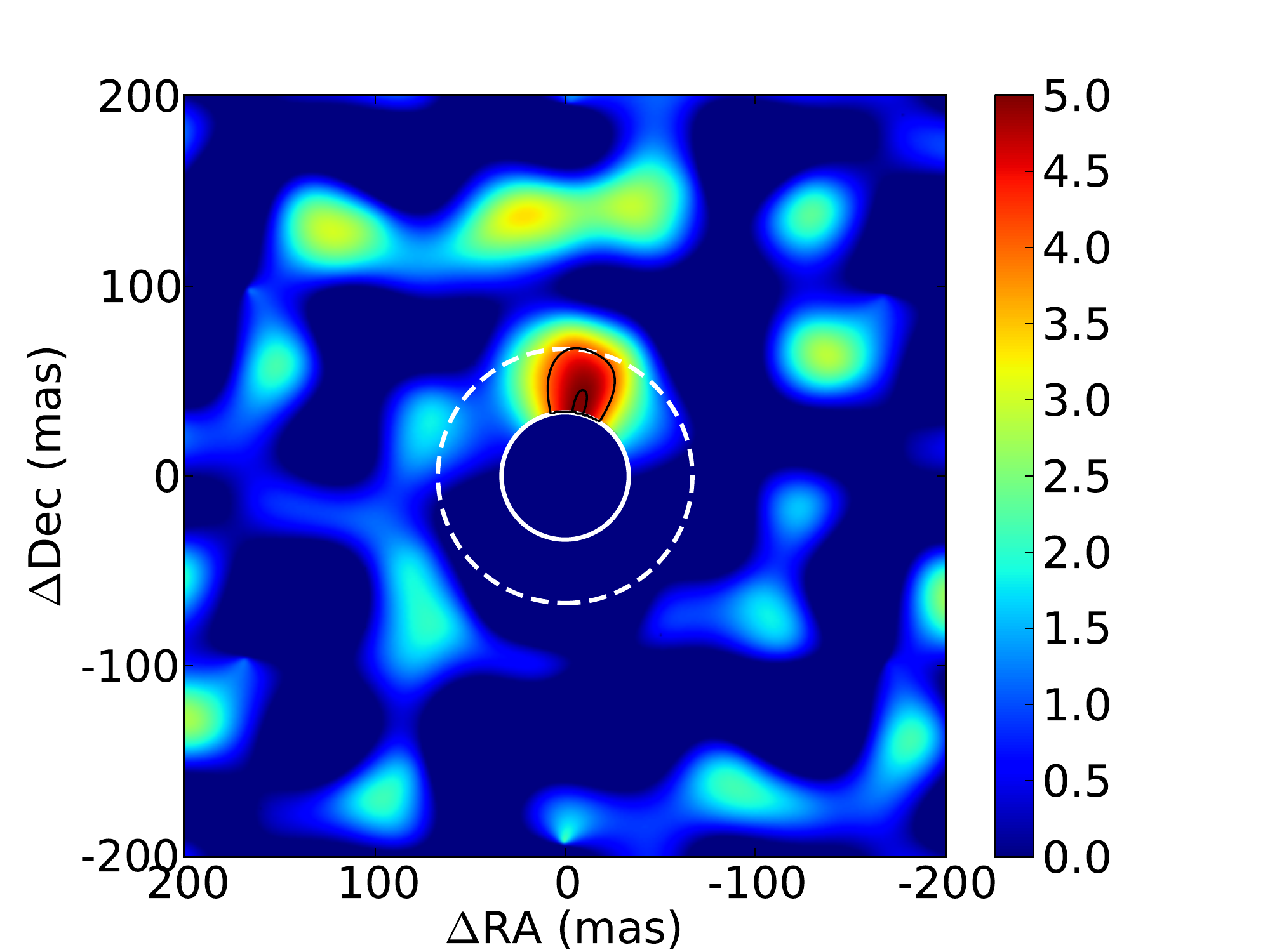}\\
   \includegraphics[height=4cm, angle=0,trim={0.15cm 0.15cm 0.0cm 0.05cm},clip]{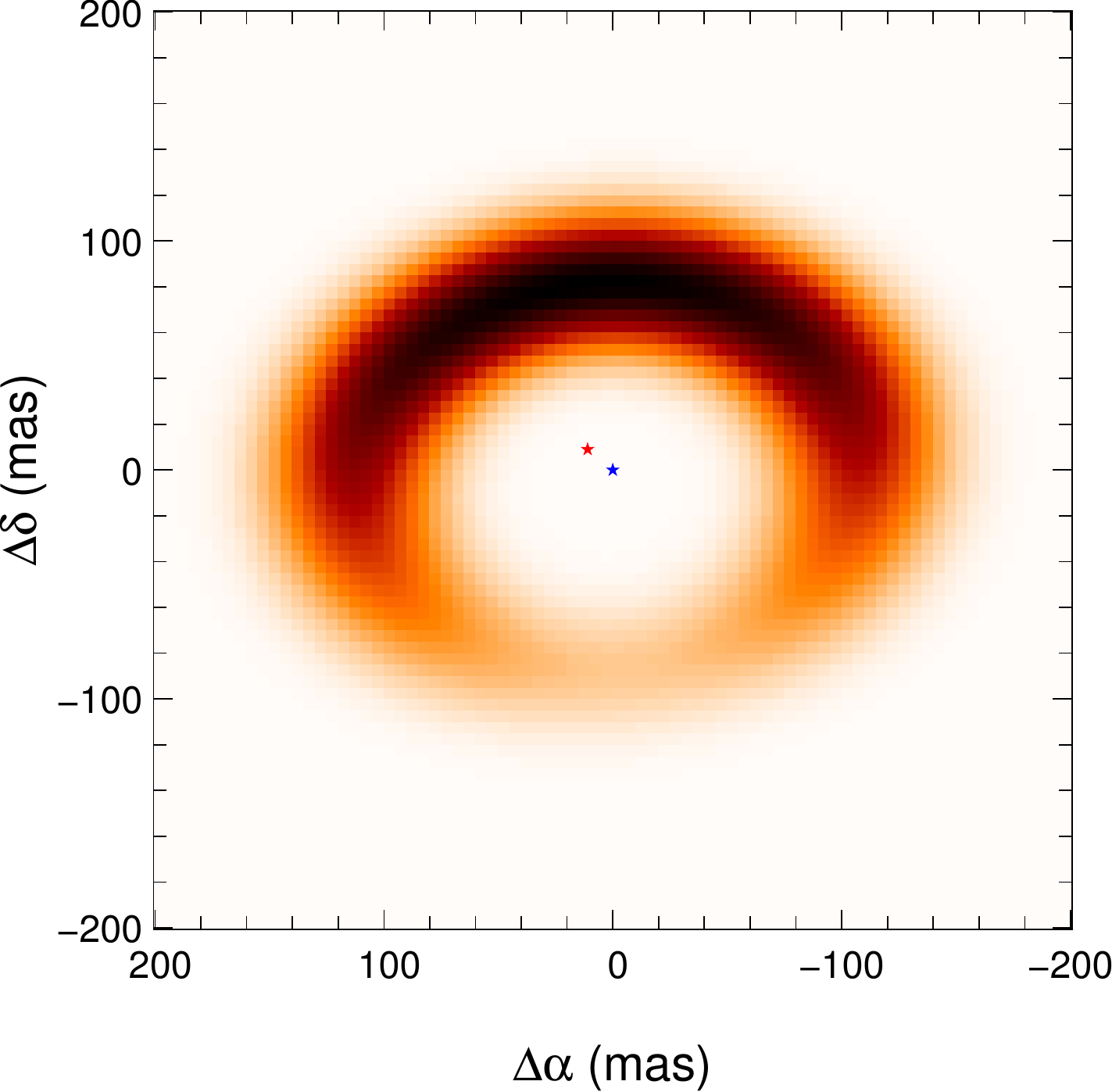} &
   \includegraphics[height=4cm, angle=0,trim={0.5cm 0.1cm 0.65cm 1.25cm},clip]{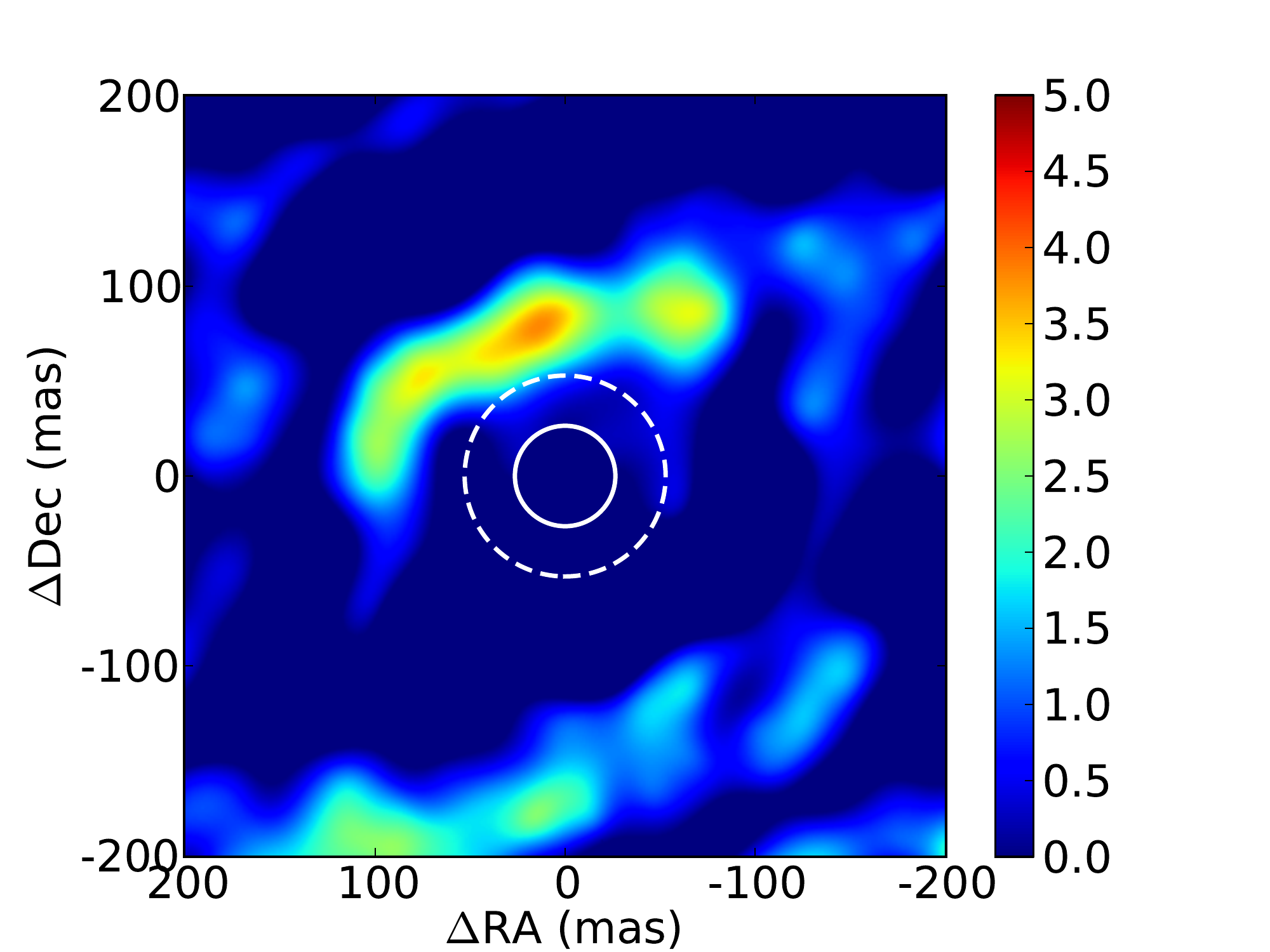} &
   \includegraphics[height=4cm, angle=0,trim={0.5cm 0.1cm 0.65cm 1.25cm},clip]{V1247Ori_KBand_131020_sigMap_phys.pdf} & \includegraphics[height=4cm, angle=0,trim={0.5cm 0.1cm 0.65cm 1.25cm},clip]{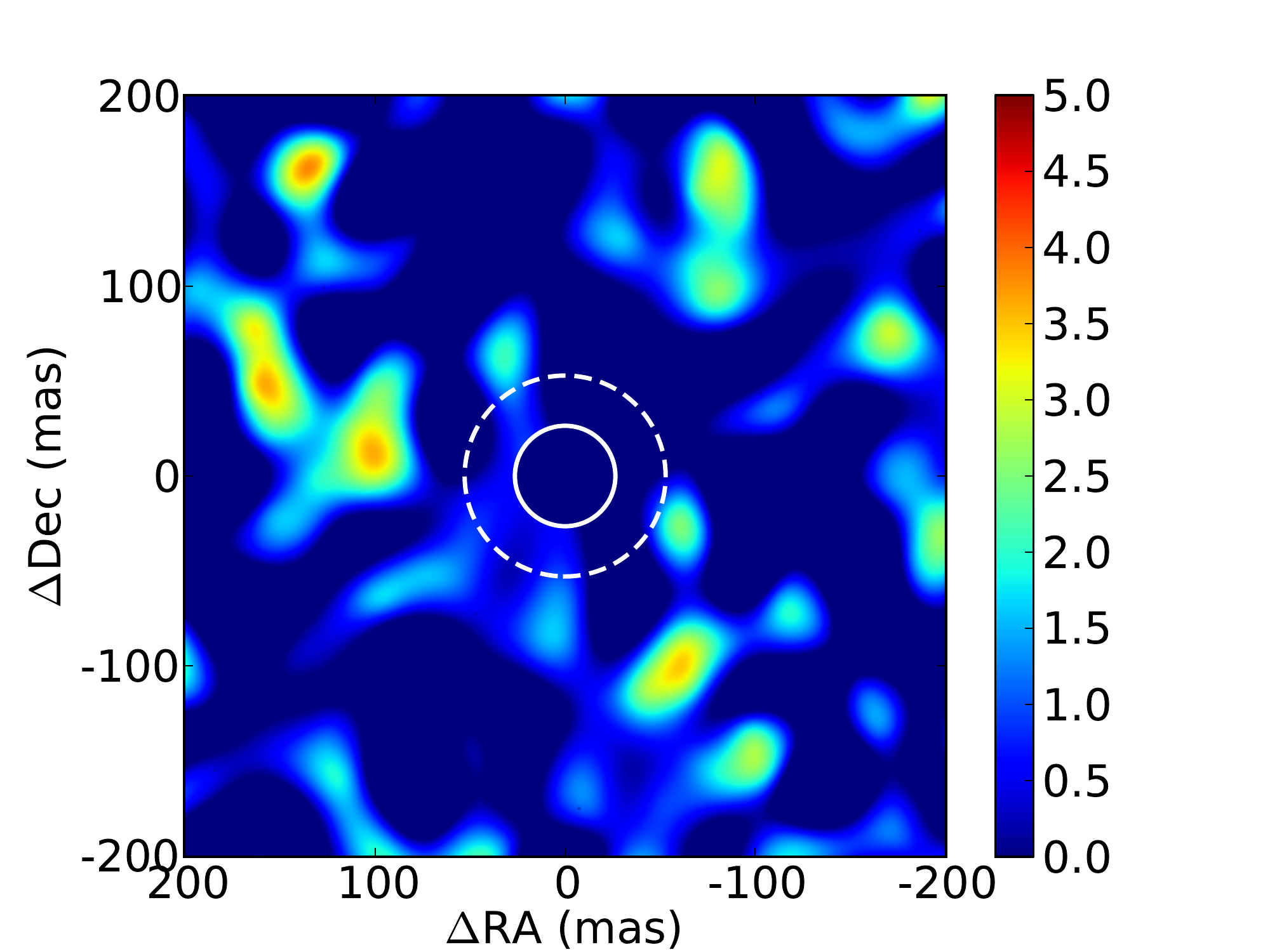}\\
        \end{array}$
\end{center}
\caption{Figure showing a comparison of significance maps for a number of models of the environment around V1247\,Ori. The first column displays the input model used to calculate model closure phases, while the second shows the resulting significance maps from the disc+companion model data sets. The real SAM data obtained for V1247\,Ori SAM is shown in the third column. Finally we show results of a binary fit from a companion-only model (fourth column). The different rows show the three epochs covered by our observations, namely 2012-01-09 (top row), 2012-12-18 (middle row), and 2013-10-20 (bottom row). The significance maps for the simulated data sets are normalised for the strongest detection to be the same as in the corresponding real data set. The disc+companion models (first and second column) reproduces the measured data (third column) more readily than companion-only scenario (fourth column). The dashed and solid circles correspond to the sample resolution limits described for the degeneracy plots, where the dashed line is $\lambda/B$ and the solid line is $\lambda/2B$ approximately defining the region under which the contrast ratio and separation of a point source are less well constrained.
The significance map colour bars show the significance of the binary fit to the data at each pixel location ($\sigma$).}
\label{fig:DiscCompComparisons}
\end{figure*}

The results of our simulations are shown in Figure \ref{fig:DiscCompSim}.  We find that the separation can indeed be significantly over-estimated in the presence of the asymmetric disc emission, namely by up to a factor of 2, if the point source and the peak emission from the disc rim are roughly aligned along a similar position angle. This effect appears in both the image reconstructions and our binary fits to the closure phase data.

We find that the precise over-estimation factor in the fitted separation of the point source is most sensitive to the disc radius, disc position angle, and disc flux. We also find dimmer companions to be more susceptible to larger shifts in their fitted separations compared to brighter point sources. Inversely the brighter point sources' contrasts are more affected by the presence of a disc than the dimmer point sources (Figure \ref{fig:DiscCompSim}, top-left panel). Fractionally however, the fitted contrast ratio is most affected when the companion is dimmer, with the effect weakly increasing approximately logarithmically as the magnitude of the companion is reduced with the worst affected by around 10\%.  When varying the disc flux contribution (Figure \ref{fig:DiscCompSim}, top-right panel) we do observe that the over-estimation effect does not shift the fitted position of the point source outside of $\lambda/B$, except when the flux becomes large enough for the disc asymmetries to dominate, and the best fit position becomes a point co-located on the disc rim.
Additionally we only observe a significant overestimation in separation within a narrow window of the position angle of the disc semi-major axis (Figure \ref{fig:DiscCompSim}, lower left panel). For our particular simulation setup, the strongest fitted separation over-estimation occurs for disc semi-major axis position angles between 90-100$^{\circ}$, but some effects are present over the range from 70-120$^{\circ}$.
The relationship between disc radius and the over-estimation is particularly complex (Figure \ref{fig:DiscCompSim}, lower-right panel), displaying a decaying sinusoidal relationship similar to a first order Bessel function with the over-estimation in the fitted separation greatest between disc radii of 70-90\,mas and weakest between 100-120\,mas. This pattern repeats between 140-150\,mas and out beyond 160\,mas.
It is also possible to understand the reason for this relationship, and for the occurrence of the overestimation effect, intuitively by considering the Fourier phases of the different components in the uv-plane. The sine-wave phases associated with a binary brightness distribution are modulated by a Bessel function that corresponds to the Fourier phases of a ring-like disc feature. Ignoring the modulation and fitting with the simple sine function will therefore result in a bias in the fit. The mean phase over all baselines determines the strength of the overestimation. Where the disc phases pass through nulls determines therefore how the strength of the overestimation is modulated for different disc radii.

In our simulations with extended disc emission, we find that the point source flux is always under-estimated compared to the true value, producing smaller contrast ratios.
The difference between the simulated contrast ratio of the point source and the retrieved, best-fit contrast ratio never exceeds more than a factor of three.

Following this parameter study on the influence of disc emission on binary fits, we attempted to fit our specific five aperture masking data sets on V1247\,Ori with this model. For this purpose, we considered computing a full grid of all model parameters, but found that this proved unfeasible due to the large number of free parameters involved. To simplify the problem we therefore fixed the parameters of the disc rim to the values  previously inferred from VLTI long-baseline interferometry \citep[disc semi-major axis\,=\,110\,mas, disc semi-major axis PA\,=\,104$^\circ$;][]{2013ApJ...768...80K}. The flux contributions of the disc were fixed using the visibility data (fraction of the total flux, $f_{\mathrm{disc}}/f_{\mathrm{total}}$\,=\,0.02, where $f_{\mathrm{total}}$ is approximately equal to $f_{\mathrm{star}}$). We then constructed synthetic data sets of each epoch, using the noise characteristics and $uv$-coverage of each night with K-band data and insert an point source with a contrast ratio of $f_{\mathrm{disc}}$\,=\,0.02$f_{\mathrm{star}}$ and adjust the position angle of the companion so the observed position angle change between the different epochs is consistent with Keplerian motion. This implies an orbital separation of 6\,au (values displayed in Table~\ref{tab:binresultsComp}). We find that including the disc rim allows us to better reproduce the structure seen in the significance maps. For instance, the model is able to reproduce the complex structure seen in the second epoch better including the artefacts to the south, and it readily reproduces the low-significance structures seen to the north-east in the third epoch (Figure~\ref{fig:DiscCompComparisons}).

Repeating these simulations for the H-band data sets, we find that we can also reproduce the measured data with the disc+companion models, although repeating the parameter study is more difficult as a result of the narrower range of disc position and orientation parameters over which the overestimation occurs in H-band.
We can reproduce the second H-band observation (2012-12-18), including the overestimation of the separation in the fit.

We resolve the inserted point source in the 2012-01-10 H-band model but at a contrast ratio below our 99\% confidence level in the real data set ($\Delta\,f_{model}\,=\,5.8,$, $\Delta\,f_{99\%}\,=\,5.6$), which provides an explanation for its absence within the significance map and reconstructed image. Our ability to detect the point source in the simulation is likely a result of under-representing the noise in the synthetic data sets. Evidence for this can be found in the third epoch of the K-band simulations where we retrieve more of the extended portions of the disc emission in the simulated data compared to the real data.
Our simulations also show that the separation is over-estimated in the VLT/NACO H-band data set as the achievable resolution in H-band with this instrument is similar to the Keck-II/NIRC2 K-band data sets.
Once the disc radius is extended enough to reach a phase null on the longest baselines, the effect on the mean phase is minimal and hence no longer produces a detectable over-estimation (Figure \ref{fig:DiscCompSim}, lower-right panel). The point source signal is therefore largely constrained by the non-zero closure phase signal on the shorter baselines.
The lack of longer baselines makes disentangling the disc phases from the companion phases in the second epoch data set even more difficult compared to the first epoch. This acts to amplify the binary signal in the second epoch H-band data, providing a potential explanation for the strong binary signal we observe. We see additional evidence for the influence of the disc rim in the visibility data in both the H and K-band data sets (see Figure~\ref{fig:PotCompanions}, far right panel), as we observe a drop in the squared visibilities, which indicates that we at least partially resolve the disc rim.

While these simulations show that a point source at $\sim$15-20\,mas moving with a Keplerian velocity can reproduce the observed data, the scenarios presented here are not exhaustive.
\begin{table*}
\centering
\caption{Companion Candidate Properties}
\label{tab:binresultsComp}
\begin{tabular}{c c c c c c c c}
\hline \hline
    Identifier & $\rho$ & PA & Sig & Semi-Major Axis & $M_H$ & $M_K$ & $ M_{c} \dot M_{c}$ \tablefootmark{a} \\
	 & [mas] & [$^{\circ}$] & [$\sigma$] & [AU] & [mag] & [mag] & [10$^{-6}$ M$^2 _J$yr$^{-1}$] \\
\hline

V1247\,Ori\,b & 43$\pm$5 & 308$\pm$3 & 6.9 & 17 & --- & 4.7$\pm$0.2 & 10$^{-3}$ \\
V1247\,Ori\,b & 40$\pm$5 & 353$\pm$3 & 7.0 & 17 & 4.8$\pm$0.2 & --- & 10$^{-3}$ \\
\hline
V1247\,Ori\,b & 18 & 308 & --- & 6 & 4.2 & 3.6 & 10$^{-3}$ \\
V1247\,Ori\,b & 15 & 353 & --- & 6 & 4.2 & 3.6 & 10$^{-3}$ \\
V1247\,Ori\,b & 13 & 45 & --- & 6 & --- & 3.6 & 10$^{-3}$ \\

\end{tabular}
\tablefoot{Columns are organised by Identifier, angular separation, position angle, significance, orbital separation based on previous observations of the disc inclination and position angle, absolute magnitude of the companion candidate, stellar accretion rate and companion candidate mass. Dereddening was performed as described in \citet{1989ApJ...345..245C}. The rows are are in chronological order with the second half detailing the model parameters used in Figure \ref{fig:DiscCompComparisons}.
\tablefoottext{a}{Values derived from \citet{0004-637X-799-1-16} assuming circumplanetary disc radii extending from 2 $R_J$ outwards.}}
\end{table*}

\subsection{SPHERE}
\label{subsec:SPHERE}

\begin{figure}
 \begin{center}
\includegraphics[width=9cm, angle=0]{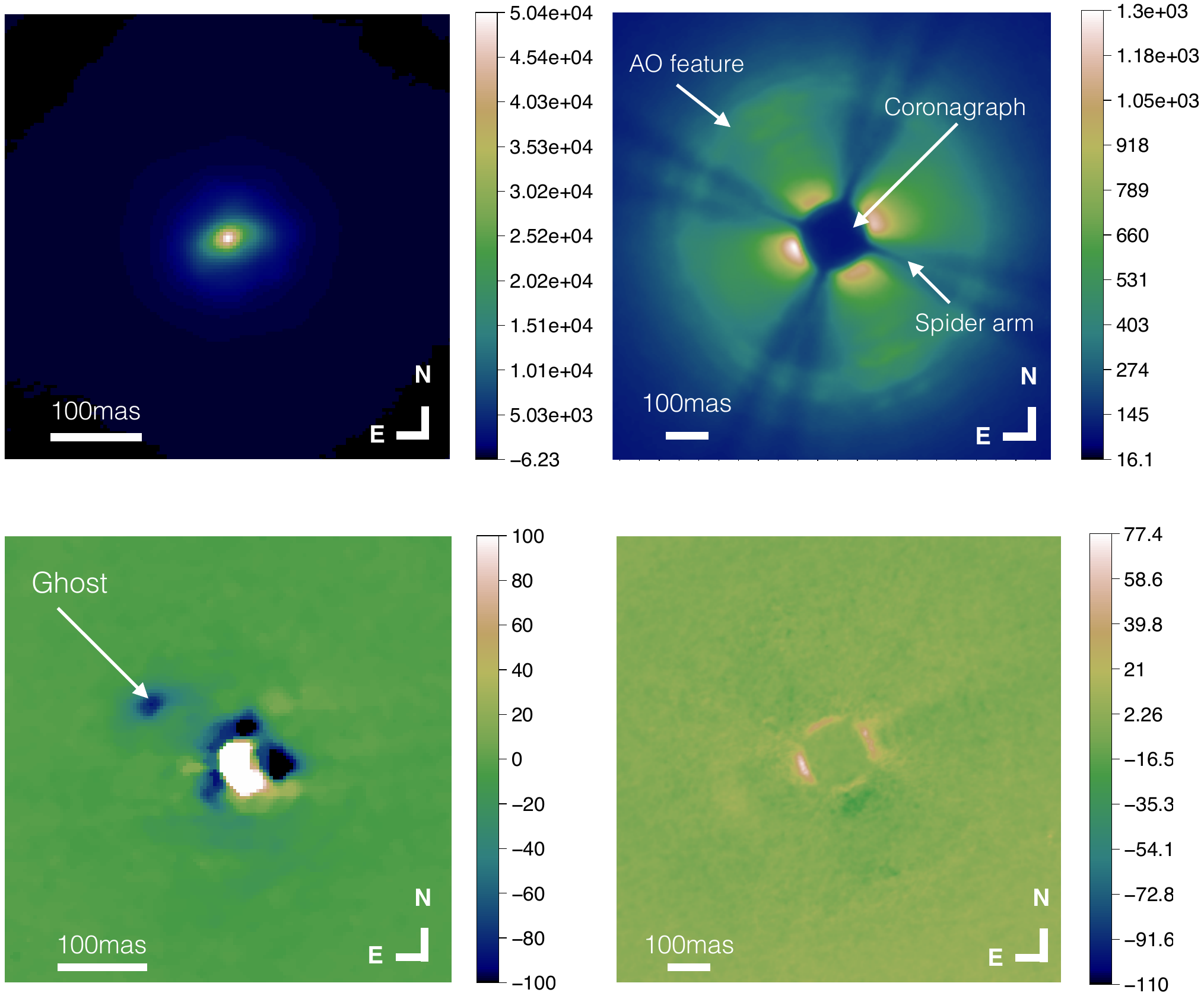}
\end{center}
\caption{SPHERE/ZIMPOL images of V1247 Ori. \textbf{Top Left:} Intensity image of the target without coronagraph. We see an elongated structure. \textbf{Top Right:} Intensity image with the coronagraph. \textbf{Bottom Left:} SDI image without the coronagraph. The ghost is located at $\sim$110mas from the central star. \textbf{Bottom right:} coronagraphic SDI image.}
\label{fig:ZIMPOLimages}
\end{figure}

The high contrast imaging enabled by SPHERE/ZIMPOL provides the opportunity to search for on-going accretion onto forming protoplanets through SDI and coronagraphic imaging in the H$\alpha$ line. In the non-coronagraphic images (Figure \ref{fig:ZIMPOLimages}) we see that the emission from the immediate area around V1247\,Ori on small separations ($<$100\,mas) differs from a point source in both filters, where the difference between the two filters is negligible.
The emission is elongated along a position angle of 108$^\circ\pm$10$^\circ$, which is consistent with the disc orientation. Likely, we see stellar light scattered from the surface of the disc.

\begin{figure}
 \begin{center}
    \includegraphics[width=8cm, angle=0]{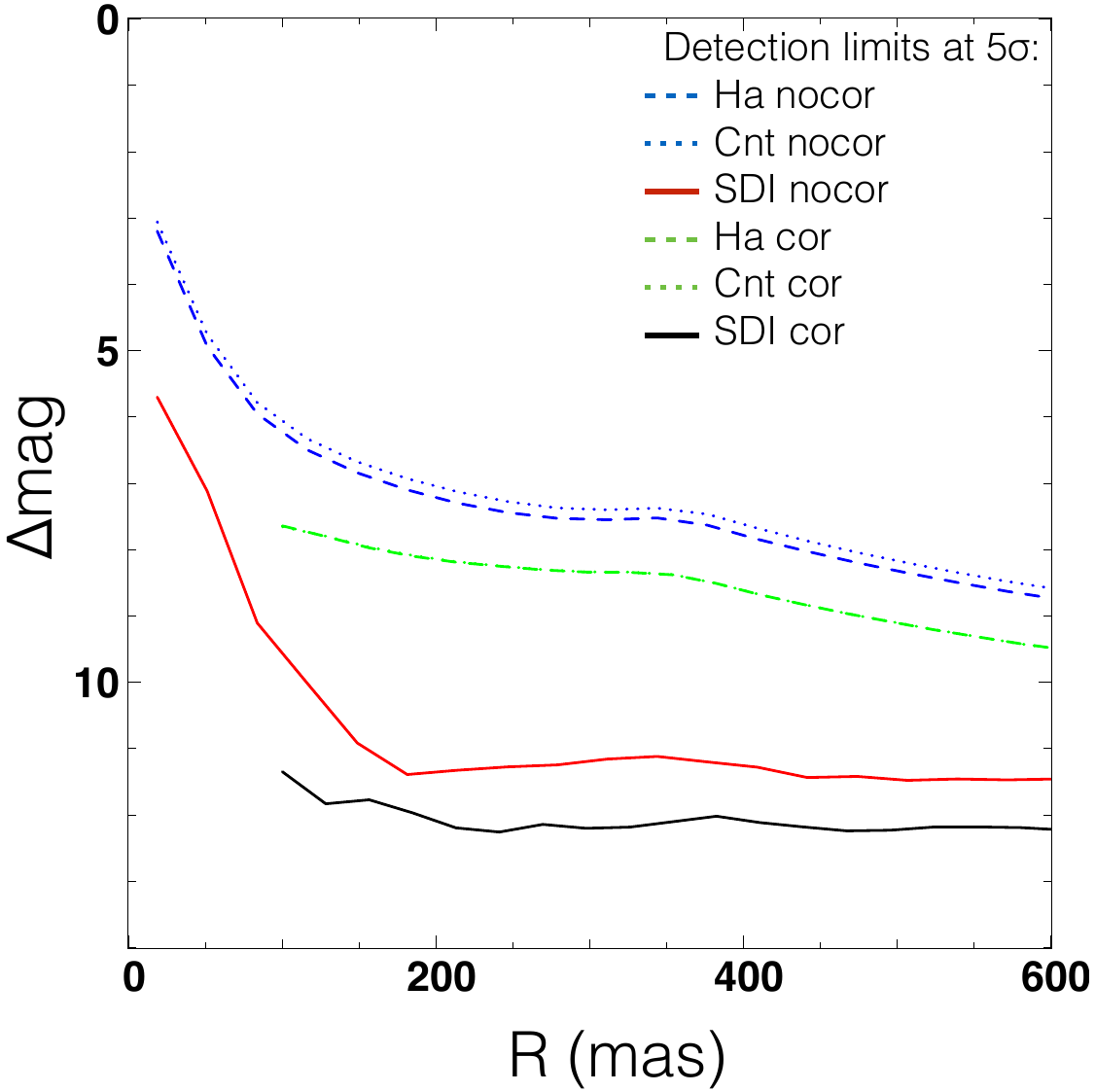}
\end{center}
\caption{5-$\sigma$ detection limits derived from the SPHERE/ZIMPOL data set and plotted as function of the angular distance to the star with coronagraph (cor) and without coronagraph (nocor). The subtraction between the two filters allow a gain of between 2 and 4 magnitudes in contrast.}
\label{fig:ZIMPOLDetlim}
\end{figure}

Our SPHERE images do not reveal any point source in the inner 100-600\,mas region around V1247\,Ori, besides the instrumental 'ghost' reflection outlined in Section~\ref{subsec:sphere}, and lack the angular resolution to detect the companion candidate discussed in Section~\ref{subsubsec:CompDisc}. Therefore, we derive detection upper limits from our images, where we avoid the area near the reflection ('ghost') in the continuum filter.
To compute the detection limits we have first defined the zero point as the peak flux of the median image without the coronagraph (and that were divided by the corresponding dit) for each filter.
We then corrected for the loss of flux due to the presence of the Lyot stop and the coronagraphic mask itself by multiplying the coronagraphic images with a factor of 1.1 so their profiles match the non-coronagraphic images profiles (see Fig.\,\ref{fig:profiles}).
We then estimated the noise in concentric annuli having a width of 18\,mas.
We multiplied the obtained value per five to obtaine the 5-$\sigma$ flux limit.
In the intensity images we derive upper limit contrasts of $\sim$11 magnitudes at 200\,mas without the coronagraph and $\sim$12 magnitudes with it (Figure~\ref{fig:ZIMPOLDetlim}). Without the coronagraph mask we reach typical contrast limits of 3-6 magnitudes in the H$\alpha$ filter for separations between 25-100\,mas. These values are consistent with the prediction by the SPHERE/ZIMPOL exposure time calculator. Applying the SDI subtraction results in a gain of $\sim$2 magnitudes on the detection limit within 100\,mas and a gain of $\sim$4 magnitudes from 100-400\,mas with this gain shrinking steadily to within $\sim$3 magnitudes beyond 400\,mas.
The SDI detection limits are derived with the assumption that the companion flux ratio between the H$\alpha$ and the continuum filters is high (f$_{H\alpha}$/f$_{cont}>$5) allowing a direct computation of the detection limit \citep{2015A&A...581A..80R}.

From the H$\alpha$ detection limits we calculate upper limits for the accretion rate onto an accreting body. Our method is identical to that outlined in \citet{2012A&A...548A..56R} and \citet{2014ApJ...781L..30C} after correcting for the stellar H$\alpha$ contribution and assuming a mass of 1\,M$_\textrm{J}$ for the accreting object.
Assuming that there exists equal extinction for V1247\,Ori\,A and a potential companion, and that the accretion-luminosity relation for low-mass T~Tauri stars also applies to a hypothetical companion, we find this corresponds to a range of $M \dot M$ values between 10$^{-3}$-10$^{-4}$ M$^2_J$yr$^{-1}$ for the non-coronagraphic data sets (seven\,-\,nine magnitudes in H$\alpha$). With the coronagraph mask in place we achieve higher contrast limits (eight\,-\,ten magnitudes in H$\alpha$) which under the same assumptions corresponds to $M \dot M$ values between 10$^{-4}$-10$^{-5}$ M$^2_J$yr$^{-1}$. These values are highly dependent on the value of extinction in R-band, $A_R$, which is likely substantially higher within the disc gap than towards the central star.
Such is indeed likely in the case of V1247\,Ori, due to the known substantial population of dusty material still resident within the gap \citep{2013ApJ...768...80K}.
We therefore rule out a companion similar in H$\alpha$ brightness to LkCa\,15\,b between separations of 100 to 600\,mas, or such a companion would need to be more deeply embedded in V1247\,Ori than found for LkCa\,15\,b.

In a similar manner we calculated the accretion luminosity of V1247\,Ori to be equivalent to a mass accretion rate of $\sim$10$^{-8}$M$_\odot$yr$^{-1}$ onto the primary star by utilising the HARPS data presented by \citet{2013ApJ...768...80K}, calculating the strength of the H$\alpha$ emission and using the parameters presented in Table~\ref{tab:targlist} along with our calculated accretion luminosity.

\subsection{SMA}

\begin{figure}
 \begin{center}
\scriptsize
    \includegraphics[width=8cm, angle=0]{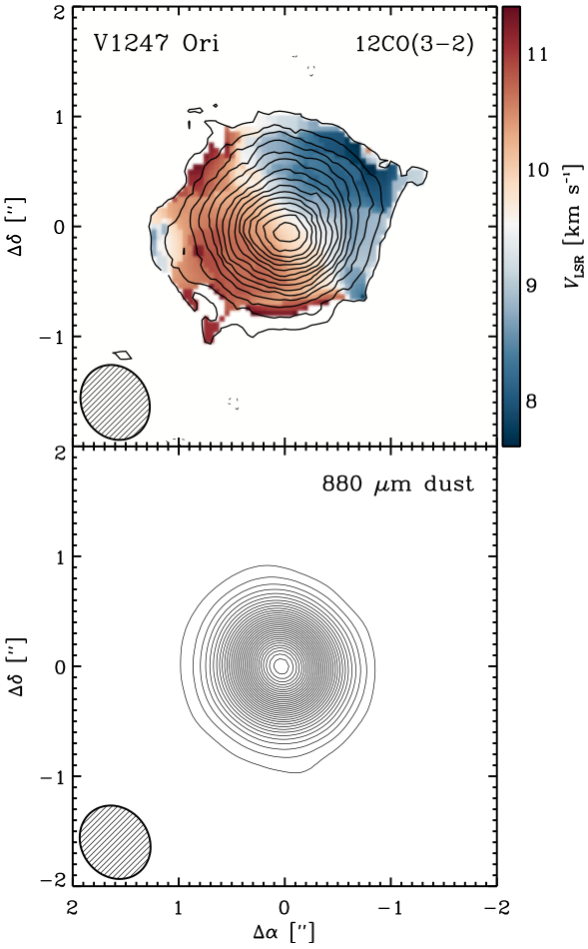} \\
\end{center}
\caption{({top}): Synthesised $4\farcs0\times 4\farcs0$ map of the CO $J$=3$-$2 integrated intensity from the disc around V1247\,Ori (contours increase at 150\,mJy km s$^{-1}$ beam$^{-1}$ intervals) overlaid on the intensity-weighted velocities (colour scale; to show sense of rotation).  ({\it bottom}) The synthesised $4\farcs0\times 4\farcs0$ 880\,$\mu$m continuum map.  Contours increase at 4.25\,mJy beam$^{-1}$ intervals.  The bottom left corners of each map show the synthesised beam dimensions.}
\label{fig:SMAimages}
\end{figure}

\begin{figure}
 \begin{center}
\scriptsize
    \includegraphics[width=8cm, angle=0]{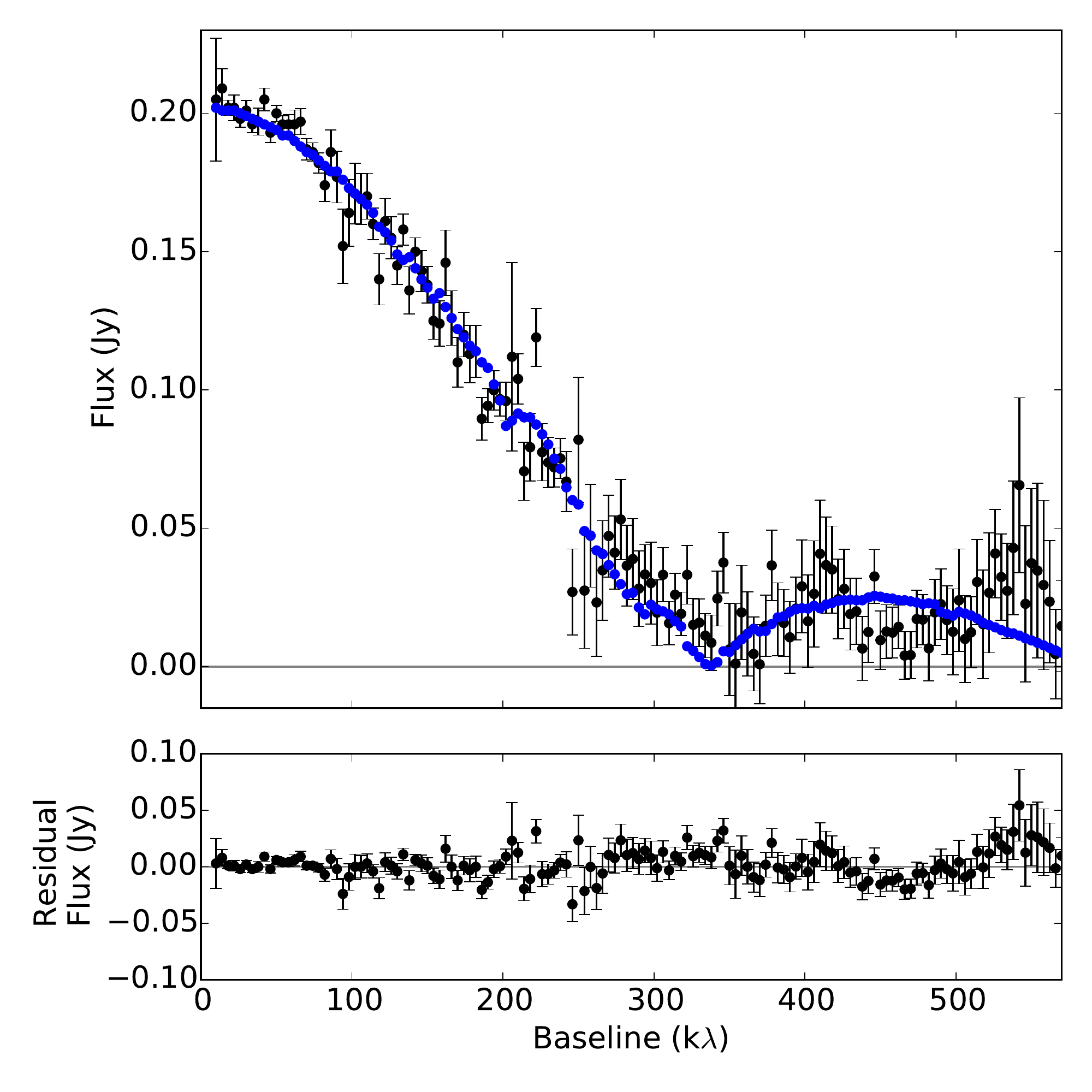} \\
\end{center}
\caption{880~$\mu m$ SMA visibility data with with the baselines deprojected.} The best fit uniform disc profile visibilities are overlaid in blue and the residual visibilities from the fit are presented below.
\label{fig:SMAvisibilities}
\end{figure}

Our SMA sub-millimetre maps on V1247\,Ori are shown in Figure~\ref{fig:SMAimages}.
We reconstructed a 880\,$\mu$m continuum image with natural weighting from all baseline data (Figure \ref{fig:SMAimages}, bottom). Here, once beam shape has been accounted for, we resolve an elongated structure that is oriented along south-east/north-west direction, roughly consistent with the disc position angle derived by our previous infrared interferometric observations \citep{2013ApJ...768...80K}. Fitting a uniform disc model to the continuum visibilities yields a disc position angle of 142$\pm$5$^\circ$. Figure \ref{fig:SMAvisibilities} shows the extracted visibilities, with the deprojected model visibilities overplotted in blue. This value is slightly higher than our estimate of the disc position angle from mid-infrared interferometry \citep[$104 \pm 15^{\circ}$;][]{2013ApJ...768...80K}, but still consistent to within 3$\sigma$. The estimated inclination value of 29$\pm$2$^\circ$ is in line with our previous estimate ($31.3\pm7.5^{\circ}$) within a single standard deviation. The integrated 880\,$\mu$m flux is 0.20$\pm$0.01\,Jy.

The SMA data covers also the $^{12}$CO(3-2) line, where we detect a typical Keplerian disc rotation profile. The position angle of the disc rotation axis (semi-minor axis along position angle 32$\pm$10$^\circ$) matches again the previous observations of the outer disc orientation \citep[$104 \pm 15^{\circ}$;][]{2013ApJ...768...80K} and our own observations with SAM in L'-band (semi-minor axis along position angle 30$\pm$15$^\circ$).

Combining our constraints on the orbital motion of the companion candidate (Section \ref{subsubsec:CompOnly}) with the disc rotation direction measurement in CO allows us also to draw conclusions to the 3-dimensional orientation of the disc.
Our aperture masking observations show that the companion candidate moves from north-west to north-east in an anti-clockwise direction, in other words with increasing position angle in time. However, this alone does not determine the orientation of the disc with respect to the line-of-sight, as the northern arc of the orbit could face either towards us as observer, or away for us.
This ambiguity is removed by our SMA disc rotation measurement, where we see the approaching part of the disc being located north-west of the star, and the receding disc part located south-east of the star.
Assuming that the disc and the companion candidate orbit in the same direction, we conclude that the northern part of the orbit is facing towards us.
Accordingly, we also expect the near side of the rim to be located to the north-east of the star. This is consistent with the strong disc rim feature that we resolve to the north-east in our L'-band SAM observations and that also affects our H+K band observations. This emission is likely tracing forward-scattered light from the near side of the rim.

%

\section{Discussion}
\label{sec:discussion}

Through invoking the presence of a close-in point source-like companion and a disc rim we can explain the unexpectedly fast changes in position angle that are observed in our multi-epoch H- and K-band aperture masking data. This scenario explains also the difficulty in detecting the companion candidate in the third epoch as the companion candidate simultaneously moves closer to the central star as it orbits within the same inclined plane as the outer disc derived from infrared long baseline interferometry while also moving out of the small range in position angle where the fit to the separation is overestimated. This replaces the previous interpretation of the asymmetries seen in \citet{2013ApJ...768...80K} of a single complex structure stretching across the gap, with two simple sources of asymmetry within the intensity distribution.
Evidence for the location and orientation of the disc rim comes from previous observations but also from observations we have presented here including a scattered-light detection in non-coronagraphic SPHERE imaging, the SMA 880\,$\mu$m continuum imaging and the $^{12}$CO(3-2) moment map. The strongest evidence shown here for the disc rim and its properties, in particular its location, comes from the L'-band SAM observations. We see the static structure to the north-north-east in all three epochs which combined into a single set produces a direct image of the rim (Figures~\ref{fig:PotDiskFeatures}~\&~\ref{fig:DiskFeaturesModel}). Fitting to this data set we find a position angle of $ 95\pm15^\circ$ which is consistent with the previous measurements of the disc reported by \citet[][104$\pm$15$^\circ$]{2013ApJ...768...80K}.

The presence of the disc makes a direct fit to the separation of the companion candidate ambiguous. To constrain the separation between V1247\,Ori and the companion candidate, we use the fitted position angles from the two strong companion-like detections (K-band 2012-01-09: H-band 2012-12-18) and the literature value for the mass of V1247\,Ori. Assuming a circular orbit within the plane of the disc we then derive an orbital separation of 6$\pm$1\,au.

Our disc+companion simulations show that in the presence of extended asymmetric disc emission, the flux of a close-in point source retrieved by fitting SAM data is always underestimated.
The difference between the input and retrieved flux can vary up to a factor $\sim 2$ (see Figure~\ref{fig:DiscCompSim}).

From the SPHERE observations we can rule out the presence of an accreting companion at separations beyond 50\,mas.
The lower limits we calculate are heavily dependent upon the assumption that the extinction within the disc is the same as towards the central star.
However this is unlikely to be the case, as earlier observations detected significant amount of optically-thin dust in the gap region \citep{2013ApJ...768...80K}.

%

\section{Conclusions}
\label{sec:conclusions}

We presented multi-wavelength, multi-epoch high-angular resolution observations of the pre-transitional disc V1247\,Ori, with the goal to identify the nature of the asymmetries that have been detected around this object.
Our SAM observations cover three epochs, with a total time between the first and last epoch of 678~days. We observe with H-, K- and L'-band filters using the Keck-II/NIRC2 instrument and the VLT/NACO instrument.
SMA observations in the 880\,$\mu$m continuum and in the $^{12}$CO(3-2) line resolved the dust emission of the disc and its rotation profile. Fitting 2-D Gaussian profiles to the continuum data allowed us to measure the orientation and inclination of the disc, yielding values of 142$\pm$5$^\circ$ and 28$\pm$2$^\circ$ respectively.

Within our L'-band SAM observations we detect a structure consistent with a disc wall across all three epochs. This disc wall is located at $\sim 54$\,au with a position angle and inclination in agreement with the values found previously by \citet{2013ApJ...768...80K}. These values are in line with the disc orientation and rotation observed within our SMA observations but is located towards a position angle which differs by $\sim$15$^\circ$ compared to the long baseline observations.

We find evidence for a bright companion candidate ($\Delta\,m_H=4.8\pm0.2\,\textrm{mag}, \Delta\,m_K=4.7\pm0.2\,\textrm{mag} $) at separation $<45$\,mas from V1247\,Ori.
We identify a new degeneracy that affects the detection of close-in point sources with aperture masking interferometry in the presence of extended disc emission. We conducted detailed simulations to study this effect, where the position of the point source was known and the parameters were then retrieved with standard interferometry fitting routines. We conclude that the presence of the disc emission can lead to an overestimation of the companion separation, in particular if the position angle of the point source and of the disc asymmetry are aligned.
For our V1247\,Ori, we have shown that the visibility and phase data measured at three epochs are consistent with a companion candidate located on a Keplerian orbit at a distance of $\sim$20\,mas (=6\,au) from the star.
We estimate a value of $M \dot M = 10^{-3}$M$_\textrm{J}$ yr$^{-1}$, which is substantially higher than the value found for the confirmed protoplanet LkCa\,15\,b and the protoplanet candidates DM\,Tau\,b and TW\,Hya\,b \citep[$10^{-5}\,M_{\odot}$yr$^{-1}$,][]{2015Natur.527..342S, 2016arXiv160803629W}.
The value found for V1247\,Ori\,b is more in line with the value found for the companion candidate identified around LkH$\alpha$\,330.

Our SPHERE spectral differential imaging does not reach the small separation deduced for the companion candidate detected with aperture masking ($\rho=12$\,mas), but it allowed us to search for an accreting pre-main sequence star at separations $\gtrsim 50$\,mas, which resulted in a non-detection.

V1247\,Ori represents a highly interesting target due to the presence of the companion candidate and the complexity of its inner circumstellar environment. Further study of this target holds the potential to answer many questions on the evolution of circumstellar discs and planet formation. Future observations with extreme-AO instruments and AO-supported aperture masking imaging on the E-ELT will remove much of the ambiguity discussed above, while additionally holding the potential to probe scales of a few au for many nearby transitional disc targets.

\begin{acknowledgements}
We want to thank Julien Rameau for useful discussion on some technical aspects of the paper.

We acknowledge support from an ERC Starting Grant (Grant Agreement No.\ 639889), STFC Rutherford Fellowship (ST/J004030/1), STFC Rutherford Grant (ST/K003445/1), Marie Sklodowska-Curie CIG grant (Ref.\ 618910), and Philip Leverhulme Prize (PLP-2013-110).

We additionally acknowledge support from NASA KPDA grants (JPL-1452321, 1474717, 1485953, 1496788).

The Submillimeter Array is a joint project between the Smithsonian Astrophysical Observatory and the Academia Sinica Institute of Astronomy and Astrophysics and is funded by the Smithsonian Institution and the Academia Sinica.
The authors wish to recognise and acknowledge the very significant cultural role and reverence that the summit of Mauna Kea has always had within the indigenous Hawaiian community. We are most fortunate to have the opportunity to conduct observations from this mountain.Some of the data
presented herein were obtained at the W.M.\ Keck Observatory, which is operated as a scientific partnership among the California
Institute of Technology, the University of California and the National Aeronautics and Space Administration. The Observatory was
made possible by the generous financial support of the W.M.\ Keck Foundation.
J.\ Kluska acknowledges support from the research council of the KU Leuven under grant number C14/17/082.
M.\ Cur\'e and S.\ Kanaan acknowledge financial support from Centro de Astrofísica de Valparaiso.
S.\ Kanaan thank the support of Fondecyt iniciacíon grant No.\ 11130702.
T.\ Muto is partially supported by JSPS KAKENHI grant No.\ 26800106.

Finally we wish to thank the referee for their feedback and suggestions. We thank them for their time and efforts in enabling us to publish this work.
\end{acknowledgements}

\bibliography{aa.bbl}{}
\bibliographystyle{aa.bst}

\begin{appendix}

\section{Supplementary figures}

\begin{figure}
 \begin{center}
\scriptsize
$\begin{array}{ @{\hspace{-1.0mm}} c @{\hspace{-10.0mm}} c}
    \includegraphics[height=4cm, angle=0]{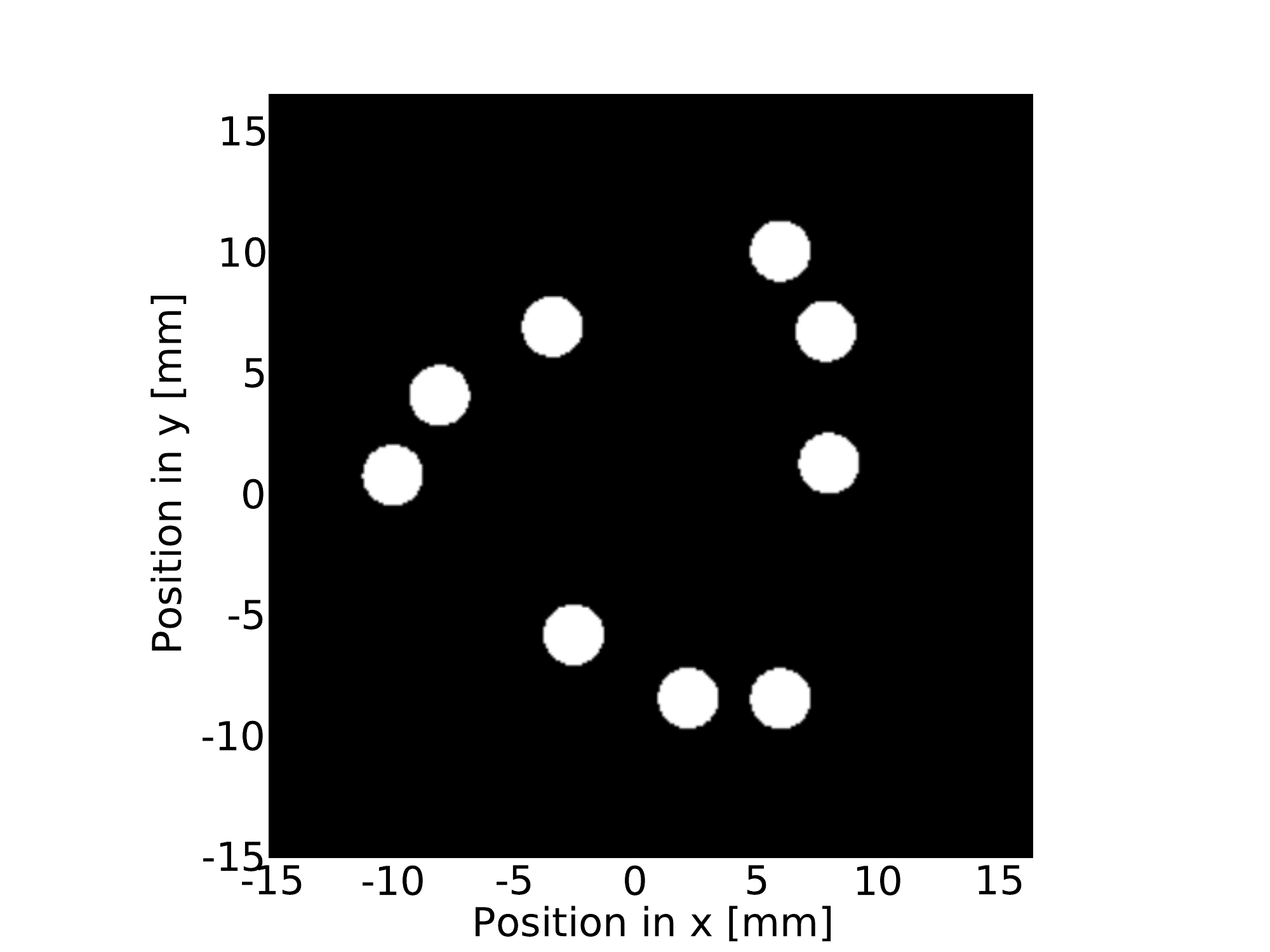} & \includegraphics[height=4cm, angle=0]{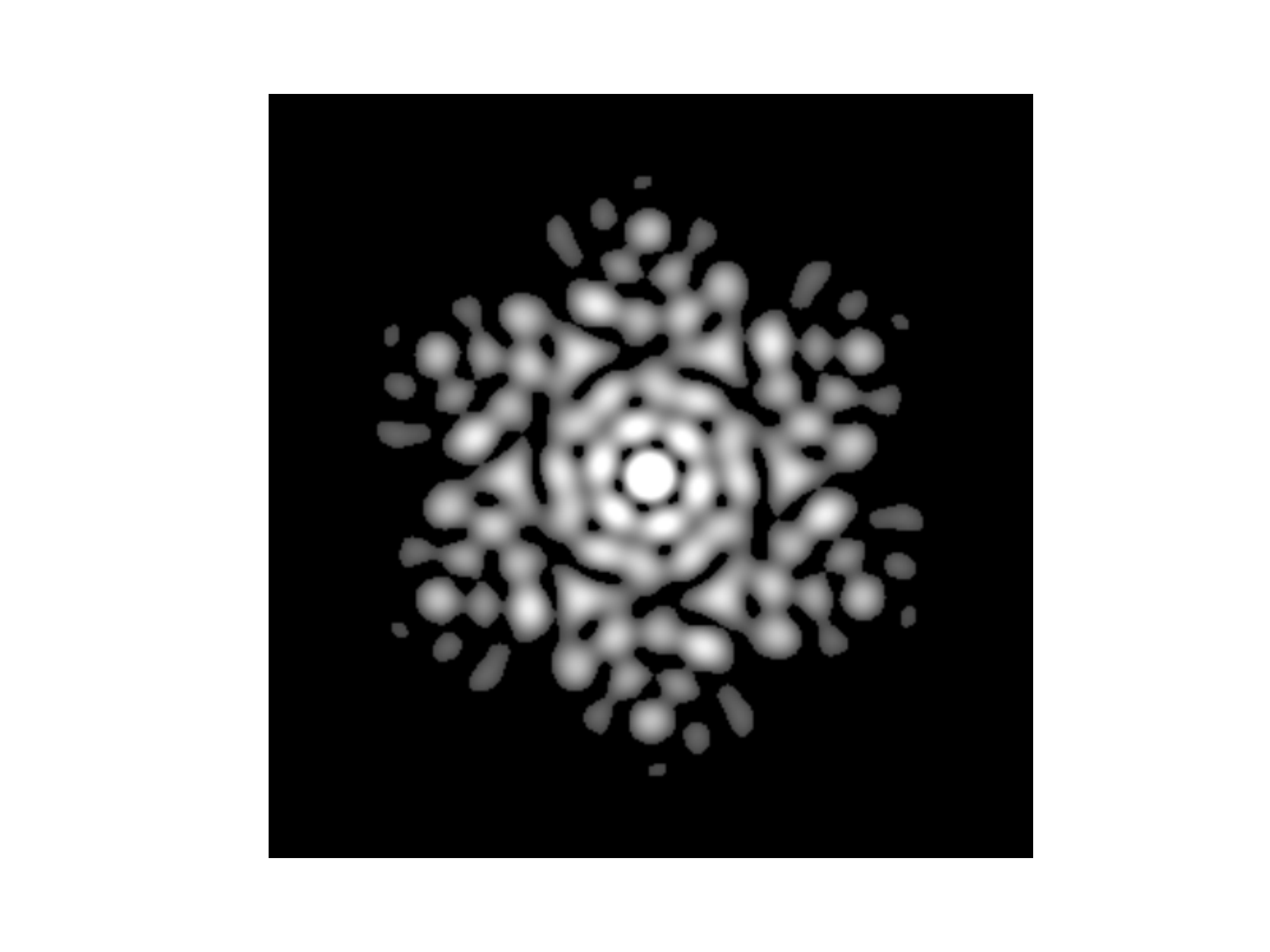}\\
 \end{array}$
\end{center}
\caption{Keck-II/NIRC2 instrument 9-hole mask details. \textbf{Left:} Mask geometry. Mask diameter approximately 32mm, mask holes 2.5mm in diameter. \textbf{Right:} Power spectrum produced by mask shown to the left.}
\label{fig:keck9hMask}
\end{figure}

\begin{figure*}
\begin{center}
\scriptsize
$\begin{array}{ @{\hspace{-1.0mm}} c @{\hspace{-5.0mm}} c @{\hspace{-5.0mm}} c}
    \includegraphics[height=4cm, angle=0]{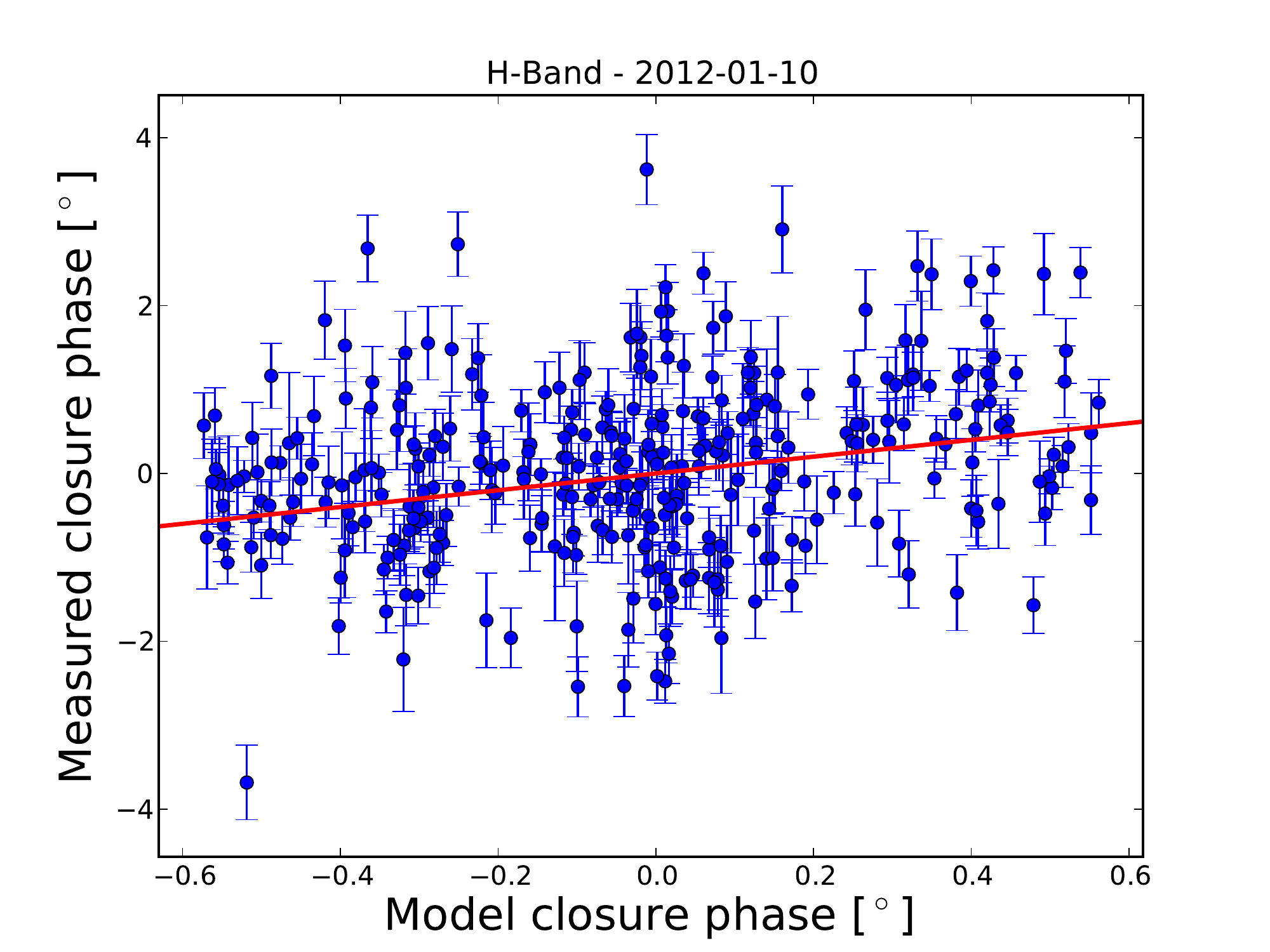} & \includegraphics[height=4cm, angle=0]{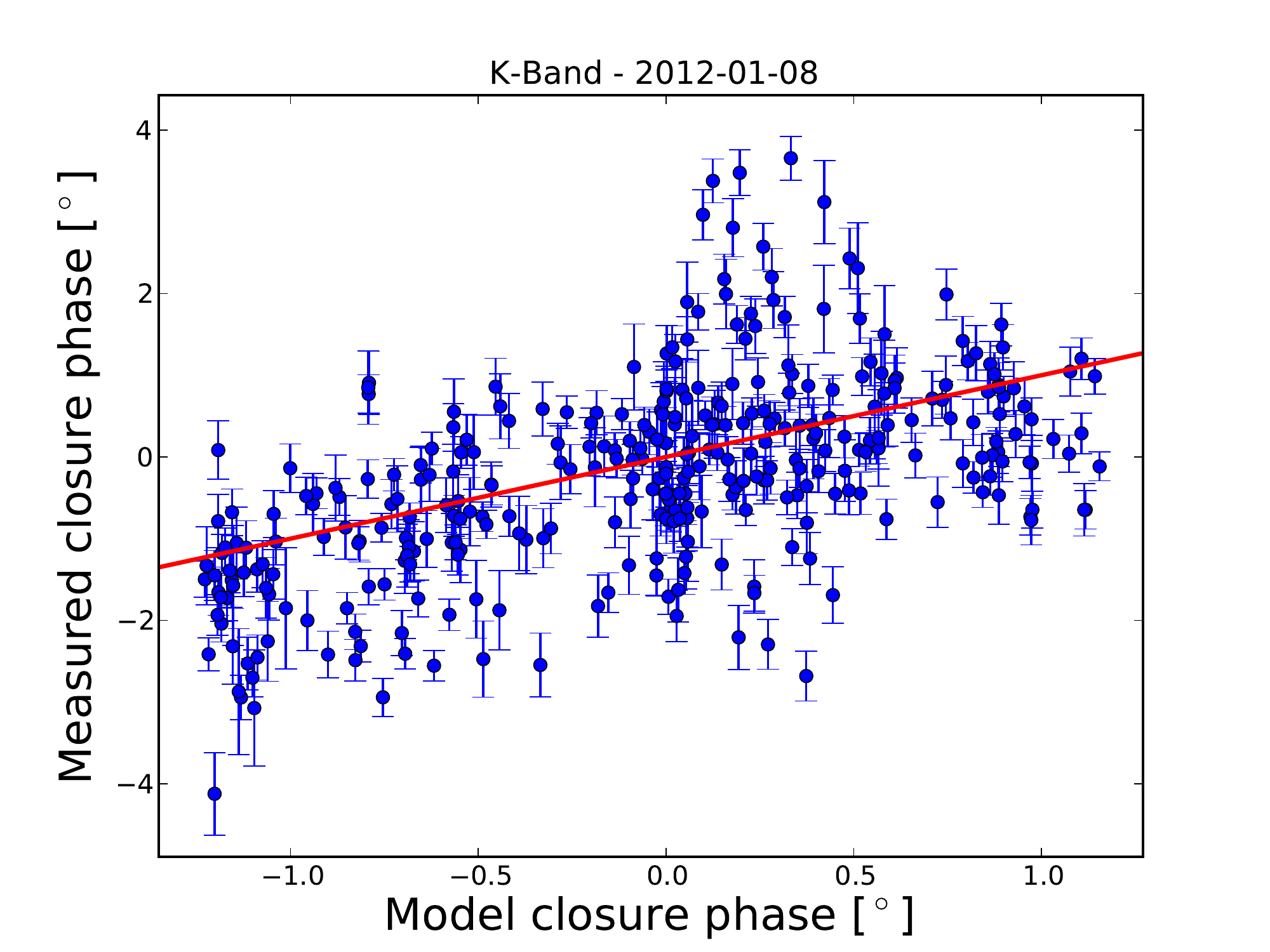} & \includegraphics[height=4cm, angle=0]{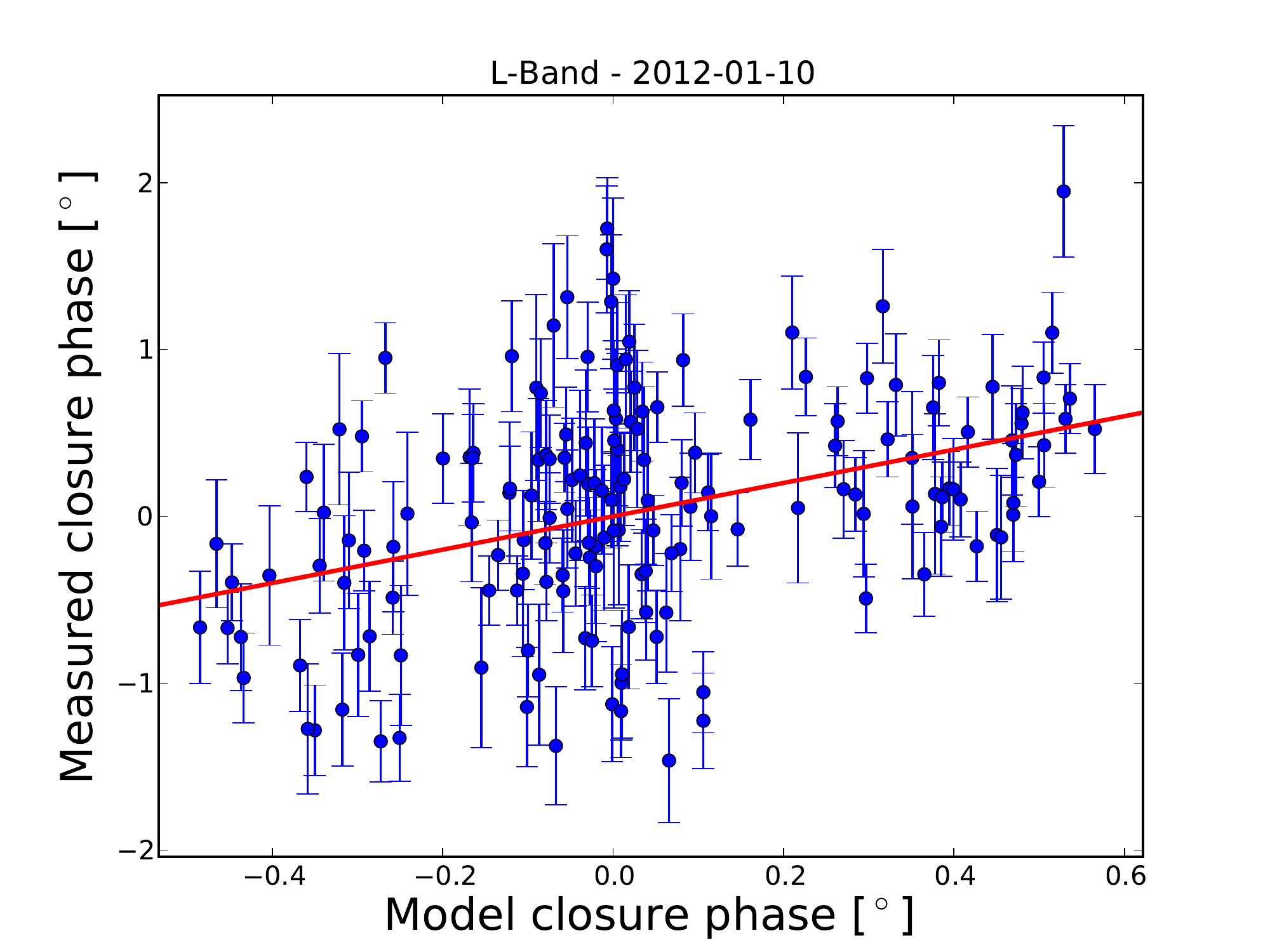}\\
    \includegraphics[height=4cm, angle=0]{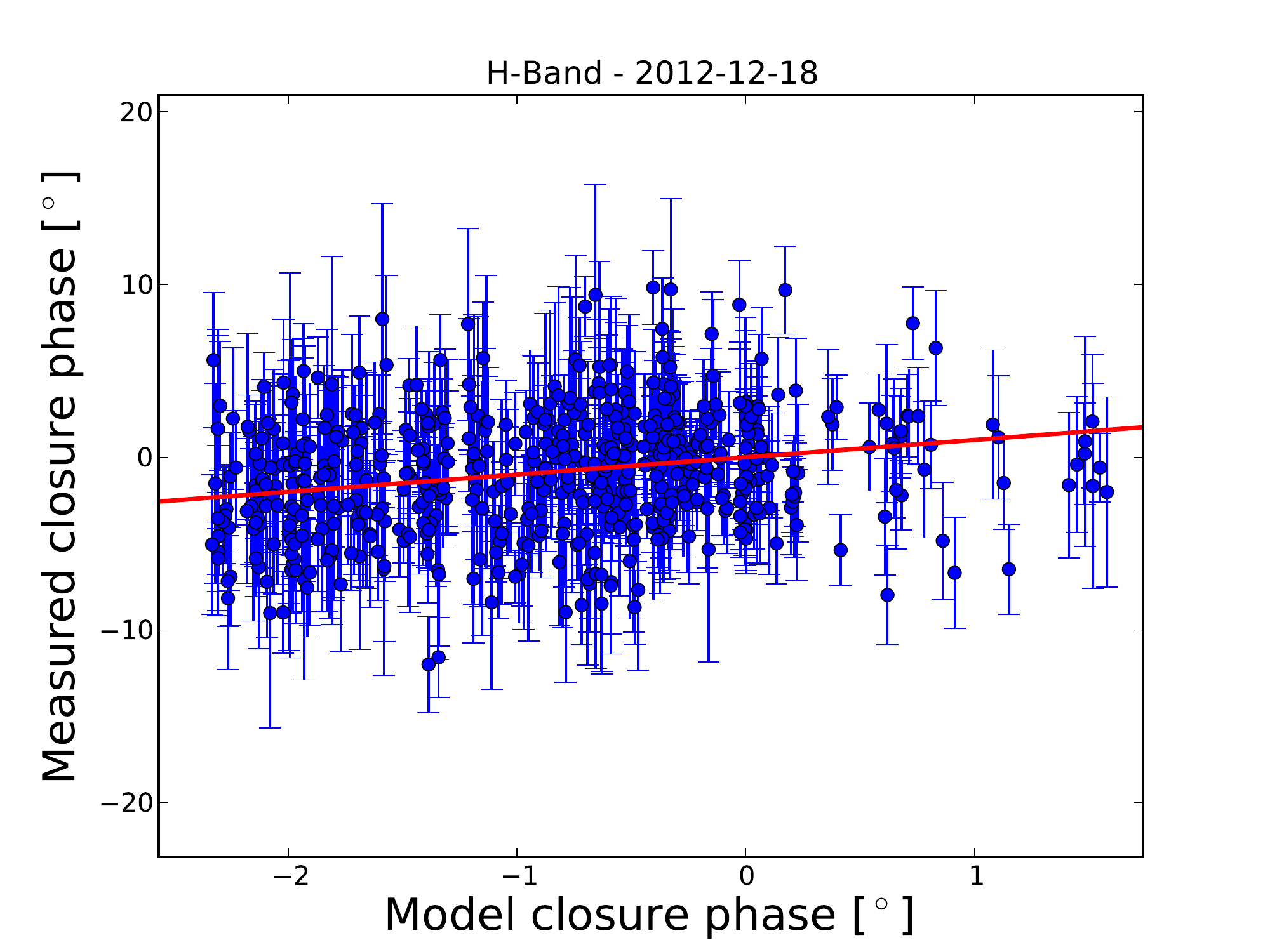} & \includegraphics[height=4cm, angle=0]{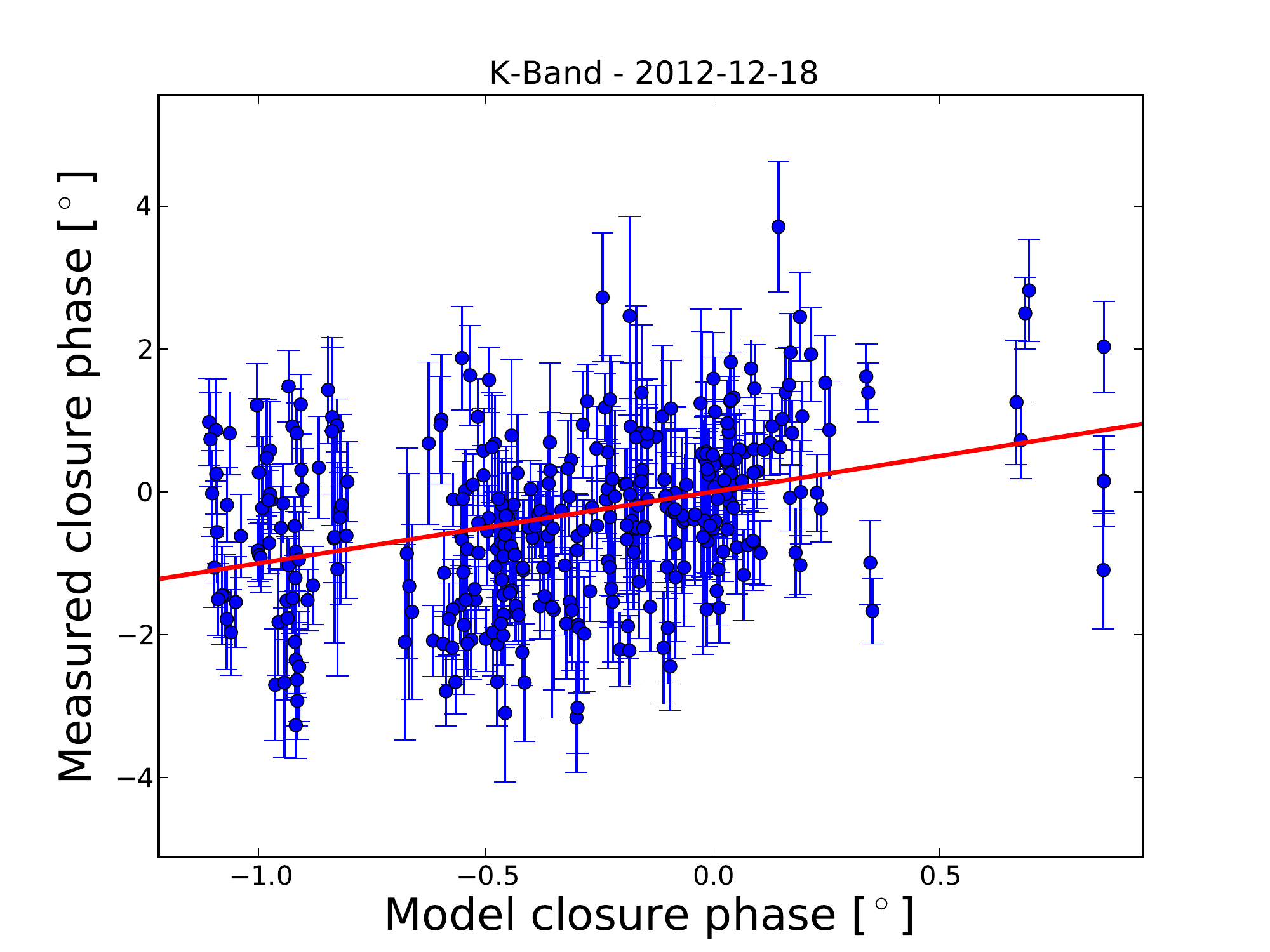} & \includegraphics[height=4cm, angle=0]{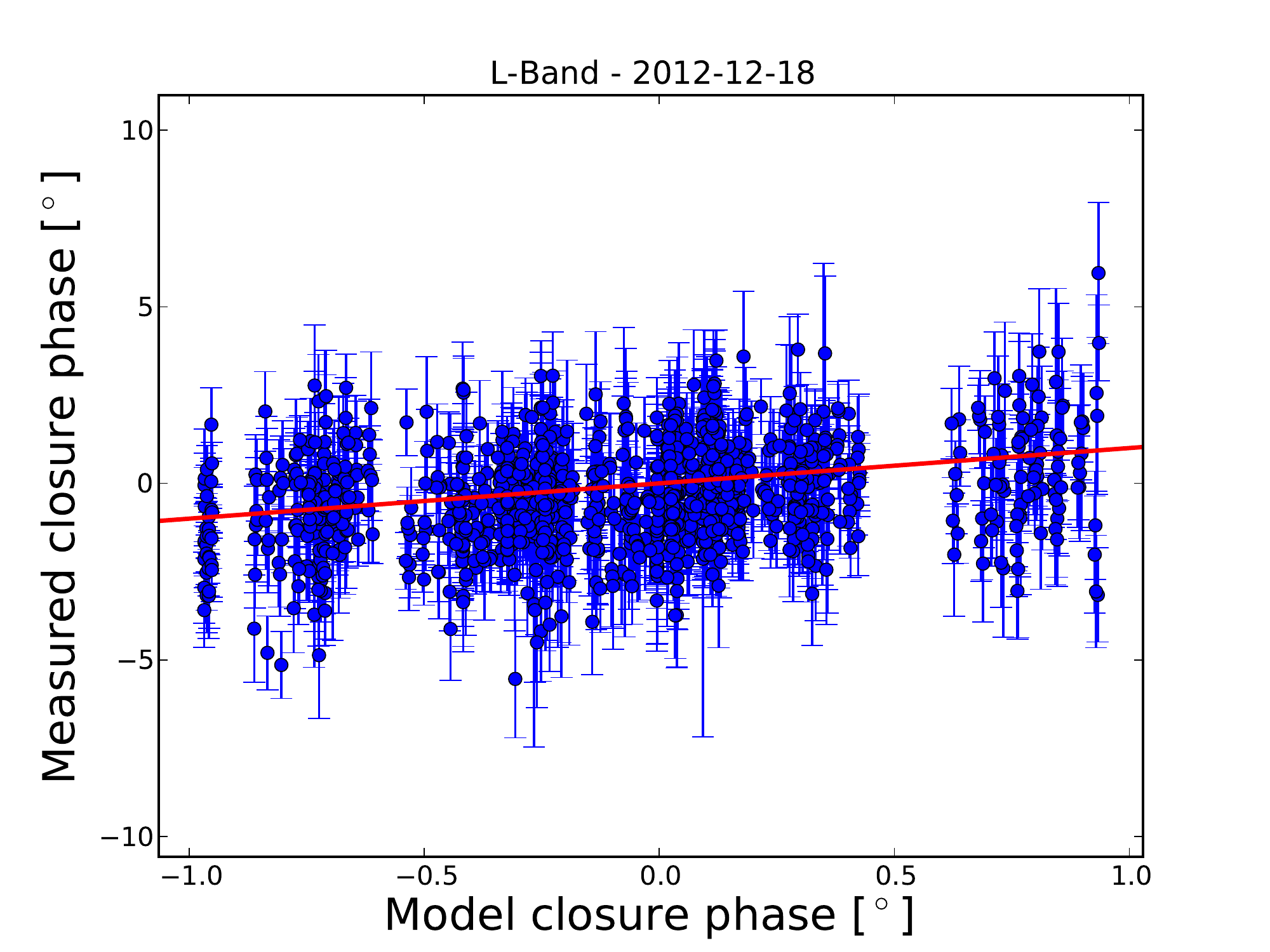}\\
     &
     \includegraphics[height=4cm, angle=0]{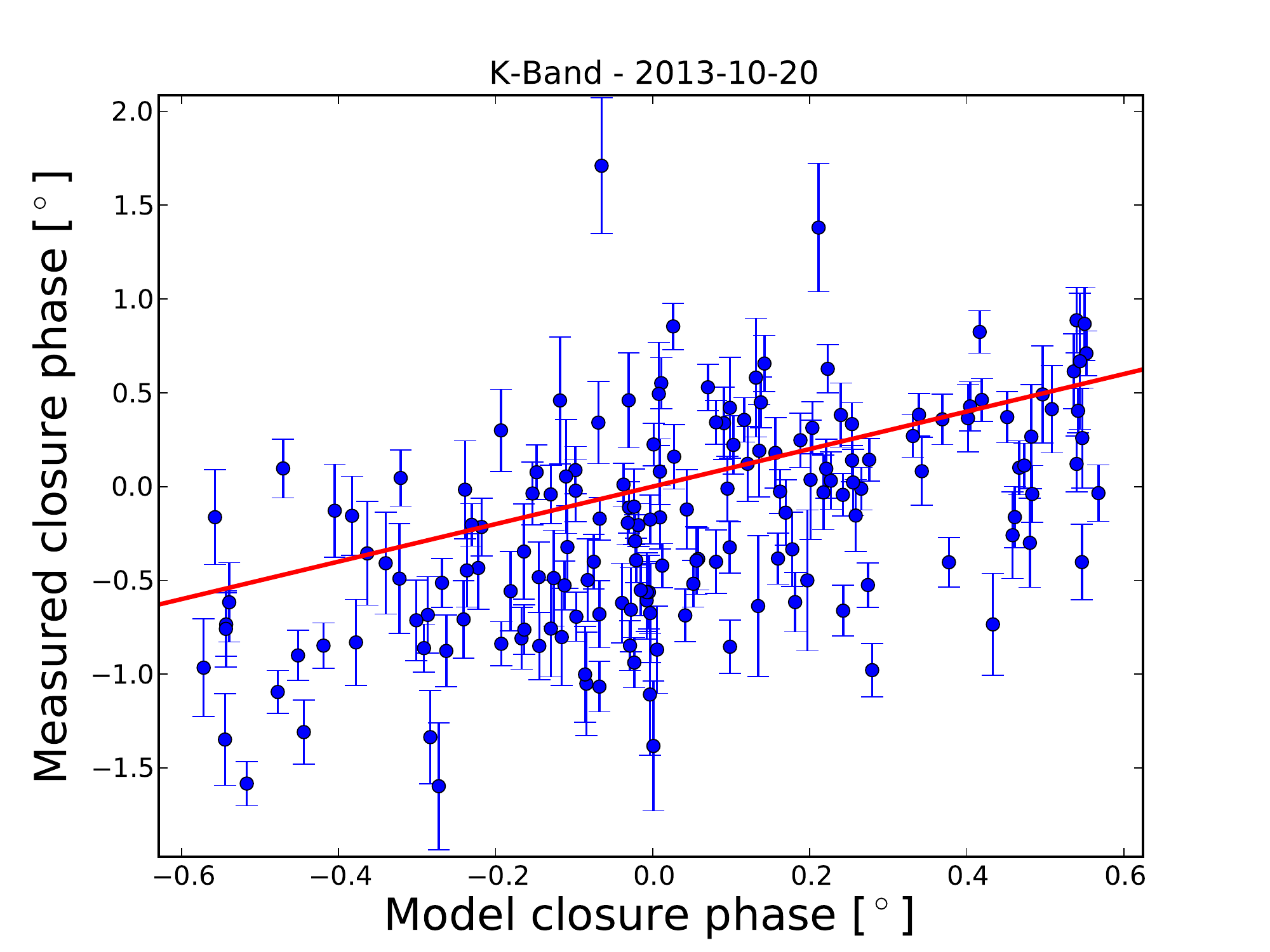} &
     \includegraphics[height=4cm, angle=0]{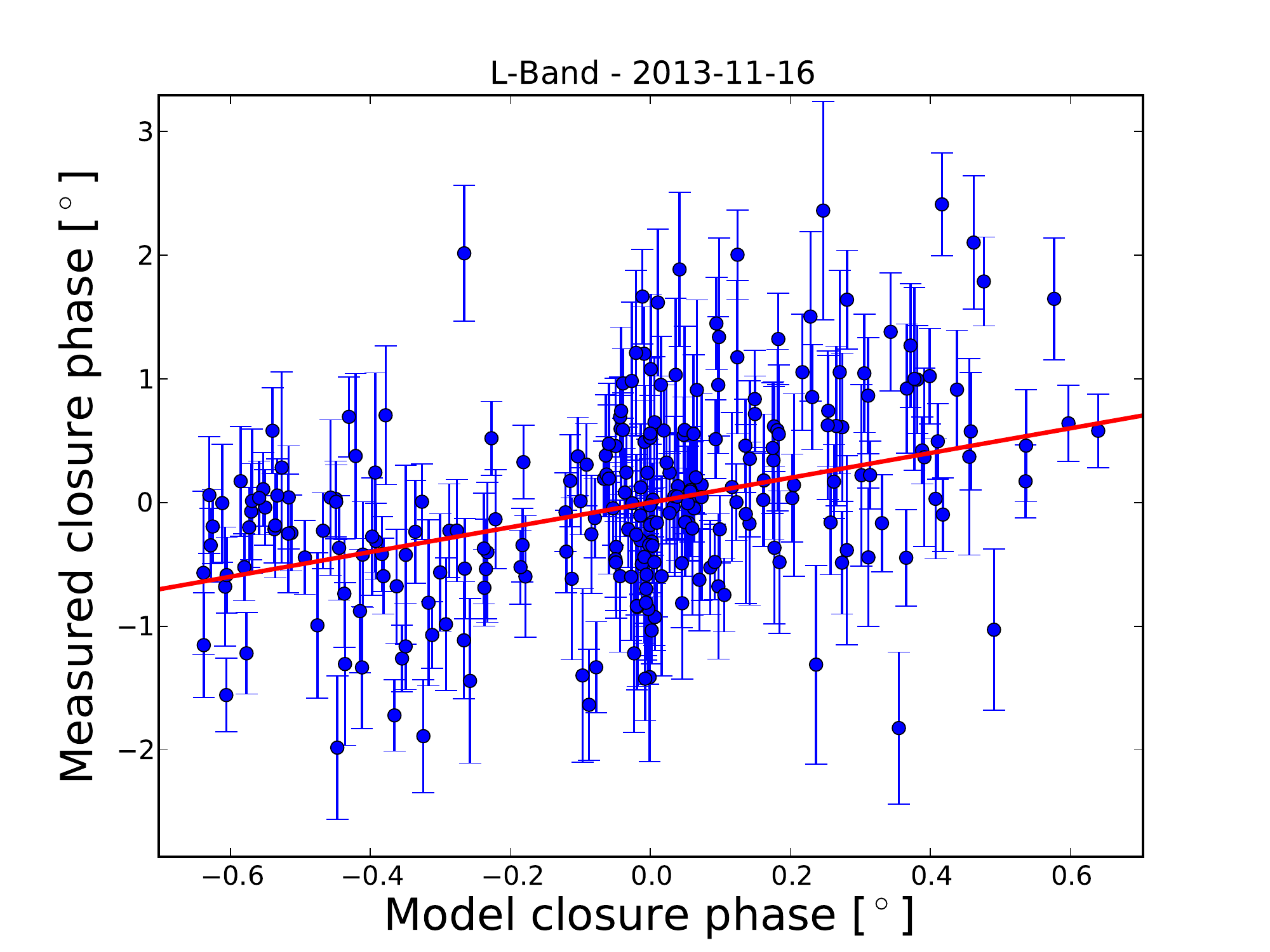} \\
 \end{array}$
\end{center}
\caption{Correlation diagram containing the V1247\,Ori measured closure phases plotted against best fit binary model closure phases. Data is arranged by epoch by row and wavelength by column. Overlaid is the best fit binary model for each data set representing a line with a gradient equal to 1.0.}
\label{fig:cpData}
\end{figure*}

\begin{figure*}
\begin{center}
\scriptsize
    \includegraphics[height=15cm, angle=0]{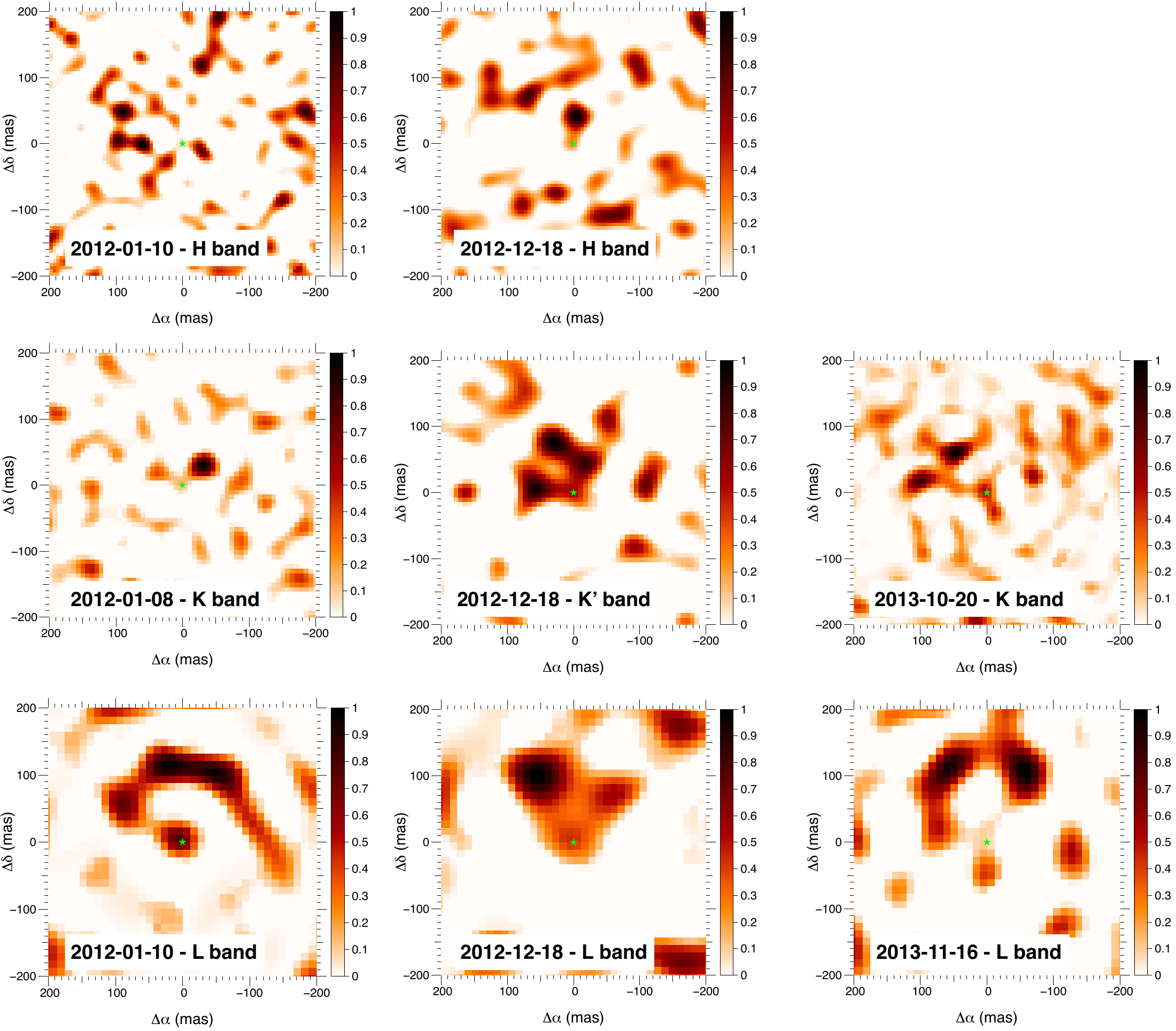}
\end{center}
\caption{Image reconstructions selected by the minimum of the global cost function $\mathscr{F}$. The central star is represented by the green star. The colour bars show the flux contained in each pixel as a fraction of the peak flux in the frame after subtracting the central star.}
\label{fig:img_f}
\end{figure*}

\begin{figure*}
\begin{center}
\scriptsize
    \includegraphics[height=6cm, angle=0]{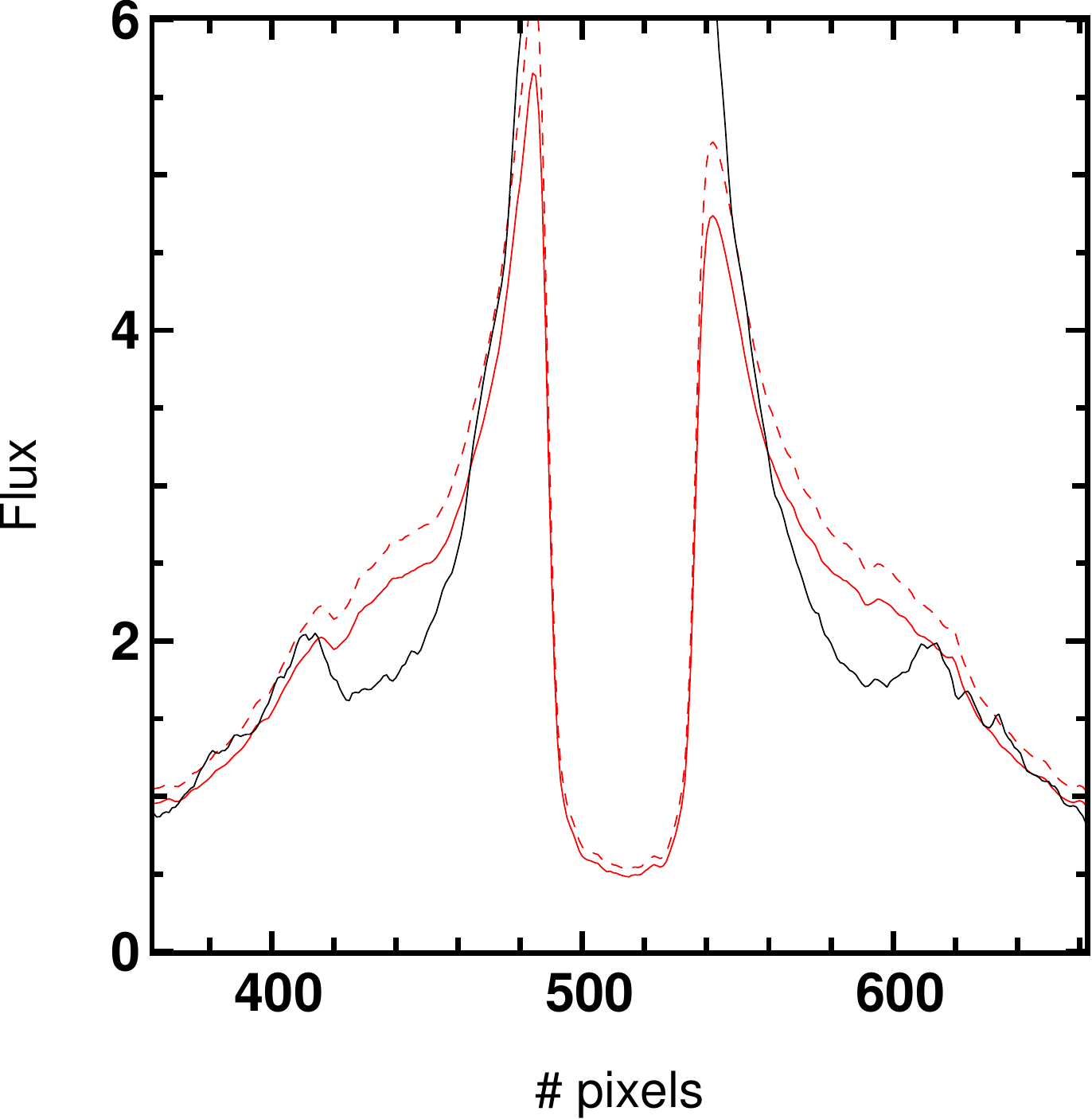}
\end{center}
\caption{Profiles of the SPHERE images with (solid red line) and without (solid black line) coronagraph. The dashed red line represents the cut of a coronagraphic image scaled to the non-coronagraphic one with a factor of 1.1.}
\label{fig:profiles}
\end{figure*}

\end{appendix}

\end{document}